\documentclass[%
 reprint,
superscriptaddress,
 amsmath,amssymb,
 aps,
prx,
]{revtex4-2}
\usepackage{physics} 
\usepackage{graphicx}% Include figure files
\usepackage{dcolumn}% Align table columns on decimal point
\usepackage{bm}% bold math
\usepackage{mathtools} % for =
\usepackage{stmaryrd}
\usepackage{amssymb} 
\usepackage[caption=false]{subfig}

\usepackage{float}

\usepackage[percent]{overpic}

\usepackage{lipsum}
\usepackage[english]{babel}

\usepackage{appendix}

\usepackage{pifont}
\newcommand{\anglevec}[1]{\vec{\theta}^{(#1)}}
\newcommand{\rotangle}[2]{\theta^{(#1)}_{#2}}
\newcommand{\avg}[1]{\langle#1\rangle}
\newcommand{\bigavg}[1]{\big\langle#1\big\rangle}
\newcommand{\Bigavg}[1]{\Big\langle#1\Big\rangle}

\newcommand{\LIF}{\epsilon}
\newcommand{\logicalZ}{\hat{\bar Z}_L}
\newcommand{\logicalZsuperop}{\mathcal{\bar Z}_L}
\newcommand{\logicalZLZ}{\mathcal{H}}

\usepackage[colorlinks=true,
            linkcolor=blue,
            citecolor=blue,
            urlcolor=blue]{hyperref}
\usepackage{xcolor}
\usepackage{tikz}
\usetikzlibrary{shapes.geometric}

\definecolor{colConventional}{HTML}{000000}
\definecolor{colPEC}         {HTML}{17A1F1}
\definecolor{colLT}          {HTML}{228833}
\definecolor{colDD}          {HTML}{013CFD}   % yellow???Maybe orange -  check.
\definecolor{colPT}          {HTML}{AA4499}
\definecolor{colPECLT}       {HTML}{ED6E0D}

\newlength{\mkrBL}

\newcommand{\iconConventional}{}

\newcommand{\iconLT}          {}

\newcommand{\iconPECLT}       {}

\usepackage{xcolor}

\usepackage[final]{microtype}
\usepackage{hyphenat}
\let\oldaffiliation\affiliation
\renewcommand{\affiliation}[1]{\oldaffiliation{\nohyphens{#1}}}

\begin{document}
\preprint{APS/123-QED}

% \title{Correlated Coherent Errors in Stabilizer Codes: \\ Role of the Encoding Eigenspace and Noise Correlations as a Resource}

\title{Correlated Coherent Errors in Stabilizer Codes: A General Cumulant Framework and 
Interference-Based Error Suppression}

\author{Rohan N Rajmohan}
\email {rohanrajmohan2028@u.northwestern.edu}
\affiliation{Department of Physics and Astronomy, Northwestern University, Evanston, Illinois 60208, USA}

\author{Antoine Brillant}
\affiliation{Pritzker School of Molecular Engineering, University of Chicago, Chicago, IL, USA}

\author{Peter Groszkowski}
\affiliation{National Center for Computational Sciences, Oak Ridge National Laboratory, Oak Ridge, TN, USA}

\author{Alireza Seif}
\affiliation{IBM Quantum, IBM T.J. Watson Research Center, Yorktown Heights, NY, USA}

\author{Jens Koch}
\affiliation{Department of Physics and Astronomy, Northwestern University, Evanston, Illinois 60208, USA}

\author{Aashish Clerk}
\email {aaclerk@uchicago.edu}
\affiliation{Pritzker School of Molecular Engineering, University of Chicago, Chicago, IL, USA}

% \begin{abstract}
% Stabilizer codes are central to quantum error correction (QEC). Motivated by experimental implementations, their behavior under coherent errors has attracted growing attention. These errors can be correlated across qubits and QEC cycles, for example due to slow frequency drifts or control crosstalk. Despite their prevalence in current hardware, the impact of these correlations remains largely unexplored. Here, we develop a cumulant expansion for assessing correlated coherent errors in arbitrary stabilizer codes. We show that their effect generally depends on the stabilizer eigenspace chosen as the codespace. Motivated by this dependence, we introduce protected stabilizer eigenspace (PROSE) encoding as an error suppression technique. We find that PROSE encoding can be combined with logical Pauli twirling, yielding a protocol that matches or outperforms other standard suppression techniques in practically relevant regimes. Remarkably, once suitable suppression techniques are applied, noise correlations, typically considered detrimental to QEC, become a resource instead. Our framework, assessment of suppression strategies, and perspective on noise correlations open new directions for understanding correlated coherent errors in stabilizer codes.
% \end{abstract}

\begin{abstract}
Coherent errors in stabilizer codes are often correlated across qubits and QEC cycles.  Having a general analytical treatment of such noise would thus be extremely valuable.  We derive here the exact logical channel induced by repeated QEC cycles under correlated coherent $Z$ noise, and 
develop a broadly general cumulant-expansion framework that yields a tractable expression for the noise-averaged logical infidelity.  Crucially, this expression is non-perturbative in the noise, and applies to arbitrary stabilizer codes and correlation structures.  It reveals a feature with no analogue in standard stochastic Pauli error models: the induced channel depends on which stabilizer eigenspace is chosen as the codespace. Exploiting this, we introduce protected stabilizer eigenspace (PROSE) encoding, an error-suppression strategy that selects the optimal codespace.  We show that this eigenspace can be efficiently identified in many relevant situations.  Further, when combined with logical Pauli twirling, PROSE matches or outperforms standard error suppression techniques (dynamical decoupling, Pauli twirling of physical qubits).  We also show that noise correlations, usually assumed to be harmful to QEC, can instead be a resource: with the right encoding, even positive correlations reduce the logical infidelity below the uncorrelated baseline. Our results offer a new, broadly applicable lens on correlated coherent noise in stabilizer codes.
\end{abstract}

\maketitle

\newpage

\section{Introduction}
\label{sec:introduction}

Quantum computing promises speedups in solving certain problems~\cite{shor_polynomial-time_1995,harrow_quantum_2009}, contingent on being able to suppress hardware errors. Quantum error correction (QEC) addresses this challenge by encoding a small number of logical qubits into many physical qubits. Perhaps the most typical approach to QEC employs stabilizer codes~\cite{gottesman_stabilizer_1997}, in which quantum information is stored in the joint eigenspace of a commutative Pauli subgroup, and errors are diagnosed by measuring group generators. This structure facilitates the simulation and analysis of  stochastic Pauli errors. Such noise models consequently serve as the standard starting point for assessing the performance of such codes \cite{calderbank_quantum_1997, raussendorf_fault-tolerant_2007, wootton_high_2012, tuckett_ultrahigh_2018}.

Stochastic Pauli error models, however, do not capture all errors relevant to current quantum computing platforms. Coherent errors, such as systematic overrotations, can produce effects that differ qualitatively from those of stochastic Pauli errors \cite{greenbaum_modeling_2017, bravyi_correcting_2018, huang_performance_2019}. Further, the rotation angles of such coherent errors may be correlated in both space and time. For example, in many platforms, dephasing arises from slowly fluctuating noise or control fields that persist over many QEC cycles and affect multiple qubits. Despite the prevalence of such correlated coherent errors in current hardware \cite{kumar_origin_2016, rower_evolution_2023, puebla_protected_2017}, their impact remains relatively underexplored. Recent studies have provided valuable insight, but these are largely numerical, or tied to specific codes and correlation structures \cite{pataki_coherent_2024, witzel_correcting_2026}. A general analytical framework is currently lacking.

\begin{figure*}[t]
  \centering
  \includegraphics[width=\textwidth]{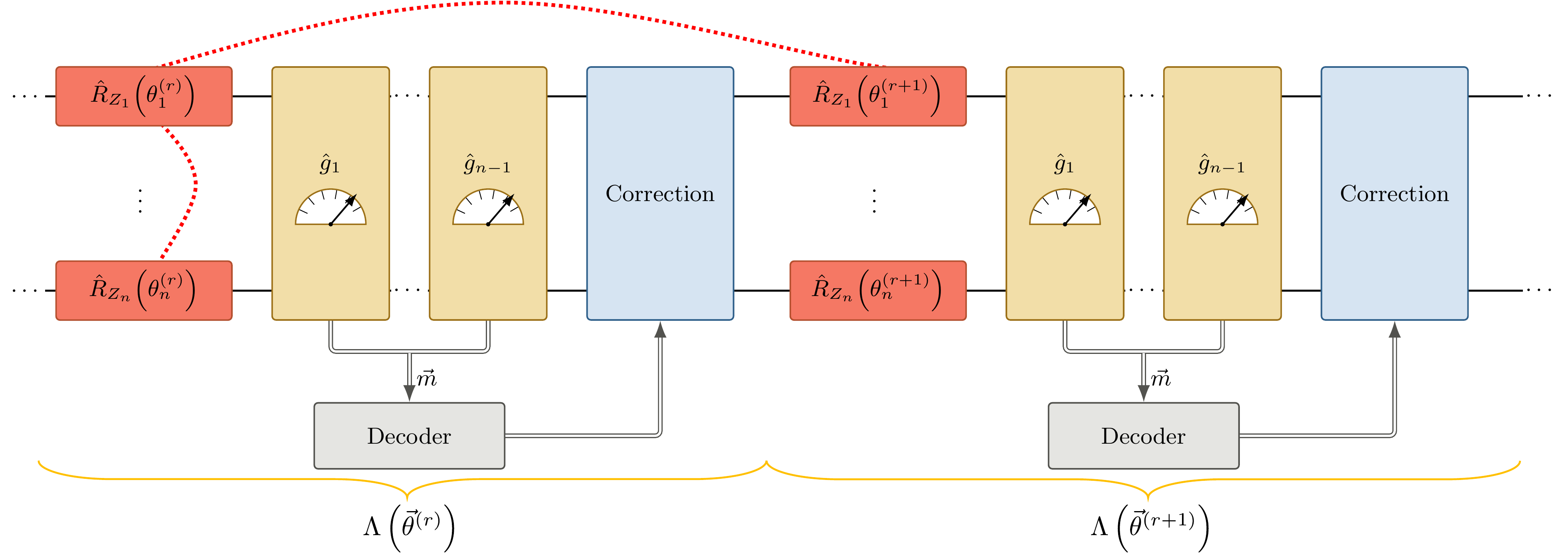}
\caption{
\textbf{Quantum-circuit representation of two consecutive QEC cycles for a stabilizer code.} In each cycle $r$, errors on the $n$ data 
qubits (red boxes) are modeled as coherent $Z$ rotations, $\hat R_{Z_i}(\theta_i^{(r)})$ [Eq.~\eqref{eq:overrotations}], where the stochastic rotation angles may be correlated both across qubits and across cycles (dashed red lines). 
The stabilizer generators \(\hat g_i\) are measured directly (yellow boxes), and the resulting syndrome $\vec{m}$ is decoded using a maximum-likelihood decoder.  This then determines the recovery operation (blue box). For a fixed realization of the correlated noise, the action of each complete QEC cycle on the data qubits, restricted to a stabilizer eigenspace, is described by an effective logical channel [Eq.~\eqref{eq:single_cyc_channel}]. The two cycles shown induce the logical channels \(\Lambda(\vec{\theta}^{(r)})\) and \(\Lambda(\vec{\theta}^{(r+1)})\), respectively.}
  \label{fig:noise_model}
\end{figure*}

In this work, we develop such a framework, in which correlated $Z$ noise acts only on data qubits, while the QEC gadget\textemdash comprising the syndrome measurement and conditional recovery\textemdash is assumed to be noiseless. This setting captures the intrinsic response of the error correction mechanism to correlated coherent errors, without additional interplay with QEC imperfections. We first derive a simple expression for the exact channel induced on the codespace by $R$ cycles of QEC, for a fixed realization of the noise. We then combine this with a cumulant-expansion method to describe the noise-averaged logical infidelity for weak noise.  This allows us to go beyond leading-order perturbation theory in the noise (as is typically done, e.g., in Ref.~\cite{witzel_correcting_2026}), and lets us treat varying degrees and patterns of noise correlations.  These results reveal that the induced channel and the logical infidelity are strongly sensitive to which stabilizer eigenspace is chosen as the codespace, providing analytical insight into observations made in Refs.~\cite{debroy_optimizing_2021,witzel_correcting_2026}. For Gaussian noise, our approach shows that the physics is sensitive to the parity of the code's $Z$ distance, but is otherwise universal across stabilizer codes.  

Our work also uses these analytic tools to develop and assess strategies for enhancing logical fidelity in the presence of correlated coherent Pauli noise.  This includes a new approach we term PROSE (PROtected Stabilizer Eigenspace) encoding, something we compare against standard techniques that suppress coherent errors: logical dynamical decoupling, logical Pauli twirling, and physical Pauli twirling. We clarify which features of correlated coherent errors each technique suppresses and identify regimes in which they outperform QEC-only.  Contrary to standard expectations, 
we find that noise correlations are not intrinsically detrimental: with a suitable choice of encoding eigenspace or suppression technique, even positive noise correlations can be exploited as a resource to reduce the logical infidelity below the uncorrelated limit.
We also show that PROSE encoding and logical Pauli twirling can be used complementarily. Combining them yields a protocol that performs comparably to, or better than, other techniques considered here, in practically relevant regimes such as the small-infidelity regime under stationary noise.

The rest of this paper is organized as follows. Section~\ref{subsec:noise_model} introduces our noise model. Section~\ref{subsec:exact_analysis} outlines the derivation of the exact channel induced on the codespace by repeated QEC cycles for a single noise realization. Section~\ref{subsec:approximation} develops a cumulant expansion of the logical infidelity and specializes it to Gaussian noise correlations. In Sec.~\ref{sec:interpretation}, we use this expansion to assess how Gaussian-correlated coherent errors affect QEC performance. Section~\ref{sec:error_suppression_techniques} introduces and analyzes error suppression strategies: PROSE encoding, logical dynamical decoupling, logical Pauli twirling, and physical Pauli twirling. In Sec.~\ref{sec:comparison}, we combine logical Pauli twirling with PROSE encoding and compare the resulting protocol with the other suppression techniques. Finally, we conclude in Sec.~\ref{sec:conclusion}.

\section{Noise model}
\label{subsec:noise_model}

We consider dephasing about the $Z$ axis of each data qubit, described by the Pauli operator $\hat Z_\ell$, where $\ell$ indexes the data qubits. Concretely, during each QEC cycle $r$, qubit $\ell$ undergoes a coherent $Z$ error
\begin{equation}
    \hat{R}_{Z_\ell}(\theta_\ell^{(r)})
    = \exp\!\left(-i\hat{Z}_\ell\,\theta_\ell^{(r)}\right).
    \label{eq:overrotations}
\end{equation}
The error angles $\theta_\ell^{(r)}$ are random and allowed to be correlated in space and time.  Our focus is on how such correlations affect QEC performance.

For a fixed noise realization, the action of the noise on $n$ data qubits during cycle $r$ is specified by the superoperator
\begin{align}
  \mathcal{N}\!\big(\anglevec{r}\big)[\cdot]
  &= \prod_{\ell=1}^{n} \mathcal{R}_{Z_\ell}\!\big(\theta_{\ell}^{(r)}\big)[\cdot].
  \label{eq:noise_single_cycle}
\end{align}
Here, we collect the angles in cycle $r$ into the vector $\vec\theta^{(r)}=(\theta_1^{(r)},\ldots,\theta_n^{(r)})$ and define
\begin{equation}
  \mathcal{R}_{Z_\ell}\!\big(\theta_\ell^{(r)}\big)[\cdot]
  = \hat{R}_{Z_\ell}\!\big(\theta_\ell^{(r)}\big)\,[\cdot]\,
    \hat{R}_{Z_\ell}^\dagger\!\big(\theta_\ell^{(r)}\big).
\end{equation}
It is useful to expand in the $Z$-type Pauli basis. To fix our notation, let $\mathbb{P}_n$ denote the $n$-qubit Pauli group, including phases, and $\overline{\mathbb{P}}_n$ denote its phase $+1$ representatives:
\begin{equation}
    \overline{\mathbb{P}}_n
    =
    \left\{
        \hat P_1\otimes\cdots\otimes \hat P_n
    \right\},
\end{equation}
where each $\hat P_\ell$ is a Pauli matrix. More generally, for any $\mathbb{T}_n\subseteq\mathbb{P}_n$, $\overline{\mathbb{T}}_n$ shall denote the corresponding set of phase $+1$ representatives. We use $\overline{\mathbb{P}}_{n,Z}\subseteq\overline{\mathbb{P}}_n$ for the $Z$-type subset, with $\hat P_\ell\in\{I,Z\}$.  With this notation, the superoperator can be expanded as:
\begin{align}
\mathcal{N}\!\big(\anglevec{r}\big)[\cdot]
&= \sum_{\mathclap{\hat E,\hat E'\in\overline{\mathbb{P}}_{n,Z}}}
c_{\hat E}(\anglevec{r})\,c_{\hat E'}^*\!(\anglevec{r})\;
\hat E\,[\cdot]\,\hat E',
\label{eq:expanded_noise_superoperator}
\end{align}
where
\begin{equation}
c_{\hat E}(\anglevec{r})
=\!\!\!\prod_{\ell\notin\mathrm{supp}(\hat E)}\!\!\!
\cos(\rotangle{r}{\ell})
\prod_{\ell\in\mathrm{supp}(\hat E)}\!\!\!
\Big(-i\sin(\rotangle{r}{\ell})\Big)
\label{eq:cE_def}
\end{equation}
is the amplitude of the Pauli $Z$ error $\hat E$ in cycle $r$, and $\mathrm{supp}(\hat E)$ is the set of qubits on which $\hat E$ acts nontrivially.

Over \(R\) QEC cycles, a single realization of the error angles is specified by $\vec{\theta}=(\anglevec{1},\ldots,\anglevec{R})$. Across QEC shots, these angles vary, and their distribution affects QEC performance. While our initial analysis will be general, we will eventually specialize to a zero-mean multivariate Gaussian distribution. In this setting, the noise is fully characterized by the covariance matrix \(\boldsymbol{\Sigma}\) with entries,
\begin{equation}
    \boldsymbol{\Sigma}_{\ell\ell^\prime}^{(r,r^\prime)}
    = \avg{\theta_{\ell}^{(r)}\,\theta_{\ell^\prime}^{(r^\prime)}},
    \label{eq:covariance_matrix}
\end{equation}
where \(\avg{\cdots}\) denotes the average over noise realizations.

Our approach allows us to treat a wide variety of possible noise correlation structures, allowing for broad insights into how the degree and patterns of such correlations impact QEC.  Note that limiting cases of our noise model yield simpler models that have been treated previously.   When \(\boldsymbol{\Sigma}\) is diagonal, the rotations are uncorrelated across qubits and cycles, and the model is equivalent to the commonly studied case of independent stochastic Pauli errors. When the error angles on each qubit are maximally correlated across QEC cycles but remain uncorrelated across qubits, $\avg{\rotangle{r}{\ell}\rotangle{r'}{\ell'}} =\delta_{\ell\ell'}\avg{(\rotangle{r}{\ell})^2}$ for all \(r,r'\), the model reproduces the quasi-static noise model of Ref.~\cite{pataki_coherent_2024}. Limiting cases of our general noise model were also studied in Ref.~\cite{witzel_correcting_2026}.

\section{Logical infidelity: exact expression and cumulant expansion\texorpdfstring{\protect\iconConventional}{}}
\label{sec:bare_code}

We now assess the performance of an $\llbracket n,1\rrbracket$ stabilizer code over $R$ QEC cycles under the noise model introduced above. The code is specified by a generating set $\{\hat g_\ell\}_{\ell=1}^{n-1}$ of the stabilizer group $\mathbb{S}$. Without loss of generality, we fix each generator's phase to $+1$: $\hat{g}_\ell\in\overline{\mathbb{P}}_n$.

We allow logical information to be encoded in an arbitrary joint eigenspace of the stabilizer generators $\mathcal{C}_{\vec g}$:
\begin{equation}
    \mathcal{C}_{\vec{g}}
    =
    \Bigl\{
        |\psi\rangle:\ \hat g_\ell|\psi\rangle=(-1)^{g_{\ell}}|\psi\rangle
        \ \ \forall \ell=1,\ldots,n-1
    \Bigr\},
    \label{eq:stabilizer_sector_def}
\end{equation}
where $\vec{g}=(g_{1},\ldots,g_{n-1})$ is the reference syndrome labeling the codespace, with $g_{\ell}\in\{0,1\}$. 

Our performance metric is defined using the effective noise-averaged channel induced by repeated cycles of QEC on data qubit states supported on $\mathcal{C}_{\vec g}$,
\begin{equation}
    \overline{\mathcal{E}}(R)= \bigavg{\mathcal{E}(R,\vec{\theta})},
\end{equation}
where $\mathcal{E}(R,\vec{\theta})$ is the effective channel for the noise realization \(\vec{\theta}\). We characterize the code's performance by the logical infidelity, defined by:
\begin{equation}
    \LIF(R)= 1-\int_{\ket{\psi}\in\mathcal{C}_{\vec{g}}} d\psi \,
    \bra{\psi}\,\overline{\mathcal{E}}(R)\!\left(\ket{\psi}\bra{\psi}\right)\ket{\psi}.
    \label{eq:average_logical_infidelity}
\end{equation}

% The remainder of this section is organized as follows. In Sec.~\ref{subsec:exact_analysis}, we
% provide a simple form for the exact $R$-cycle channel $\mathcal{E}(R,\vec{\theta}\,)$ for a fixed noise realization $\vec{\theta}$ and use it to obtain $\LIF(R)$. In certain cases, this exact channel provides a more efficient route to characterizing the data qubit evolution than a full numerical simulation of the QEC protocol. In Sec.~\ref{subsec:approximation}, we develop a cumulant expansion of $\LIF(R)$ to gain analytical insight into the effect of noise correlations in the weak-noise regime. Unlike a direct Taylor expansion in $\LIF(R)$, this approach remains accurate beyond the small $\LIF(R)$ regime. Upon specializing to Gaussian correlations, we obtain a form in which the dependence on the covariance matrix $\boldsymbol{\Sigma}$ is explicit. 

\subsection{Exact computation of $\LIF(R)$}
\label{subsec:exact_analysis}

We derive a simple yet exact expression for $\LIF(R)$, generalizing prior results for repetition codes \cite{greenbaum_modeling_2017,huang_performance_2019}.   

\subsubsection{Single-cycle evolution}

We first compute the channel induced by a single QEC cycle on the $n$ data qubits supported on the codespace $\mathcal{C}_{\vec g}$. For a single noise realization, the data qubits first evolve under the noise process in Eq.~\eqref{eq:noise_single_cycle}. Errors are then diagnosed by measuring stabilizer generators $\hat g_\ell$, yielding outcomes $\lambda_\ell=(-1)^{m_\ell}$ with $m_\ell\in\{0,1\}$. The collection $\vec m=(m_1,\ldots,m_{n-1})$ is  the measured syndrome. 

Conditioned on $\vec{m}$, the decoder applies a correction. Since the effective channel depends on which errors are mapped to logical errors by this correction step, we must specify a concrete decoding rule. As is standard, we use a maximum-likelihood decoder that assumes an uncorrelated, stochastic Pauli $Z$ error model, with phase-flip probabilities set by the marginal error probabilities of our noise model. Concretely, in cycle $r$, the probability of a $Z$ error on data qubit $\ell$ is taken to be $\avg{\sin^2\small(\rotangle{r}{\ell}\small)}$.

Under this decoding rule, a subset of the errors compatible with each measured syndrome $\vec m$ is corrected. Since the initial state has reference syndrome $\vec g$, the Pauli $Z$ errors compatible with $\vec m$ lie in the
syndrome-specific set $\overline{\mathbb{P}}_{n,Z}^{\vec m\oplus\vec g}$. The symbol $\oplus$ denotes component-wise addition modulo $2$ and the syndrome-specific sets are defined as
\begin{equation}
\overline{\mathbb{P}}_{n,Z}^{\vec s}
=
\{\hat E\in\overline{\mathbb{P}}_{n,Z}\;|\;\vec s_{\hat E}=\vec s\},
\label{eq:compatible_set}
\end{equation}
where $\vec{s}_{\hat{E}}$ is the syndrome of the error $\hat E$,
\begin{equation}
(\vec s_{\hat E})_\ell \;=\;
\begin{cases}
0, & [\hat g_\ell,\hat E]=0,\\[2pt]
1, & \{\hat g_\ell,\hat E\}=0,
\end{cases}
\qquad \ell=1,\dots,n-1 .
\label{eq:syndrome_def}
\end{equation}
Since $\vec s_{\hat E}$ depends only on the commutation relations between $\hat E$ and the stabilizer generators, the syndrome-specific sets are independent of the choice of encoding eigenspace. Each syndrome-specific set partitions into two equivalence classes of errors. Errors within a class differ by multiplication by a $Z$ stabilizer $\hat S_Z\in\mathbb{S}_Z$, where
\begin{equation}
    \mathbb{S}_Z=\mathbb{S}\cap\mathbb{P}_{n,Z}
\end{equation}
is the subgroup of $Z$ stabilizers. Applying a correction operator from one class corrects all errors in that class while errors in the other class map to a logical error. Under the maximum-likelihood rule, the decoder applies a correction operator $\hat R_{\vec m\oplus\vec g}$ from the more probable class of $\overline{\mathbb{P}}_{n,Z}^{\vec m\oplus\vec g}$ for each measured syndrome $\vec{m}$.

With this correction rule, the single-cycle channel induced on data-qubit states $\hat{\rho}$ supported on $\mathcal{C}_{\vec g}$ is
\begin{equation}
    \Lambda(\anglevec{1})[\hat\rho]
    =
    \sum_{\vec m}
    \hat R_{\vec m\oplus\vec g}
    \hat\Pi_{\vec m}
    \mathcal{N}(\anglevec{1})[\hat\rho]
    \hat\Pi_{\vec m}
    \hat R_{\vec m\oplus\vec g}^{\dagger},
    \label{eq:single_cycle_direct_measurement}
\end{equation}
where
\begin{equation}
    \hat\Pi_{\vec m}
    =
    \prod_{\ell=1}^{n-1}
    \frac{\hat I+(-1)^{m_\ell}\hat g_\ell}{2}
\end{equation}
is the projector associated with the measured syndrome $\vec m$. As shown in App.~\ref{app:computing_the_single_cycle_channel}, this channel simplifies to
\begin{equation}
\!\!\!\!\Lambda(\anglevec{1})
\!=\!
\chi^\text{II}(\anglevec{1})\mathcal{I}
\!+\!\chi^{\text{ZI},\mathrm{im}}(\anglevec{1})\logicalZLZ
\!+\!\chi^\text{ZZ}(\anglevec{1})\bar{\mathcal{Z}}_L ,
\label{eq:single_cyc_channel}
\end{equation}
with
\begin{equation}
\begin{aligned}
\mathcal{I}[\cdot] &= \hat{I}[\cdot]\hat{I},\\
\bar{\mathcal{Z}}_L[\cdot] &= \logicalZ[\cdot]\logicalZ,\\
\logicalZLZ[\cdot] &= -i[\logicalZ,\cdot\,]\label{eq:operators_in_the_single_cycle_channel}.
\end{aligned}
\end{equation}
Here $\hat{\bar Z}_L$ is a representative from the single logically nontrivial class of operators in $\overline{\mathbb{N}_Z(\mathbb{S})}$, where
\begin{equation}
    \mathbb{N}_Z(\mathbb{S})
    =
    \mathbb{N}(\mathbb{S})\cap \mathbb{P}_{n,Z}
\end{equation}
is the set of $Z$-type Pauli operators in the normalizer $\mathbb{N}(\mathbb{S})$. For notational convenience, we choose the logical states such that this representative is the logical $Z$ operator. The coefficients of the channel are given by:
\begin{align}
\chi^{\text{ZZ}}(\anglevec{1})
&=\sum_{\vec s}\Bigl|\sum_{\hat E\in\mathbb{\bar{P}}_{n,Z}^{\vec s,\times}}
(-1)^{\phi_{\hat{R}_{\vec{s}}\hat E}}\,c_{\hat E}(\anglevec{1})\Bigr|^{2},
\label{eq:explicit_process_matrix_elements}\\
\chi^{\text{ZI},\mathrm{im}}(\anglevec{1})
&=\Im\Bigl[
\sum_{\vec s}\!\!\sum_{\substack{\hat E\in\mathbb{\bar{P}}_{n,Z}^{\vec s,\checkmark}\\
\hat E'\in\mathbb{\bar{P}}_{n,Z}^{\vec s,\times}}}\!\!\!\!\!\!
(-1)^{\phi_{\hat{E}\hat{E'}}}\!c_{\hat E}(\anglevec{1})c^*_{\hat E'}(\anglevec{1})
\Bigr],
\notag
\end{align}
where \(\Im[\cdot]\) denotes the imaginary part and the phases $\phi_{\hat{P}}$ are discussed below. The set $\overline{\mathbb{P}}_{n,Z}^{\vec s,\checkmark}$ contains the errors in $\overline{\mathbb{P}}_{n,Z}^{\vec s}$ corrected by the correction step, while $\overline{\mathbb{P}}_{n,Z}^{\vec s,\times}$ contains those mapped to a logical $Z$ error. Trace preservation relates the logical Pauli-diagonal coefficients by:
\begin{equation}
\chi^{\text{II}}(\anglevec{1})=1-\chi^\mathrm{ZZ}(\anglevec{1}).
\label{eq:PauliDiagonalCoeffs}
\end{equation}

\subsubsection{Encoding eigenspace dependence}
\label{sec:encoding_eigenspace_dependence}
The phases $(-1)^{\phi_{\hat P}}$ above, defined for each $\hat P\!\!\in\!\! \overline{\mathbb{N}_Z(\mathbb{S})}$, encode an important feature of the channel. They are defined as follows. Each $\hat P\!\!\in \!\!\overline{\mathbb{N}_Z(\mathbb{S})}$ decomposes as:
\begin{equation}
    \hat{P}=\hat S_{Z_{\hat P}}\logicalZ^{\nu_{\hat{P}}},
    \label{eq:decomposition_rule}
\end{equation}
where $\hat{S}_{Z_{\hat{P}}}\in\mathbb{S}_Z$ is the $Z$ stabilizer relating $\hat{P}$ to either $\hat{I}$ or the logical-$Z$ representative $\hat{\bar Z}_L$, depending on whether $\hat{P}$ is logically trivial ($\nu_{\hat{P}}=0$) or nontrivial ($\nu_{\hat{P}}=1$). The phase factor $(-1)^{\phi_{\hat P}}=\pm 1$ is the eigenvalue of the stabilizer $\hat S_{Z_{\hat P}}$ associated with the chosen codespace $\mathcal{C}_{\vec{g}}$:
\begin{equation}
    \hat S_{Z_{\hat P}}\ket{\psi}
    =
    (-1)^{\phi_{\hat P}}\ket{\psi},
    \qquad
    \forall\,\ket{\psi}\in\mathcal{C}_{\vec g}.
    \label{eq:phase_definition}
\end{equation}
These phases thus encode a dependence of the channel on the stabilizer eigenspace chosen as the codespace.

This dependence is striking, with no analogue in stochastic Pauli $Z$ noise models. It arises here because, under coherent Pauli errors, the amplitudes $c_{\hat{E}}(\anglevec{1})$ [Eq.~\eqref{eq:cE_def}] associated with different errors in the same syndrome-specific set add coherently to yield the probabilities that set the channel coefficients. The logical Pauli diagonal coefficient $\chi^{\mathrm{ZZ}}(\anglevec{1})$ is set by interference among the amplitudes of uncorrectable errors within a syndrome-specific set, whereas the off-diagonal coefficient $\chi^{\mathrm{ZI},\mathrm{im}}(\anglevec{1})$ is set by interference between the amplitudes of pairs of correctable and uncorrectable errors within a syndrome-specific set. These amplitudes interfere constructively or destructively depending on the phases.

Viewed this way, it is immediately clear why this eigenspace-dependence is absent in the phase-flip repetition codes studied in Refs.~\cite{greenbaum_modeling_2017,huang_performance_2019}. These codes have no nontrivial $Z$ stabilizers, so $\hat S_{Z_{\hat P}}=\hat I$ and the phases $(-1)^{\phi_{\hat P}}=+1$ regardless of the chosen codespace.

% Thus, the channel's dependence on the encoding choice stems from the presence of either multiple correctable or uncorrectable errors within a syndrome-specific set whose amplitudes interfere. This explains why this dependence is absent in the phase-flip repetition codes analyzed in Refs.~\cite{greenbaum_modeling_2017,huang_performance_2019}. These codes have no nontrivial $Z$ stabilizers, and each syndrome-specific set thus contains only one correctable error and one uncorrectable error.

\subsubsection{From the single-cycle channel to $\LIF(R)$}

The single-cycle channel of the first QEC cycle preserves the codespace. Consequently, the single-cycle channels of subsequent cycles take the same form. Composing these channels yields the channel induced by $R$ QEC cycles for a fixed noise realization $\vec{\theta}$ [see App.~\ref{app:computing_the_R_cycle_channel}]. Averaging this channel over the error angles yields the noise-averaged $R$-cycle channel, and hence the logical infidelity:
\begin{align}
\LIF(R)
&=\frac{1-\!\Re\big(\avg{\Gamma(R,\vec{\theta}\,) }\big)}{3},\label{eq:LIF_in_terms_of_Gamma}
\end{align}
where 
\begin{equation}
\Gamma(R,\vec{\theta}\,) =
\prod_{r=1}^R\!\Big(1-2\,\chi^\text{ZZ}(\anglevec{r})
+2i\,\chi^{\text{ZI},\mathrm{im}}(\anglevec{r})\Big).
\label{eq:Gamma_coefficient}
\end{equation}
This expression is exact. Throughout this paper, we use it to numerically compute the logical infidelity:  we sample random error angles $\vec{\theta}$, evaluate $\Gamma(R,\vec{\theta}\,)$ for each realization, and average over the samples. This approach avoids the repeated circuit shots required in full numerical simulations of the QEC protocol.

\subsection{Weak-noise approximation of $\LIF(R)$}
\label{subsec:approximation}

The effects of noise correlations on $\LIF(R)$ are encoded in $\avg{\Gamma(R,\vec{\theta}\,)}$, but extracting analytical insight from this quantity is generally difficult. Here, we develop a cumulant expansion of $\LIF(R)$ that makes correlation effects transparent in the weak-noise regime. Unlike a direct Taylor expansion of $\LIF(R)$ in the noise strength, our approach remains accurate for a wide range of correlation structures, even beyond the small $\LIF(R)$ regime. Upon specializing to Gaussian correlations, we obtain a simplified expression in which the dependence of $\LIF(R)$ on the covariance matrix $\boldsymbol{\Sigma}$ is explicit. We illustrate below through representative examples that our treatment is applicable at practically relevant noise strengths
\cite{place_new_2021,tuokkola_methods_2025,aghababaie-beni_quantum_2025,eickbusch_demonstration_2025}.

\subsubsection{Scaling and approximation of the channel coefficients}

To approximate $\avg{\Gamma(R,\vec{\theta}\,)}$, we first identify the scaling of the coefficients $\chi^{\mathrm{ZZ}}(\anglevec{r})$ and
$\chi^{\mathrm{ZI},\mathrm{im}}(\anglevec{r})$ with the noise strength. These coefficients depend on the distribution of errors into $\overline{\mathbb{P}}_{n,Z}^{\vec s,\checkmark}$ and
$\overline{\mathbb{P}}_{n,Z}^{\vec s,\times}$ [see definition below Eq.~\eqref{eq:PauliDiagonalCoeffs}], which is set by the code and the marginal error probabilities provided to the decoder. However, in the weak-noise regime of interest, 
the scaling can be determined solely from the weights of errors in these sets. Under the stochastic Pauli $Z$ error model assumed by our decoder, an error of weight $w$ has probability proportional to the product of $w$ small phase-flip probabilities. The maximum likelihood decoder thus preferentially corrects equivalence classes with lower weight errors.

The range of error weights which the decoder can correct depends on the code's $Z$ distance, defined as:
\begin{equation}
    d_Z = \min\Bigl\{\, w(\hat{P}) \;:\; \hat{P}\in \overline{\mathbb{N}_Z(\mathbb{S})}\setminus {\mathbb{S}}_Z \Bigr\},
    \label{eq:Z_distance}
\end{equation}
where $w(\hat{P})$ denotes the number of qubits on which $\hat P$ acts nontrivially. We focus on the common case in which $d_Z$ is odd. The even-$d_Z$ case is discussed briefly in Sec.~\ref{subsubsec:even_dZ} and exhibits qualitatively different behavior. For odd $d_Z$, all Pauli $Z$ errors of weight $w\le (d_Z-1)/2$ are jointly correctable. When the per-qubit marginal error probabilities are sufficiently small and comparable, the decoder corrects all such low weight errors in each cycle, but maps some errors of weight $(d_Z+1)/2$ to a logical error. The leading contribution to $\chi^{\text{ZZ}}(\anglevec{r})$ then comes from these errors, which we term \textit{dominant uncorrectable errors}. Approximating $\chi^{\text{ZZ}}(\anglevec{r})$ by these contributions,
\begin{equation}
\chi^{\text{ZZ}}(\anglevec{r})
\!\!\approx\!\!
\sum_{\vec s}\Biggl(\!\!\!\!\!\!\!\!\!
\sum_{\substack{\hat E\in\overline{\mathbb{P}}_{n,Z}^{\vec s,\times}:\\ w(\hat E)=(d_Z+1)/2}}
\!\!\!\!\!\!\!\!\!\!\!(-1)^{\phi_{\hat{R}_{\vec{s}}\hat E}}
\Big(\!\!\!\!
\prod_{\ell\in\text{supp}(\hat{E})}\!\!\!\!\rotangle{r}{\ell}\Big)
\Biggr)^{2}
\!\!\!+\mathcal{O}(\theta^{d_Z+3}),
\label{eq:chiZZ_leading}
\end{equation}
where $\theta$ denotes the characteristic scale of the error angles. We have also expanded the products of sines and cosines to leading-order in $\theta$.

In contrast, the coefficient $\chi^{\mathrm{ZI},\mathrm{im}}(\anglevec{r})$ depends on the products of amplitudes of correctable and uncorrectable errors belonging to the same syndrome-specific set. The product of any such pair of errors is a logical $Z$ representative and therefore has weight at least $d_Z$. Hence, the leading-order contributions to $\chi^{\mathrm{ZI},\mathrm{im}}(\anglevec{r})$ are at least order $\theta^{d_Z}$. For odd $d_Z$, terms of this order are present and set the leading-order behavior. As shown in App.~\ref{app:leading_order_expansions_of_single_cycle_channel_coefficients}, $\chi^{\mathrm{ZI},\mathrm{im}}(\anglevec{r})$ may therefore be approximated as:
\begin{equation}
    \chi^{\text{ZI},\mathrm{im}}(\anglevec{r})\approx f(d_Z)\!\!\!\!\!\!\!\!\!\!\!
\sum_{\substack{
\hat Z_L\in\overline{\mathbb{N}_Z(\mathbb{S})}\setminus{\mathbb{S}}_Z:
\\ \ w(\hat Z_L)=d_Z}}
\!\!\!\!\!\!\!\!\!\!
(-1)^{\phi_{\hat Z_L}}\!
\Big(\!\!\!\!\!\!\!\prod_{\ell\in\text{supp}(\hat{Z}_L)}\!\!\!\!\rotangle{r}{\ell}\Big)+\mathcal{O}(\theta^{d_Z+2}),
\label{eq:chiZI_leading}
\end{equation}
again expanding to leading-order in $\theta$, and
\begin{equation}
    f(d_Z)=\frac{d_Z+1}{2d_Z}\binom{d_Z}{(d_Z+1)/2}.
\end{equation}
Thus, in the weak-noise regime, $\chi^\text{ZZ}(\anglevec{r})$ scales as $\theta^{d_Z+1}$, whereas $\chi^\text{ZI,im}(\anglevec{r})$ scales as $\theta^{d_Z}$.  

\subsubsection{Second-order cumulant expansion of $\LIF(R)$}

Having established how both $\chi^\mathrm{ZZ}$ and $\chi^\mathrm{ZI,\rm{im}}$ scale for weak noise $\theta \ll 1$, we next approximate each single-cycle contribution to $\Gamma(R,\vec{\theta}\,) $ as
\begin{align}
1\!\!-\!\!2\chi^\text{ZZ}(\anglevec{r})
\!\!+\!\!2i\chi^{\text{ZI},\mathrm{im}}(\anglevec{r})
\!\approx\!
\exp\!\Big(
&\!\!-2\chi^\text{ZZ}(\anglevec{r})\label{eq:exp_approx_cycle}\\
&\!\!+2i\chi^{\text{ZI},\mathrm{im}}(\anglevec{r})
\Big),\notag
\end{align}
where the exponent is accurate up to corrections of order $\mathcal{O}({\theta^{2d_Z}})$ in its real part and $\mathcal{O}(\theta^{2d_Z+1})$ in its imaginary part. For practically relevant QEC codes, since $d_Z\geq 3$, these terms are higher-order compared to the retained terms, and may be neglected. 

Neglecting these higher-order terms gives a simple interpretation to the channel coefficients that helps build intuition for our results. To the accuracy of Eq.~\eqref{eq:exp_approx_cycle}, the single-cycle channel in Eq.~\eqref{eq:single_cyc_channel} can be written as
\begin{flalign}
    \hspace{-2.8mm}\Lambda(\anglevec{r})
    \!\!\approx\!
    &\mathcal{R}_{\logicalZ}\!\!
    \big(\chi^{\mathrm{ZI},\mathrm{im}}(\anglevec{r})\big)
    \!\big(
      \chi^{\mathrm{II}}(\anglevec{r})\mathcal{I}
    \!+\!\chi^{\mathrm{ZZ}}(\anglevec{r})\bar{\mathcal Z}_L
    \big),
    \label{eq:approximate_induced_channel}
\end{flalign}
where
\begin{equation}
\mathcal{R}_{\logicalZ}(\vartheta)[\cdot]
=
\exp\big({-i\logicalZ \vartheta}\big)\,[\cdot]\,
\exp\big({i\logicalZ \vartheta}\big)
\end{equation}
describes a logical $Z$ rotation. This form explicitly shows the generally non-unitary nature of the effective channel and separates the residual coherent and incoherent logical errors. In particular, $\chi^{\mathrm{ZI},\mathrm{im}}(\anglevec{r})$ plays the role of the coherent logical error angle in cycle $r$, whereas $\chi^{\mathrm{ZZ}}(\anglevec{r})$ is the probability of an incoherent logical error.

The approximation in Eq.~\eqref{eq:exp_approx_cycle} converts the product in \(\Gamma(R,\vec{\theta}\,) \) into a sum in the exponent, simplifying $\LIF(R)$ to:
\begin{equation}
    \LIF(R)\!\approx\!\frac{1\!-\Re\left(\!\Bigavg{\exp\!\!\Big(\sum\limits_{r=1}^R
\!\!-2\chi^\text{ZZ}\!(\anglevec{r})
\!+\!2i\chi^{\text{ZI},\mathrm{im}}\!(\anglevec{r})\Big)}\right)}{3}
\label{eq:LIF_exp_bare}
\end{equation}
Viewing $(\chi^\mathrm{ZZ}\!\!-\!i \chi^\mathrm{ZI,{im}})$ as a random variable, we can approximately perform the noise average in this expression via a second-order cumulant expansion. This yields:
\begin{equation}
    \LIF(R)
    \approx
    \frac{1-\Re\!\Big(\exp\!\big(\kappa_{1}(R)
    +{\kappa_{2}(R)}/{2}\big)\Big)}{3},
    \label{eq:bare_LIF_in_terms_of_cumulants}
\end{equation}
where $\Re[\cdot]$ denotes the real part, and $\kappa_1(R)$ and $\kappa_2(R)$ are the first and second cumulants. 

Although this cumulant-expansion method applies broadly, we now specialize to Gaussian correlated noise, where the random error angles $\vec{\theta}$ have a zero-mean multivariate Gaussian distribution with covariance $\boldsymbol{\Sigma}$ [Eq.~\eqref{eq:covariance_matrix}]. Due to the symmetry of this distribution under inversion, $\vec{\theta}\mapsto -\vec{\theta}$, the cumulants simplify to:
\begin{subequations}\label{eq:kappa_BR}
\begin{align}
\kappa_{1}(R)
&=
-2\sum_{r=1}^R
\avg{\chi^\mathrm{ZZ}(\anglevec{r})},
\label{eq:kappa1_BR}
\\
\kappa_{2}(R)
&=
\,\,4\!\!\sum_{r,r'=1}^R
\avg{
\chi^\mathrm{ZZ}(\anglevec{r})
\chi^\mathrm{ZZ}(\anglevec{r'})
}_{c}
\notag\\
&
\,\,\,-\!4\!\!\sum_{r,r'=1}^R
\avg{
\chi^{\mathrm{ZI},\mathrm{im}}(\anglevec{r})
\chi^{\mathrm{ZI},\mathrm{im}}(\anglevec{r'})
},
\label{eq:kappa2_BR}
\end{align}
\end{subequations}
where $\avg{\xi(\anglevec{r})\xi(\anglevec{r'})}_c$ is the connected correlator:
\begin{flalign}
 \avg{\xi(\anglevec{r})\xi(\anglevec{r'})}_c&=\avg{\xi(\anglevec{r})\xi(\anglevec{r'})}\\
 &-\avg{\xi\!(\anglevec{r})}\avg{\xi\!(\anglevec{r'})}\notag.
\end{flalign}
Since both cumulants are real for Gaussian correlated noise, the explicit $\Re(\cdots)$ in Eq.~\eqref{eq:bare_LIF_in_terms_of_cumulants} can be omitted.

We stress that this cumulant approximation differs from a naive leading-order expansion in the noise. Having the cumulants in the exponential effectively performs a resummation of terms that would emerge in a naive perturbation theory.  There is no inconsistency with our having dropped subleading corrections in Eq.~\eqref{eq:exp_approx_cycle}:  we are focusing on the most experimentally relevant regime where errors within a single cycle are weak, but where errors may accumulate to have a significant effect over many cycles. In App.~\ref{app:accuracy_of_the_second_order_cumulant_expansion} we compare against a direct cumulant expansion of Eq.~\eqref{eq:LIF_in_terms_of_Gamma}, avoiding the approximation in Eq.~\eqref{eq:exp_approx_cycle}. For weak noise, such an approach only provides modest quantitative gains, so we retain the approximation in Eq.~\eqref{eq:exp_approx_cycle} to make the underlying physics more transparent.

% This cumulant expansion is preferable to a direct Taylor expansion of $\LIF(R)$. Although subleading corrections to the exponent have been neglected in Eq.~\eqref{eq:exp_approx_cycle}, the expansion partially resums the accumulation of the retained single-cycle contributions over many cycles, thereby capturing their scaling with higher powers of $R$. As a result, the approximation remains accurate beyond the small $\LIF(R)$ regime in some cases.

\subsubsection{Interpretation and scaling of the cumulants}

We now interpret the contributions to $\LIF(R)$ captured by the cumulants and discuss their scaling. The sign of each cumulant determines whether it increases or reduces $\LIF(R)$. The first cumulant is non-positive and therefore increases $\LIF(R)$. This cumulant has a simple physical interpretation. As discussed below Eq.~\eqref{eq:approximate_induced_channel}, $\chi^\mathrm{ZZ}(\anglevec{r})$ is the probability of an incoherent logical error in cycle $r$ for a fixed noise realization. Therefore, $\sum_{r=1}^R \avg{\chi^\mathrm{ZZ}(\anglevec{r})}$ is (approximately) the noise-averaged probability that a single incoherent logical error occurs over $R$ cycles,  and $\kappa_1(R)$ captures the associated increase in infidelity $\LIF(R)$.

We now turn to the second cumulant, where inter-cycle correlations first enter. We decompose this cumulant as:
\begin{align}
   \kappa_{2}(R)
   &=
   \kappa_{2}^\mathrm{coh}(R)
   +
   \kappa_{2}^\mathrm{incoh}(R),
\end{align}
where
\begin{subequations}
\label{eq:second_cumulants_sum}
\begin{align}
\kappa_{2}^\mathrm{coh}(R)
&=
\!-4\,\Bigavg{\Big(\sum_{r=1}^R
\chi^{\mathrm{ZI},\mathrm{im}}(\anglevec{r})\Big)^2},
\label{eq:second_cumulants_sum_coh}
\\
\kappa_{2}^\mathrm{incoh}(R)
&=
4\,\Bigavg{\Big(\sum_{r=1}^R
\chi^{\mathrm{ZZ}}(\anglevec{r})\Big)^2}_c .
\label{eq:second_cumulants_sum_incoh}
\end{align}
\end{subequations}
Since $\kappa_{2}^\mathrm{coh}(R)\leq 0$ whereas $\kappa_{2}^\mathrm{incoh}(R)\geq 0$, they have opposite effects on the logical infidelity. 

To build intuition for these different contributions, we again turn to the interpretation of the single-cycle channel coefficients. The coefficient $\chi^{\mathrm{ZI},\mathrm{im}}(\anglevec{r})$ describes the coherent logical error in cycle $r$, whereas $\chi^{\mathrm{ZZ}}(\anglevec{r})$ describes the incoherent one, both to the accuracy of Eq.~\eqref{eq:exp_approx_cycle}. To this accuracy, $\kappa_{2}^\mathrm{coh}(R)$ thus captures how coherent logical $Z$ errors accumulate across cycles, increasing the infidelity $\LIF(R)$, and $\kappa_{2}^\mathrm{incoh}(R)$ captures how incoherent logical $Z$ errors in different cycles partially cancel, reducing $\LIF(R)$, both relative to the estimate provided by $\kappa_1(R)$. Motivated by this interpretation, we term these contributions the coherent and incoherent contributions to the second cumulant, respectively. 

We now discuss the scaling of these cumulants for a fixed odd-$d_Z$ code with the number of cycles $R$, the noise strength $\sigma^2$ (the scale of the error-angle variances), and the number of cycles $R_c$, out of $R$, over which inter-cycle correlations remain appreciable on the scale set by $\sigma^2$. From the scaling of $\chi^\mathrm{ZZ}$ and $\chi^\mathrm{ZI,im}$[Eqs.~\eqref{eq:chiZZ_leading} and \eqref{eq:chiZI_leading}],
\begin{subequations}
\label{eq:cumulant_scalings}
\begin{align}
    \kappa_1(R)
    &\sim R\sigma^{d_Z+1},
    \label{eq:kappa1_scaling}
    \\
    \kappa_2^\mathrm{coh}(R)
    &\sim RR_c\sigma^{2d_Z},
    \label{eq:kappa2_coh_scaling}
    \\
    \kappa_2^\mathrm{incoh}(R)
    &\sim RR_c\sigma^{2d_Z+2}.
    \label{eq:kappa2_incoh_scaling}
\end{align}
\end{subequations}
For weak noise, the coherent contribution $\kappa_2^\mathrm{coh}(R)$ dominates the second cumulant. Its relevance is controlled by $R_c$: it exceeds the higher-order corrections neglected in Eq.~\eqref{eq:exp_approx_cycle} (of order $R\sigma^{2d_Z}$; see App.~\ref{app:accuracy_of_the_second_order_cumulant_expansion}) only for long-ranged inter-cycle correlations, $R_c\gg1$. In this regime, the second cumulant primarily captures the increase in $\LIF(R)$ beyond the estimate of $\kappa_1(R)$ due to the accumulation of coherent logical errors. For short-ranged correlations, $R_c\sim1$, it is comparable to the neglected corrections and should be dropped at this level of approximation.

We stress that code-dependent factors, such as the number of weight-$d_Z$ logical-$Z$ representatives and the number of dominant uncorrectable errors in each syndrome-specific set, may be incorporated for a more quantitative understanding of the scaling at finite noise strength. Here, however, we retain a code-agnostic perspective.

\subsubsection{Approximation of the cumulants to leading-order in $\sigma$}

The second-order cumulant expansion of $\LIF(R)$ lends some insight into how the incoherent and coherent logical errors from each cycle contribute to the logical infidelity. However, its dependence on averages of single-cycle channel coefficients and their products [Eqs.~\eqref{eq:kappa1_BR} and \eqref{eq:kappa2_BR}] complicates a detailed analysis of correlation effects.

For more insight, we may approximate each cumulant to leading-order in $\sigma$ using Eqs.~\eqref{eq:chiZZ_leading} and \eqref{eq:chiZI_leading}. This yields:
% \vspace{-0.5\baselineskip}
\begin{widetext}
% \vspace{-0.5\baselineskip}
\begingroup
\setlength{\jot}{1.0em}
\begin{subequations}\label{eq:bare_model_cumulants}
\begin{align}
\kappa_{1}(R)
&\approx
-2\sum_{r,\vec{s}}
\sum_{\substack{
\hat E,\hat E'\in\mathbb{\overline{P}}_{n,Z}^{\vec s,\times}:\\
w(\hat E),w(\hat E')=(d_Z+1)/2}}
\hspace{-4mm}(-1)^{\phi_{\hat E\hat E'}}
\left\langle
\,\,\,\,\,\,\,\,\,\,
\prod_{\mathclap{\ell\in\mathrm{supp}(\hat E)}}\rotangle{r}{\ell}
\;\quad
\prod_{\mathclap{\ell'\in\mathrm{supp}(\hat E')}}\rotangle{r}{\ell'}
\right\rangle
\,+\mathcal{O}(R\sigma^{d_Z+3})
&&\label{eq:bare_model_first_cumulant}
\\
\kappa_{2}^{\mathrm{coh}}(R)
&\approx
-4f(d_Z)^2
\sum_{r,r'}
\sum_{\substack{
\hat Z_L,\hat Z'_L\in
\overline{\mathbb{N}_Z(\mathbb{S})}\setminus{\mathbb{S}}_Z:\\
w(\hat Z_L),w(\hat Z'_L)=d_Z}}
\hspace{-3mm}(-1)^{\phi_{\hat Z_L\hat Z'_L}}
\left\langle
\,\,\,\,\,\,\,\,\,\,
\prod_{\mathclap{\ell\in\mathrm{supp}(\hat Z_L)}}\rotangle{r}{\ell}
\qquad
\prod_{\mathclap{\ell'\in\mathrm{supp}(\hat Z'_L)}}\rotangle{r'}{\ell'}
\right\rangle
+\mathcal{O}(RR_c\sigma^{2d_Z+2})
&&\label{eq:bare_model_second_cumulant_coh}
\\
\kappa_{2}^{\mathrm{incoh}}(R)
&\approx
4\sum_{\substack{r,r'\\ \vec{s},\vec{s}'}}
\!\!\!\!\!\!\!\!\!\!\!\!
\sum_{\substack{
\hat E,\hat F\in\mathbb{\overline{P}}_{n,Z}^{\vec s,\times}\\
\hat E',\hat F'\in\mathbb{\overline{P}}_{n,Z}^{\vec s',\times}\\
w(\hat E),\ldots,w(\hat F')=(d_Z+1)/2}}
\hspace{-10mm}(-1)^{\phi_{\hat E\hat F\hat E'\hat F'}}
\left\langle
\,\,\,\,\,\,\,\,\,\,
\prod_{\mathclap{\ell\in\mathrm{supp}(\hat E)}}\rotangle{r}{\ell}
\;\quad
\prod_{\mathclap{m\in\mathrm{supp}(\hat F)}}\rotangle{r}{\vphantom{\ell}m}
\;\quad
\prod_{\mathclap{\ell'\in\mathrm{supp}(\hat E')}}\rotangle{r'}{\ell'}
\;\quad
\prod_{\mathclap{m'\in\mathrm{supp}(\hat F')}}\rotangle{r'}{\vphantom{\ell}m'}
\right\rangle\!\vphantom{\Big(A\Big)}_{c}% <-- {} lifts c off the tall delimiter
+\mathcal{O}(RR_c\sigma^{2d_Z+4}),
&&\label{eq:bare_model_second_cumulant_incoh}
\end{align}
\end{subequations}
\endgroup
% \vspace{-0.5\baselineskip}
\end{widetext}
% \vspace{-0.5\baselineskip}
where the phase factors have been simplified by repeatedly applying Eq.~\eqref{eq:phase_combination}. This weak-noise approximation allows transparent analysis of noise correlations: since $\vec{\theta}$ is Gaussian, Wick's theorem relates each average above directly to the covariance $\boldsymbol{\Sigma}$. Although $\kappa_2^\mathrm{incoh}(R)$ is higher-order in $\sigma$ compared to $\kappa_2^\mathrm{coh}(R)$ and is dropped when $\kappa_2^\mathrm{coh}(R)$ dominates, we provide its leading-order approximation because it becomes the dominant second cumulant contribution whenever the coherent contribution is suppressed [see Sec.~\ref{sec:logical_twirling}].

%%%%%%%%%%%%%%%%%%%%%%%%%%%%%%%%%%%
\section{Impact of Correlated Coherent Errors on QEC Performance}
\label{sec:interpretation}

The previous section established our main technical result: an analytic expression for the logical infidelity after $R$ cycles of QEC in terms of the cumulants [Eq.~\eqref{eq:bare_LIF_in_terms_of_cumulants}], together with approximations of the cumulants to leading-order in noise strength [Eq.~\eqref{eq:bare_model_cumulants}], which relate directly to the covariance matrix $\boldsymbol{\Sigma}$. Before noise averaging, each cycle results in both incoherent and coherent logical $Z$ errors.  The first cumulant $\kappa_1(R)$ describes the accumulation of incoherent errors, whereas the second cumulant $\kappa_2(R)$ accounts for how inter-cycle correlations modify the contribution of both incoherent and coherent logical errors. We now use this result to identify significant effects of correlated coherent errors that are largely universal across stabilizer codes.

% \subsubsection{Correlations that strongly affect $\LIF(R)$}

% %We first identify the correlations that most strongly influence $\LIF(R)$ in the weak-noise regime. 
% For the $\kappa_1$, the most-relevant correlations for weak noise are intra-cycle correlations between error angles supported on dominant uncorrectable errors within the same syndrome-specific set. Inter-cycle correlations only enter through the second cumulant. The leading-order contribution here is $\kappa_2^\mathrm{coh}(R)$ [Eq.~\eqref{eq:bare_model_second_cumulant_coh}]. This term depends on both intra-cycle correlations between error angles within the support of a single weight-$d_Z$ logical $Z$ representative and inter-cycle correlations between error angles in the supports of pairs of weight-$d_Z$ logical-$Z$ representatives.

\subsubsection{Effect of inter-cycle correlations}
\label{sec:features_common_across_odd_dZ_codes}

We find that inter-cycle correlations are universally detrimental across odd-$d_Z$ codes at the leading-order in noise strength. Concretely, to isolate their contribution, we compare the logical infidelity under correlated noise with covariance matrix $\boldsymbol{\Sigma}$ to that under its Markovianized counterpart $\boldsymbol{\widetilde\Sigma}$,
\begin{equation}
    \widetilde{\boldsymbol{\Sigma}}^{(r,r')}_{\ell\ell'}
    =
    \delta_{rr'}\boldsymbol{\Sigma}^{(r,r)}_{\ell\ell'} ,
    \label{eq:markovianization}
\end{equation}
which retains the intra-cycle covariances of $\boldsymbol{\Sigma}$ but removes all inter-cycle ones. Since $\kappa_1(R)$ depends only on the intra-cycle covariances, the logical infidelities under the two models differ primarily through the second cumulant. To leading-order, this difference is the coherent contribution $\kappa_2^\text{coh}(R)$ [Eq.~\eqref{eq:bare_model_second_cumulant_coh}]. Since $\kappa_2^\mathrm{coh}(R)\leq 0$ for any code, encoding, or correlation structure, long-ranged inter-cycle correlations increase $\LIF(R)$ relative to the Markovianized model at leading-order in the noise strength.

This increase has a simple physical origin. Without inter-cycle correlations, the coherent component of the logical error cancels on noise-averaging. Inter-cycle correlations prevent this cancellation, increasing $\LIF(R)$.

\subsubsection{Positive correlations with the conventional encoding}
\label{sec:positive_correlations_odd_dz_conventional_encoding}
Another code-independent feature emerges for the conventional encoding choice: positive noise correlations always serve to {\it increase} the infidelity, consistent with generic expectations \cite{clemens_quantum_2004,klesse_quantum_2005,novais_surface_2013}.  By conventional encoding, we mean that the codespace is the $+1$ eigenspace of all stabilizers. With this choice, the phases $(-1)^{\phi_{\hat P}}$ that control interference in the channel coefficients are all unity.  
The leading-order terms for $\kappa_1,\kappa_2$ [Eqs.~\eqref{eq:bare_model_first_cumulant} and \eqref{eq:bare_model_second_cumulant_coh}] then become polynomials in the entries of $\boldsymbol{\Sigma}$ with negative coefficients. Two conclusions then follow, both to the leading-order in noise strength\textemdash positive noise correlations increase $\LIF(R)$ compared to the uncorrelated case, and increasing their strength increases $\LIF(R)$ \textit{monotonically}.

\begin{figure}[t]
    \centering
    \includegraphics[width=\linewidth]{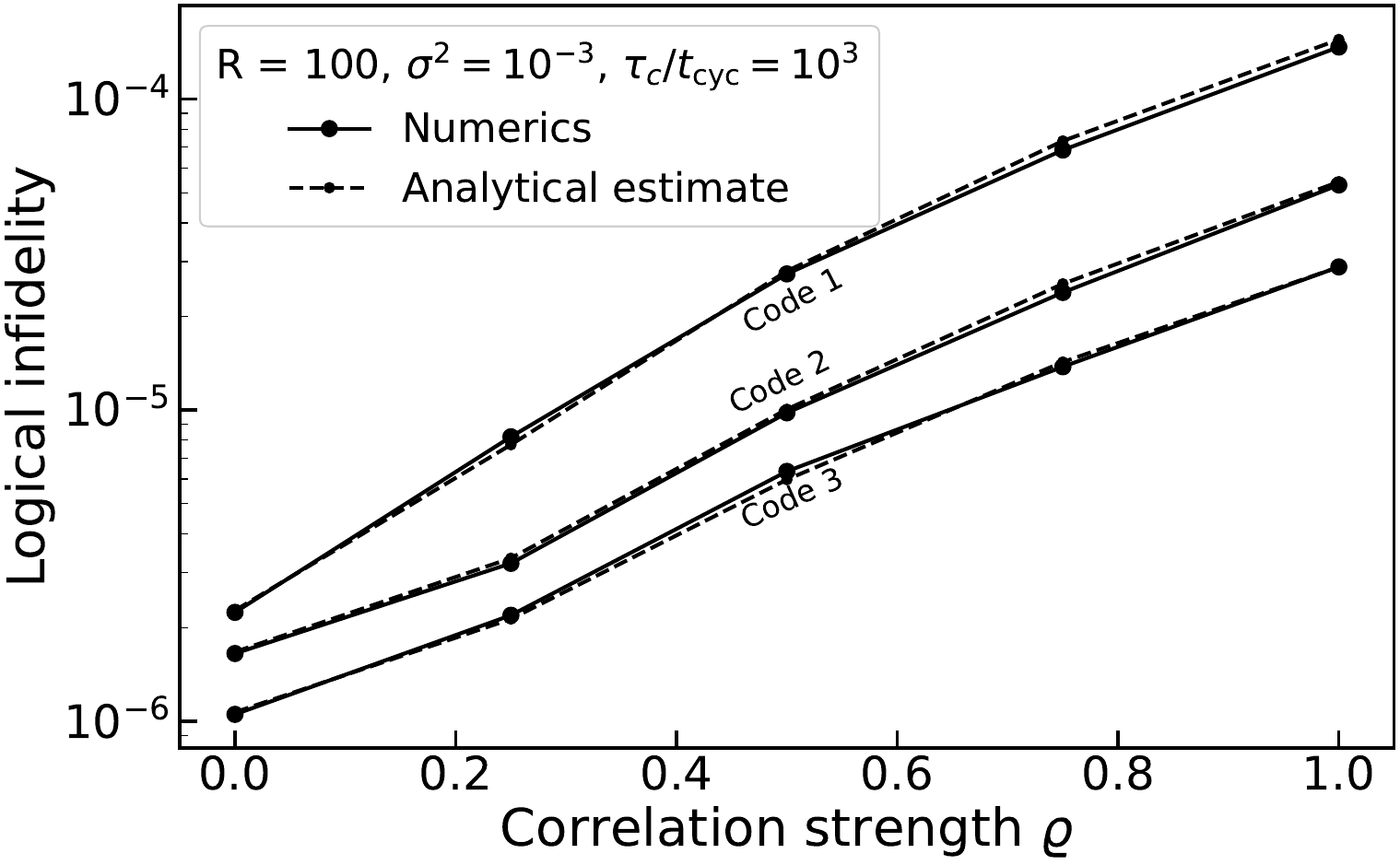}
    \caption{%
\textbf{Infidelity of random odd-$d_Z$ CSS codes with the conventional syndrome-zero encoding under increasing strength of positive noise correlations.}
We plot the logical infidelity $\LIF(R)$ after $R=100$ cycles for three random
$\llbracket 9,1\rrbracket$ CSS codes [see App.~\ref{app:numerical_details}] with $d_Z=5$, using the conventional syndrome-zero encoding and the noise model in Eq.~\eqref{eq:lorentzian_noise_correlations}, with noise strength $\sigma^2$ and inter-cycle correlation range $\tau_c/t_{\rm cyc}$ fixed while inter-qubit correlation strength $\varrho$ is varied over $[0,1]$. Over this range, noise correlations are positive, and $\LIF(R)$ increases monotonically with $\varrho$. Solid lines show
numerical results, obtained by noise-averaging the exact expression in
Eq.~\eqref{eq:LIF_in_terms_of_Gamma}; error bars are within marker size. Dashed
lines show analytical estimates from a second-order cumulant
expansion, using approximations of the cumulants $\kappa_1(R)$ and
$\kappa_2^\mathrm{coh}(R)$ from Eq.~\eqref{eq:bare_model_cumulants}.%
}
    \label{fig:conventional_encoding_random_CSS_code}
\end{figure}

To illustrate this effect, in Fig.~\ref{fig:conventional_encoding_random_CSS_code} we consider three random $\llbracket 9,1 \rrbracket$ CSS codes~\cite{calderbank_good_1996} with $d_Z=5$, specified by the check matrices in App.~\ref{app:numerical_details}. We use the conventional encoding described above, with noise described by:
\begin{equation}
    \boldsymbol{\Sigma}^{(r,r')}_{\ell\ell'}=\sigma^2
    \big((1-\varrho)\delta_{\ell\ell'}+\varrho\big)
    \exp\Big(-\abs{r-r'}\frac{t_\mathrm{cyc}}{\tau_c}\Big).
    \label{eq:lorentzian_noise_correlations}
\end{equation}
Here $\sigma$ sets the overall noise strength, $\varrho$ controls the strength of inter-qubit correlations, and $\tau_c/t_\mathrm{cyc}$ sets the inter-cycle correlation range. 
%This model allows us to tune both the strength and the range of noise correlations. 
In Fig.~\ref{fig:conventional_encoding_random_CSS_code} we vary $\varrho$ over $[0,1]$ (ensuring positive noise correlations) and fix $\tau_c/t_\mathrm{cyc}$. As expected from our general perturbative argument, $\LIF(R)$ thus increases with increasing $\varrho$.
\subsubsection{Parity-dependent effect of inter-cycle correlations}
\label{subsubsec:even_dZ}

The effect of inter-cycle correlations in the less commonly used even-$d_Z$ codes contrasts sharply with the odd-$d_Z$ case: in even-$d_Z$ codes, these correlations are beneficial at the leading-order. This contrast arises from differences in the single-cycle channel for a single noise realization. For odd $d_Z$, $\chi^\mathrm{ZI,im}\!\!\sim\!\!\theta^{d_Z}$ and $\chi^\mathrm{ZZ}\!\!\sim\!\!\theta^{d_Z+1}$ [Eqs.~\eqref{eq:chiZZ_leading} and \eqref{eq:chiZI_leading}], whereas for even $d_Z$, $\chi^\mathrm{ZI,im}\!\!\sim\!\!\theta^{d_Z+1}$ or higher while $\chi^\mathrm{ZZ}\!\!\sim\!\!\theta^{d_Z}$.

Due to $\chi^\mathrm{ZI,im}$ being suppressed compared to $\chi^\mathrm{ZZ}$ for even $d_Z$, the second cumulant contributions scale as $RR_c\sigma^{2d_Z+2}$ or higher (coherent) and $RR_c\sigma^{2d_Z}$ (incoherent). So, for weak noise, the incoherent contribution dominates\textemdash reversing the odd-$d_Z$ hierarchy. This reversal has three consequences for codes with even $d_Z$:
\begin{enumerate}
    \item The ratio of the second cumulant to the first, which captures the significance of inter-cycle correlations, is suppressed by $\sigma^2$ relative to a code of odd distance $d_Z-1$.
    \item The residual effect of these correlations is beneficial: at the leading-order, inter-cycle correlations reduce $\LIF(R)$ relative to the Markovianized model [Eq.~\eqref{eq:markovianization}]. 
    \item     With the conventional encoding and positive noise correlations, correlations help: $\LIF(R)$ is monotonically {\it decreasing} with the strength and range of inter-cycle correlations.   
\end{enumerate}
Together, these consequences suggest that long-range inter-cycle noise correlations may make it worthwhile to raise the $Z$-distance of a code from an odd value to the next even value, despite the additional hardware overhead.

The suppression underlying this reversal in inter-cycle correlation effects follows from the structure of the coefficient $\chi^\mathrm{ZI,im}$, which collects imaginary parts of products of correctable- and uncorrectable-error amplitudes within a syndrome-specific set [Eq.~\eqref{eq:explicit_process_matrix_elements}]. Such products arise at order $\theta^{d_Z}$ only when two errors with disjoint supports multiply to a weight-$d_Z$ logical $Z$ operator. For even $d_Z$, these order-$\theta^{d_Z}$ products are real and drop out of $\chi^\mathrm{ZI,im}$. Thus, for even $d_Z$, $\chi^\mathrm{ZI,im}$ scales as $\theta^{d_Z+1}$ or higher, while $\chi^\mathrm{ZZ}\!\!\sim\!\!\theta^{d_Z}$. Taken to the extreme, in codes whose logical $Z$ representatives are all even-weight (e.g., even-qubit repetition codes), $\chi^\mathrm{ZI,im}\!=\!0$. In these codes, coherent $Z$ errors on the physical qubits leave only incoherent logical errors [see App.~\ref{app:imaginary_channel_coefficient}], extending observations in Ref.~\cite{huang_performance_2019} beyond even-qubit repetition codes.

\begin{figure}[t]
    \centering{%
        \includegraphics[width=0.95\linewidth]{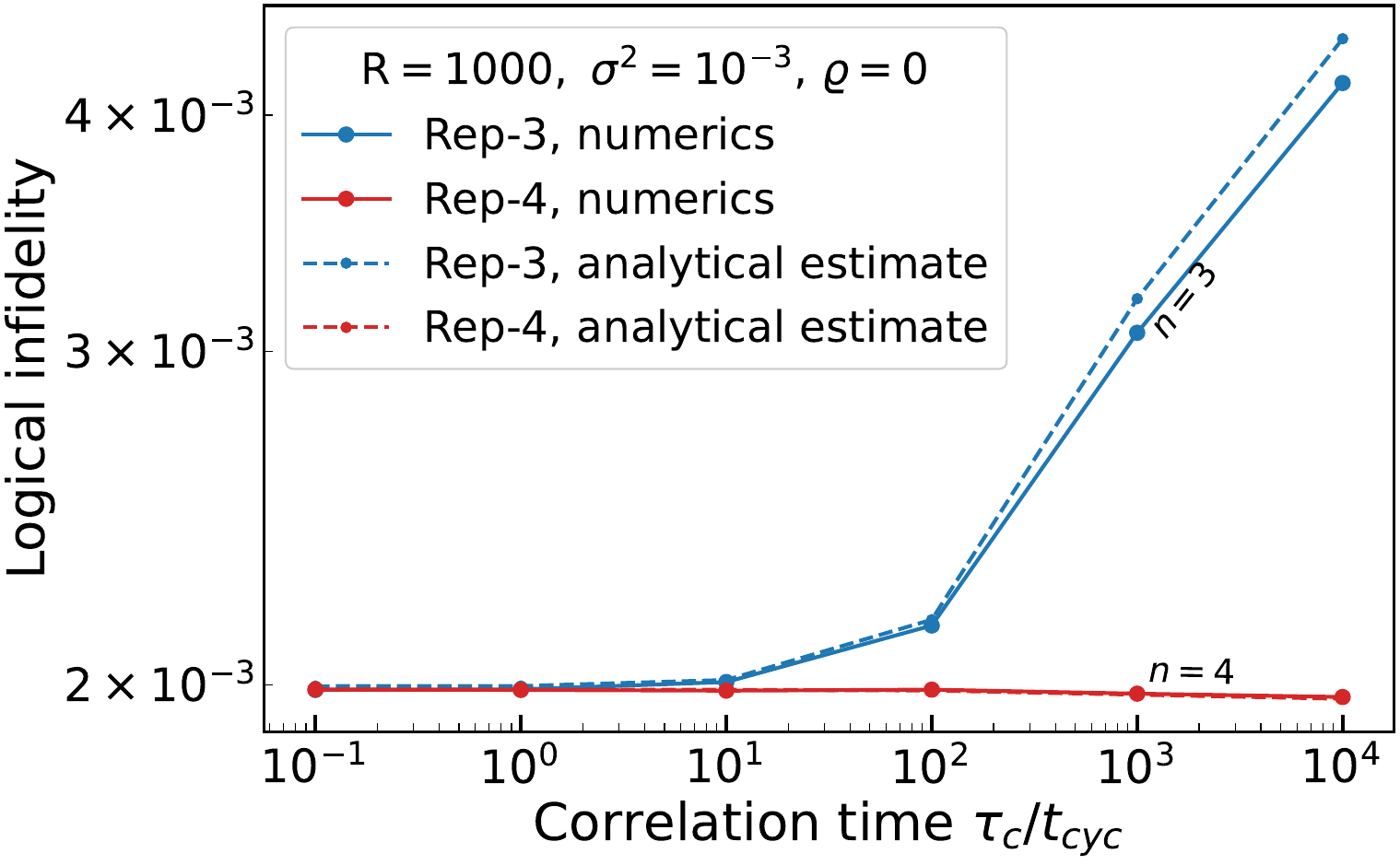}
        }
        \caption{%
\textbf{Impact of inter-cycle noise correlation range on infidelity $\LIF(R)$ and sensitivity to the parity of the code's $Z$ distance.} We plot $\LIF(R)$ after $R=1000$ cycles for the three-qubit (Rep-3; blue, circles) and four-qubit (Rep-4; red, squares) phase-flip repetition codes, under the noise model in Eq.~\eqref{eq:lorentzian_noise_correlations}, with noise strength $\sigma^2$ and inter-qubit correlation strength $\varrho$ fixed. The inter-cycle correlation
range $\tau_c/t_\mathrm{cyc}$ is varied. For $\tau_c/t_\mathrm{cyc}\rightarrow0$, the two codes perform similarly. Increasing $\tau_c/t_\mathrm{cyc}$ increases $\LIF(R)$ for the Rep-3 code, whereas for the Rep-4 code, $\LIF(R)$ decreases slightly. Solid lines show numerical results, obtained by noise-averaging the exact expression in Eq.~\eqref{eq:LIF_in_terms_of_Gamma}; error bars are within marker size. Dashed lines show analytical estimates from a second-order cumulant expansion, using leading-order approximations in Eq.~\eqref{eq:bare_model_cumulants}: $\kappa_1(R)$ and $\kappa_2^\mathrm{coh}(R)$ for Rep-3, and $\kappa_1(R)$ and $\kappa_2^\mathrm{incoh}(R)$ for Rep-4.}
\label{fig:odd_even_distinction_rep_code}
\end{figure}

Figure~\ref{fig:odd_even_distinction_rep_code} illustrates the above consequences of the $Z$ distance's parity using three- and four-qubit phase-flip repetition codes under noise specified by Eq.~\eqref{eq:lorentzian_noise_correlations}. We show the suppression and reversal in effects of inter-cycle correlation on going from odd to even $Z$ distance, and the monotonic dependence of $\LIF(R)$ on the correlation range.

\section{Error Suppression Techniques}
\label{sec:error_suppression_techniques}

We now use our framework for describing correlated coherent errors to understand how they might be suppressed (for a given code and noise structure).  
We consider four error suppression techniques.  The first and principal method directly exploits the sensitivity of infidelity to the choice of encoding subspace.  This method involves using an optimal eigenspace, something we term the protected stabilizer eigenspace (PROSE). The remaining three strategies are variants of standard techniques for suppressing coherent errors: dynamical decoupling~\cite{viola_dynamical_1999} and Pauli twirling~\cite{wallman_noise_2016,hashim_randomized_2021}, both adapted to QEC settings~\cite{paz-silva_optimally_2013,han_protecting_2025,kasatkin_quantum_2026,cai_mitigating_2020,beale_randomized_2023,jain_improved_2023}. Logical dynamical decoupling (LDD) interleaves an optimized, deterministic sequence of logical Pauli gates between QEC cycles to suppress coherent logical errors. Logical Pauli twirling (LT) instead randomizes this sequence, twirling each single-cycle channel at the logical level. Physical Pauli twirling (PT) in contrast twirls this channel at the physical-qubit level, applying random sequences of $n$-qubit Pauli gates.

We find that PROSE encoding and LT increase logical fidelity in the presence of correlated coherent noise in complementary ways.  We thus focus on these techniques, and show in Sec.~\ref{sec:comparison} how they can be combined to yield an even more powerful approach. 
Analyses of LDD and PT are presented in App.~\ref{app:dynamical_decoupling} and App.~\ref{app:physical_Pauli_twirling}.
% , respectively. This focus is motivated by the complementary ways in which PROSE encoding and LT suppress logical infidelity. In Sec.~\ref{sec:comparison}, we exploit this complementarity by combining them into a single protocol that outperforms either technique alone and PT, while performing nearly as well as PROSE encoding combined with LDD. Beyond such suppression, motivated by recent observations for Pauli-twirled Clifford circuits~\cite{brillant_noise_2026}, we examine whether these techniques can render certain noise correlations beneficial\textemdash  in particular, the positive correlations that are otherwise detrimental in odd-$d_Z$ codes when the conventional encoding is used.

\subsection{Protected stabilizer eigenspace (PROSE) encoding}
\label{sec:prose_encoding}

The logical infidelity will in general depend on which stabilizer eigenspace is chosen as the codespace [Sec.~\ref{sec:bare_code}]. This suggests a natural error suppression strategy: encode logical information in the stabilizer eigenspace that minimizes the logical infidelity.  We call this eigenspace the \textit{protected stabilizer eigenspace (PROSE)}. 

Whether such a PROSE exists depends on the noise correlation structure. For Gaussian noise, we find that correlations across qubits are necessary. If the noise is uncorrelated across qubits i.e., $\boldsymbol{\Sigma}^{(r,r')}_{\ell\ell'}\propto\delta_{\ell\ell'}$, the (noise-averaged) logical infidelity is independent of the encoding eigenspace to all orders in the noise strength (App.~\ref{app:when_does_the_eigenspace_matter}) \footnote{In non-Gaussian noise models (e.g., a common error angle shared across all qubits), the encoding dependence may still persist. Beyond the Gaussian case, App.~\ref{app:when_does_the_eigenspace_matter} extends the analysis of when this encoding dependence vanishes to any distribution symmetric under inversion of the error angles on a single qubit. Both this vanishing criterion and the associated notion of ``correlations across qubits'' required for this dependence to persist should be understood as holding within this symmetry class.}. When inter-qubit correlations are present, we find that using PROSE is especially effective for odd-$d_Z$ codes under noise with positive correlations. In this setting, the conventional encoding (all stabilizers $+1$) is the worst possible choice, as error amplitudes add constructively.  In contrast, the PROSE maximizes destructive interference of these amplitudes.  

\subsubsection{Efficiently identifying the PROSE}
\label{sec:efficiently_finding_the_PROSE}

Finding the PROSE may appear nontrivial: it will generically depend on the number of QEC cycles $R$, and having to evaluate each possible eigenspace over many-cycle evolution seems daunting.  The problem simplifies, however, in the standard setting where the noise is stationary and $R_c\sigma^{d_Z-1}\ll 1$ (i.e.,~sufficiently weak noise, short inter-cycle correlation range, and/or few QEC cycles).  In this regime $\kappa_2$ is negligible, and the logical infidelity can be approximated as
\begin{equation}
\LIF(R)\approx\frac{1-\exp\left(-2
\avg{\chi^\mathrm{ZZ}(\anglevec{1})}R\right)}{3}.
\label{eq:LIF_when_PROSE_can_be_efficiently_found}
\end{equation}
In this regime, we can thus find the PROSE just by analyzing a single cycle, a much simpler task. 

Eq.~\eqref{eq:LIF_when_PROSE_can_be_efficiently_found} also provides useful intuition.  The quantity $R\avg{\chi^\mathrm{ZZ}(\anglevec{1})}$ is the noise-averaged probability of a single incoherent logical error over $R$ cycles. We have:
\begin{flalign}
    \avg{\chi^\mathrm{ZZ}(\anglevec{1})}
&=\sum_{\vec s}\!\!
\sum_{\substack{\hat E\in\mathbb{\overline{P}}_{n,Z}^{\vec s,\times}}}
\Bigavg{\abs{c_{\hat{E}}(\anglevec{1})}^2} \label{eq:first_cumulant_opened_up}\\
&+\sum_{\vec s}\!\!
\sum_{\substack{\hat E\neq\hat E'\in\mathbb{\overline{P}}_{n,Z}^{\vec s,\times}}}
\!\!\!\!\!(-1)^{\phi_{\hat E\hat E'}}
\Bigavg{c_{\hat{E}}(\anglevec{1})c^*_{\hat{E}'}(\anglevec{1})}.
\notag
\end{flalign}
We see that all the encoding-dependence lies in the terms involving pairs of distinct errors within the same syndrome-specific set. These errors are related by $Z$ stabilizers, and the codespace-associated eigenvalue $(-1)^{\phi_{\hat E\hat E'}}$ of these stabilizers governs how the amplitudes of these errors interfere. PROSE encoding optimizes their destructive interference, reducing the noise-averaged probability of incoherent logical errors $\avg{\chi^\mathrm{ZZ}(\anglevec{1})}$. 

This decomposition also explains why encoding choice does not influence the logical infidelity in situations where the noise has no inter-qubit correlations (within the accuracy of Eq.~\eqref{eq:LIF_when_PROSE_can_be_efficiently_found}). Without such correlations, the distribution of the error angles $\anglevec{1}$ factorizes across qubits and is symmetric under inversion of any single angle $\rotangle{1}{\ell}$. Any term in $\avg{\chi^\mathrm{ZZ}(\anglevec{1})}$ that is odd in some $\rotangle{1}{\ell}$ therefore vanishes under noise-averaging. Since $c_{\hat E}(\anglevec{1})$ is odd in $\rotangle{1}{\ell}$ for every $\ell\in\mathrm{supp}(\hat E)$ [Eq.~\eqref{eq:cE_def}], the average $\avg{c_{\hat E}(\anglevec{1})c^*_{\hat E'}(\anglevec{1})}$ survives only when $\mathrm{supp}(\hat E)=\mathrm{supp}(\hat E')$ i.e., when $\hat E=\hat E'$. The error-pair terms, all with $\hat E\neq\hat E'$, thus vanish, leaving only the encoding-independent single-error terms and no encoding dependence to exploit. App.~\ref{app:when_does_the_eigenspace_matter} extends this result to arbitrary parameter regimes. 

\begin{figure}[t]
    \centering
    \captionsetup[subfloat]{position=top,justification=raggedright,singlelinecheck=false}
    \subfloat[]{%
        \includegraphics[width=\linewidth]{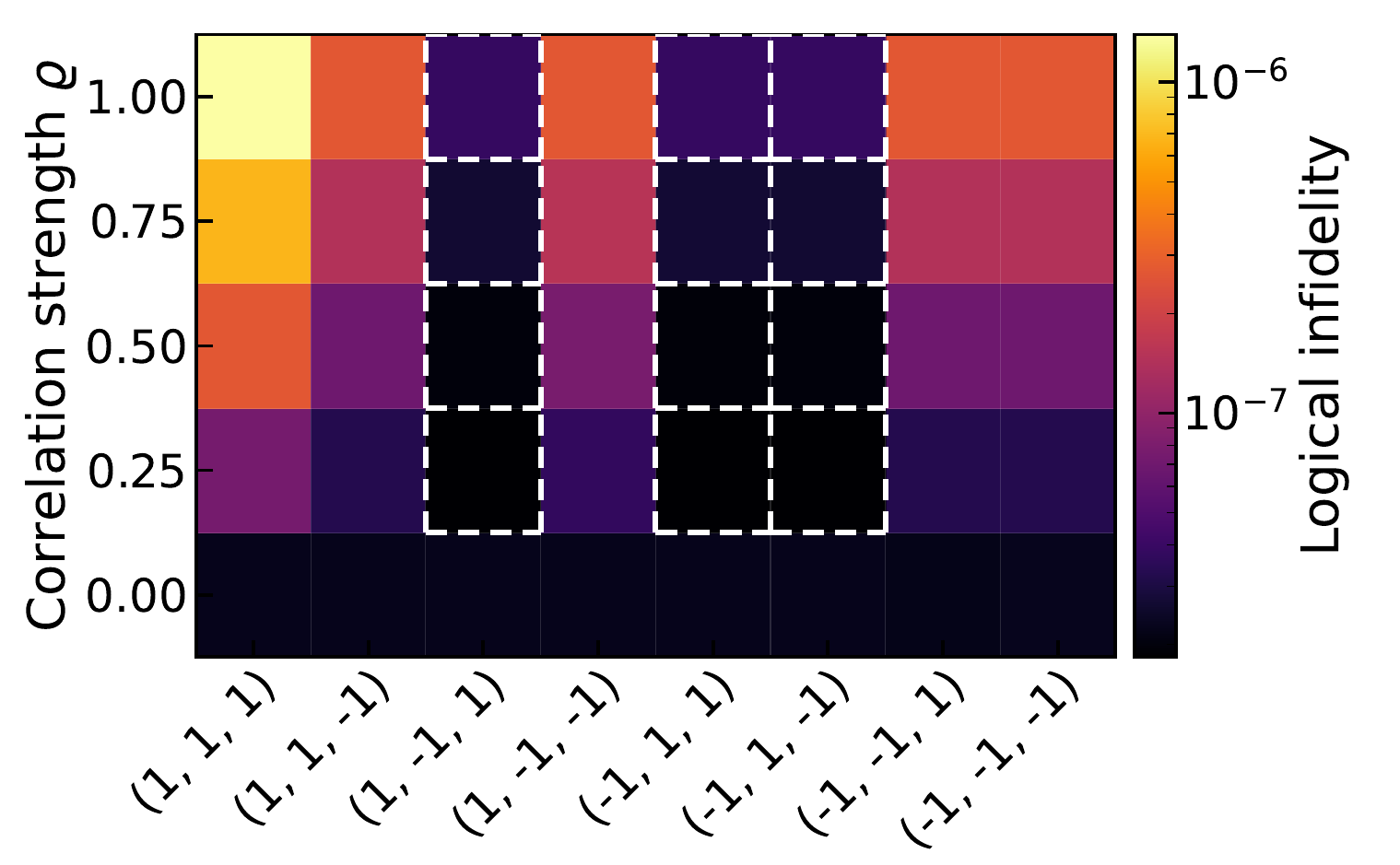}
        \label{fig:encoding_eigenspace_search}
        }
    \vspace{0.5em}
    \subfloat[]
    {
    \includegraphics[width=\linewidth]{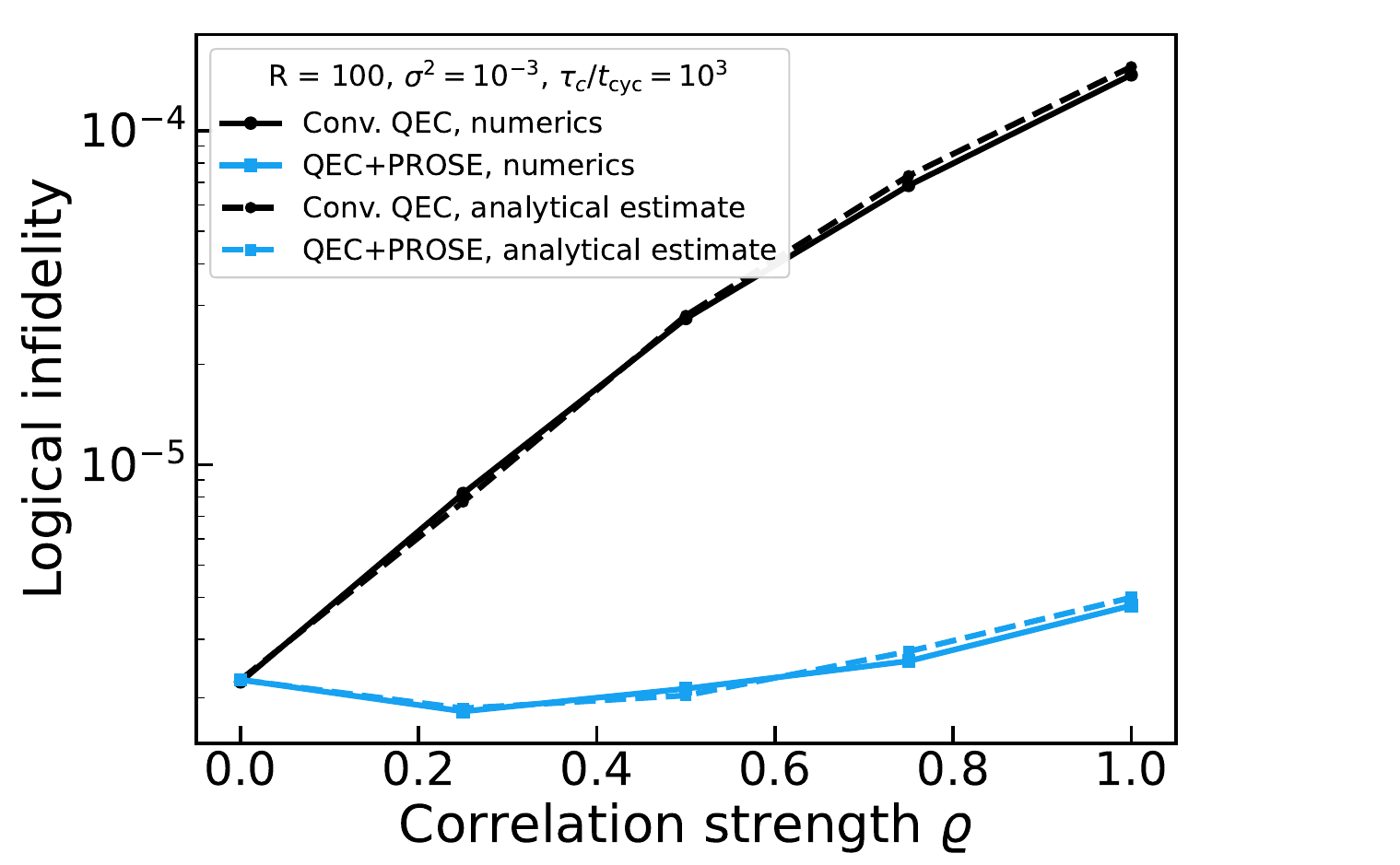}
    \label{fig:performance_enhancement_after_EO}
    }
\caption{%
\textbf{Identifying PROSE encodings and corresponding fidelity improvement. }
Results here are for one of the $\llbracket 9,1\rrbracket$ CSS codes [see App.~\ref{app:numerical_details}] and noise model from Fig.~\ref{fig:conventional_encoding_random_CSS_code}.  Noise strength $\sigma^2$ and inter-cycle correlation range $\tau_c/t_\mathrm{cyc}$ are fixed, and inter-qubit correlation strength $\varrho$ varies over $[0,1]$.
(a)~Logical infidelity for a single QEC cycle for different eigenspaces of the three $Z$ stabilizers, as a function of $\varrho$.
The eigenvalues of the $X$ stabilizer generators in the codespace are fixed to $+1$.
There are $3$ PROSEs (i.e.,~optimal encodings) for each $\varrho>0$, indicated by dashed white lines; at $\varrho=0$ all encodings perform equivalently.
(b)~Logical infidelity $\LIF(R)$ for $R=100$ as $\varrho$ is varied, comparing conventional encoding (black circles) and PROSE encoding (blue squares); the latter yields a significant improvement. Solid lines
show the results of  noise-averaging the exact expression 
Eq.~\eqref{eq:LIF_in_terms_of_Gamma}; error bars are within marker size, dashed
lines show the analytical estimate from a second-order cumulant
expansion, using the leading-order approximations 
in Eq.~\eqref{eq:bare_model_cumulants}.%
}
    \label{fig:encoding_eigenspace_optimization}
\end{figure}

Figure~\ref{fig:encoding_eigenspace_optimization} illustrates both how one identifies the PROSE, and the resulting performance gain it yields over  $100$ QEC cycles, using one of the $d_Z=5$ CSS codes and noise model from Fig.~\ref{fig:conventional_encoding_random_CSS_code} [see App.~\ref{app:numerical_details}]. 
The noise is stationary and $R_c\sigma^{d_Z-1}\ll1$, so the PROSE can be identified from a single-cycle comparison of encodings, shown in Fig.~\ref{fig:encoding_eigenspace_optimization}\subref{fig:encoding_eigenspace_search}. For a CSS code, only the codespace-associated $Z$ stabilizer eigenvalues affect $\LIF(R)$, so we fix the $X$ stabilizer eigenvalues to $+1$ and compare the eight resulting eigenspaces as $\varrho$ varies.
We see that as the inter-qubit correlation strength $\varrho$ increases, the logical infidelity becomes encoding dependent, with the conventional encoding performing the worst, as expected for positively correlated noise. Three PROSEs that perform near-equivalently (up to sampling error) emerge, marked by dashed white lines. 
% Their near-equivalent performance can be understood heuristically by using a lookup table for the code to assess how each encoding choice affects interference between the amplitudes of the dominant uncorrectable errors. For the correlation model in Eq.~\eqref{eq:lorentzian_noise_correlations}, all such interference-like contributions have comparable magnitude, so the infidelity is largely controlled by how many of these error amplitudes are made to interfere destructively by a given encoding choice. The three marked eigenspaces make the same number of these error amplitudes interfere destructively, and thus perform equivalently at least at leading-order in noise strength. Encoding into any one of these stabilizer eigenspaces suppresses the logical infidelity, as shown in Fig.~\ref{fig:encoding_eigenspace_optimization}\subref{fig:performance_enhancement_after_EO}.
Using any of these optimal eigenspaces yields a strong advantage over $100$ cycles, as shown in Fig.~\ref{fig:encoding_eigenspace_optimization}\subref{fig:performance_enhancement_after_EO}.

\subsubsection{Turning noise correlations into a resource}
\label{sec:noise_correlations_as_a_resource}

Noise correlations, particularly positive ones, are generally considered detrimental to QEC [see e.g.,~our discussion in Sec.~\ref{sec:positive_correlations_odd_dz_conventional_encoding}]. Fig.~\ref{fig:encoding_eigenspace_optimization}\subref{fig:performance_enhancement_after_EO} however shows that with PROSE encoding, even positive correlations can \emph{reduce} the logical infidelity below its value for uncorrelated noise of equal strength. This means that, with an appropriate choice of stabilizer eigenspace, noise correlations can become a resource. We now provide a simple yet striking example of this phenomenon.

\begin{figure}[t]
    \centering
    \begin{minipage}[c]{0.33\linewidth}
        \centering
        \includegraphics[width=\linewidth]{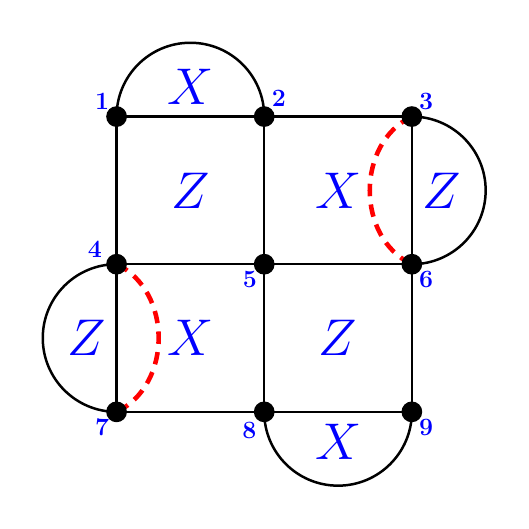}
    \end{minipage}\hfill
    \begin{minipage}[c]{0.67\linewidth}
        \centering
        \includegraphics[width=\linewidth]{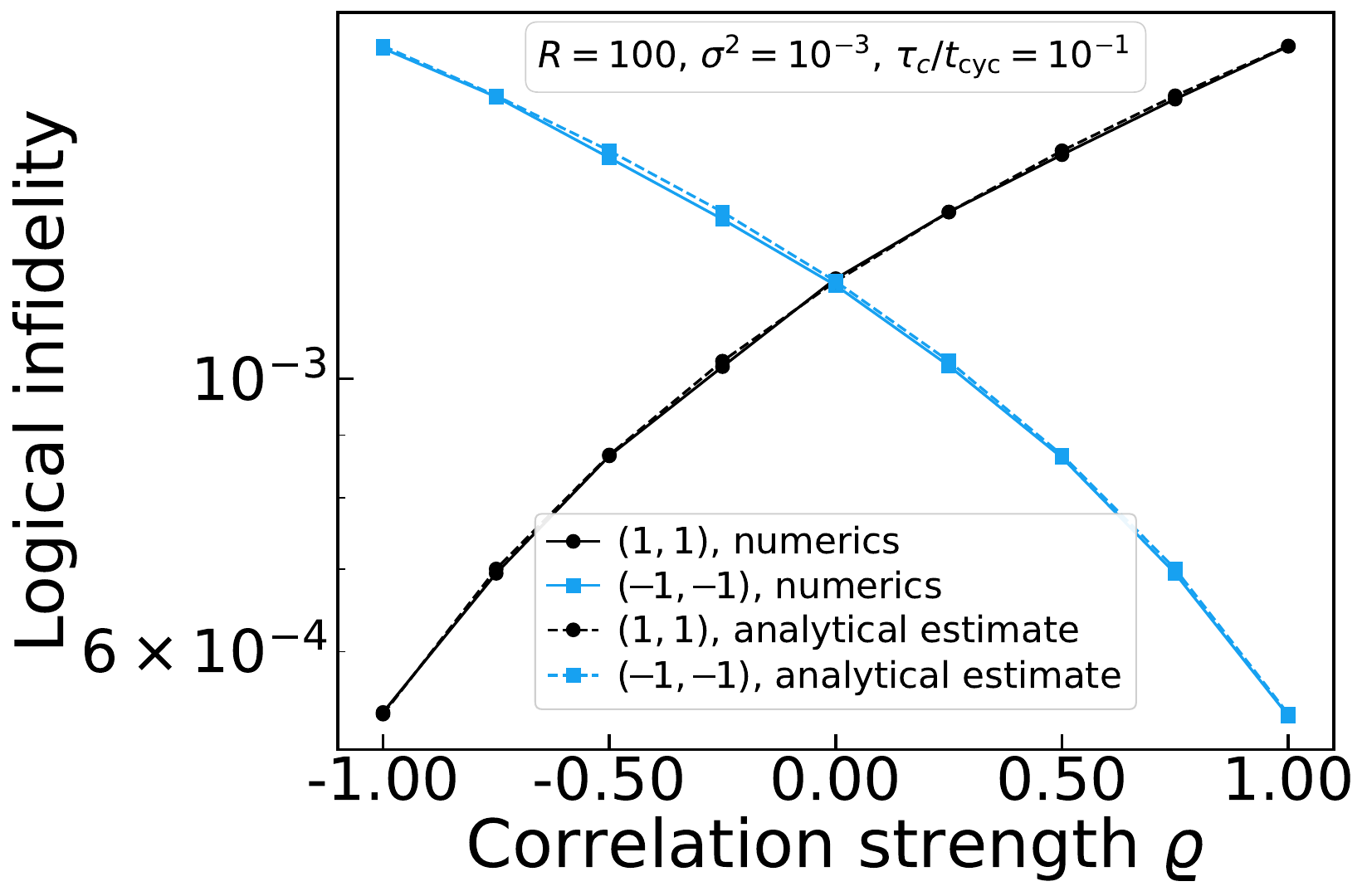}
    \end{minipage}
           \caption{%
\textbf{Impact of correlations with and without PROSE encoding.}
% \textbf{A suitable choice of encoding eigenspace can turn noise correlations into a resource.} 
Left: A distance-$3$ rotated surface code, with $X$ and $Z$ stabilizer generators and qubits labeled. Right: Logical infidelity $\LIF(R)$ after $R=100$ QEC cycles for this code, under a simplified version of the noise model in Eq.~\eqref{eq:lorentzian_noise_correlations} with intra-cycle correlations only between the qubit pairs $(3,6)$ and $(4,7)$, both controlled by $\varrho$. We vary $\varrho$ while holding the noise strength $\sigma^2$ and inter-cycle correlation range $\tau_c/t_\mathrm{cyc}$ fixed. For each value of $\varrho$, the PROSE is one of two codespaces distinguished by the eigenvalues of $\hat Z_3\hat
Z_6$ and $\hat Z_4\hat Z_7$, with all other stabilizer eigenvalues fixed to $+1$. For $\varrho>0$, the $(-1,-1)$ eigenspace (blue, squares) lowers the logical infidelity below the uncorrelated value; for $\varrho<0$, the $(+1,+1)$ eigenspace (black, circles) does the same. Solid lines show numerical results, obtained by noise-averaging the exact expression in Eq.~\eqref{eq:LIF_in_terms_of_Gamma}; error bars within marker size. Dashed lines show the analytical estimate from a second-order cumulant expansion, using the leading-order approximations in Eq.~\eqref{eq:bare_model_cumulants}.%
}
    \label{fig:encoding_eigenspace_can_make_some_intra_cycle_correlations_a_resource}
\end{figure}

Consider $100$ QEC cycles of a distance-$3$ rotated surface code~\cite{dennis_topological_2002,fowler_surface_2012}, a CSS code with $X$ and $Z$ stabilizer generators as shown in Fig.~\ref{fig:encoding_eigenspace_can_make_some_intra_cycle_correlations_a_resource}. For analytical transparency, we adopt a simplified version of the noise model in Eq.~\eqref{eq:lorentzian_noise_correlations}, retaining intra-cycle correlations only between qubits $3$ and $6$ and between qubits $4$ and $7$, each of strength $\varrho$, setting $\tau_c/t_\mathrm{cyc}=0.1$.

Since the noise is stationary and inter-cycle correlations are negligible, Eq.~\eqref{eq:LIF_when_PROSE_can_be_efficiently_found} yields a good approximation for the logical infidelity. The quantity $\avg{\chi^\mathrm{ZZ}(\anglevec{1})}$ can in turn be approximated by contributions of the dominant uncorrectable errors [Eq.~\eqref{eq:bare_model_first_cumulant}]. Most dominant uncorrectable errors within a syndrome-specific set are related by the stabilizers $\hat{Z}_3\hat{Z}_6$ or $\hat{Z}_4\hat{Z}_7$. The interference of their amplitudes is thus primarily controlled by which eigenspace of these two stabilizers defines the codespace [see below Eq.~\eqref{eq:first_cumulant_opened_up}]. By choosing the appropriate eigenspace, intra-cycle correlations present in this noise model can be leveraged to reduce the infidelity below the uncorrelated limit. We thus have a concrete example of how noise correlations can be exploited as a resource.

\subsubsection{Relation to previous work}

The idea to suppress logical infidelity under coherent errors by exploiting its encoding-dependence has only recently begun to be explored. We briefly relate PROSE encoding to two closely related works.

Ref.~\cite{witzel_correcting_2026} studies correlated coherent errors along one axis and comparable stochastic Pauli noise along an orthogonal axis, in a perfect QEC gadget. Rather than applying active corrections, it uses virtual Pauli-frame updates, so that the orthogonal stochastic noise component suppresses coherent logical errors by effectively randomizing the encoding. PROSE encoding instead optimizes the encoding and applies corrections to preserve it, requires no orthogonal noise component, and minimizes the total logical infidelity over encoding choices rather than suppressing coherent logical errors alone. Further, as we show in Sec.~\ref{sec:comparison}, PROSE encoding can be paired with a coherent logical error suppression technique such as logical Pauli twirling to lower the infidelity below what coherent logical error suppression alone achieves, across a wide range of correlation structures and parameter regimes.

Ref.~\cite{debroy_optimizing_2021} is more directly related. Using Shor's code, it experimentally demonstrated that encoding in a protected stabilizer eigenspace improves performance under a simple coherent-error model with a common rotation angle on all data qubits. Our work generalizes this technique to arbitrary codes and noise distributions, shows how the PROSE can be identified efficiently more generally, and compares it in detail against standard coherent error suppression techniques. We also identify a complementary pairing of PROSE with one of these techniques to further improve performance [see Sec.~\ref{sec:comparison}].

\subsection{Logical Pauli twirling \texorpdfstring{\protect\iconLT}{}}
\label{sec:logical_twirling}

Optimizing the encoding does not eliminate coherent logical errors, which can still accumulate over cycles, increasing the logical infidelity above the estimate set by the incoherent errors alone. To effectively suppress these errors, we consider a different error suppression method, namely logical Pauli twirling (LT)~\cite{cai_mitigating_2020,beale_randomized_2023}.

\subsubsection{Quantifying noise- and logical Pauli-averaged performance}

In the QEC+LT protocol, in each cycle $r$, a logical Pauli gate $\hat{P}_\mathrm{L}^{(r)}\in
\{\hat{I},\hat{\bar X}_L,\logicalZ,i\hat{\bar X}_L\logicalZ\}$, chosen independently and uniformly at random, is inserted before the noisy evolution. Here, $\hat{\bar X}_L$ denotes a representative of the logical $X$ operator \footnote{For notational convenience, we take the nontrivial logical operator in $\mathbb{N}_Z(\mathbb S)$ to define the logical $Z$ axis. Averaging over the logical Pauli ensemble $\{\hat{I},\hat{\bar X}_L,\logicalZ,i\hat{\bar X}_L\logicalZ\}$ then simply removes the $\logicalZLZ$ term in the single-cycle channel. More generally, coherent logical errors may lie along an arbitrary logical axis $\hat{\bar P}_L$, with the single-cycle channel $\Lambda(\anglevec{r})=\chi^\mathrm{II}(\anglevec{r})\mathcal{I}+\chi^\mathrm{ZI,im}(\anglevec{r})\mathcal{H}+\chi^\mathrm{ZZ}(\anglevec{r})\mathcal{\bar P}_L$. Two twirling protocols are then possible: averaging over an ensemble with equal total weight on logical operators that commute and anticommute with $\hat{\bar P}_L$, which straightforwardly generalizes the case considered here; or averaging over the logical Pauli ensemble, which removes the $\logicalZLZ$ term but also modifies the $\mathcal{\bar P}_L$ term by removing logical Pauli off-diagonal terms within $\mathcal{\bar P}$. For errors along logical $Z$, as considered here, the two coincide, so we do not pursue this distinction further.}. Then, the same logical Pauli gate is applied after the correction step. These gates can be absorbed into the initial state preparation and the correction steps. They do not alter the action of the QEC gadget and merely dress the induced logical channel of Eq.~\eqref{eq:single_cyc_channel}. For a single noise realization and a fixed sequence of logical Paulis, the data-qubit evolution in cycle $r$ is now described by the channel:
\begin{equation}
\hspace{-2mm}{\mathcal P}_\mathrm{L}^{(r)}\!
\Big(
\chi^\text{II}(\anglevec{r})\mathcal{I}
\!+\!\chi^{\text{ZI},\mathrm{im}}(\anglevec{r})\logicalZLZ
\!+\!\chi^\text{ZZ}(\anglevec{r})\bar{\mathcal{Z}}_L
\Big)\!
{\mathcal P}_\mathrm{L}^{(r)},
\label{eq:LT_single_cycle_def}
\end{equation}
where ${\mathcal P}_\mathrm{L}^{(r)}[\cdot]=\hat{P}_\mathrm{L}^{(r)}[\cdot]\hat{ P}_\mathrm{L}^{(r)}$ and the other operators and channel coefficients are as defined in Eqs.~\eqref{eq:operators_in_the_single_cycle_channel} and \eqref{eq:explicit_process_matrix_elements}. Conjugation by ${\mathcal P}_\mathrm{L}^{(r)}$ leaves the $\mathcal I$ and $\bar{\mathcal Z}_L$ terms invariant, but flips the sign of the $\logicalZLZ$ term when $\hat{P}_\mathrm{L}^{(r)}$ anti-commutes with $\hat{Z}_L$. This amounts to randomly flipping the sign of the angle of the coherent logical error in each QEC cycle.

To quantify the effect of this randomization on QEC performance after $R$ cycles, we average the channel describing the data-qubit evolution during these $R$ cycles over both the random logical Pauli gates and the noise. We first average over the logical Pauli gates. Since these gates are chosen independently, this average factorizes over cycles. Averaging each single-cycle channel yields:
\begin{align}
\Lambda_{\mathrm{LT}}(\anglevec{r})
&=\frac{1}{4}\sum_{\hat{P}_L^{(r)}}
{\mathcal P}_\mathrm{L}^{(r)}
\Lambda(\anglevec{r})
{\mathcal P}_\mathrm{L}^{(r)}\\
&=
\chi^\text{II}(\anglevec{r})\mathcal{I}
+
\chi^\text{ZZ}(\anglevec{r})\bar{\mathcal Z}_L.
\notag
\end{align}
% Logical twirling therefore tailors each single-cycle channel into a stochastic logical Pauli channel. 

Composing these twirled single-cycle channels produces the logical Pauli-twirled $R$-cycle channel for a single noise realization. We then average this channel over the noise, and use it to compute the logical infidelity as defined in Eq.~\eqref{eq:average_logical_infidelity}, quantifying QEC+LT performance:
\begin{equation}
    \LIF_\mathrm{LT}(R)=\frac{1-\avg{\Gamma_\mathrm{LT}(R,\vec{\theta}\,) }}{3},
    \label{eq:QEC_plus_LT_LIF}
\end{equation}
where
\begin{equation}
\Gamma_\mathrm{LT}(R,\vec{\theta}\,) =
\prod_{r=1}^R
\left(
1-2\chi^\text{ZZ}(\anglevec{r})
\right).
\label{eq:Gamma_LT}
\end{equation}

\subsubsection{Comparison with QEC-only and QEC+LDD}
\label{sec:LT_versus_QEC_only}

Using $ \LIF_\mathrm{LT}(R)$, we now compare QEC+LT to QEC-only and QEC+LDD. For weak noise, the comparison simplifies by approximating $\Gamma_\mathrm{LT}(R,\vec{\theta})$ as in Eq.~\eqref{eq:exp_approx_cycle}, yielding:
\begin{equation}
    \LIF_\mathrm{LT}(R)\approx\frac{1-\Bigavg{
\exp\!\Big(
-2\sum_{r=1}^R
\chi^\text{ZZ}(\anglevec{r})
\Big)
}}{3}.
\label{eq:LIF_LT_approx_exp}
\end{equation}

First, we compare with QEC-only. In principle, applying LT could \textit{increase} the logical infidelity, just as Pauli twirling on a single qubit can. For weak noise, however, LT increases the logical infidelity at most negligibly, and generally decreases it strongly. Concretely, to the accuracy of the above approximation and the corresponding approximation in Eq.~\eqref{eq:LIF_exp_bare}, we have
\begin{flalign}
    \LIF(R)-\LIF_\mathrm{LT}(R)&\approx \frac{1}{3}\Bigavg{\exp\!\Big(
-2\sum_{r=1}^R
\chi^\text{ZZ}(\anglevec{r})
\Big)\label{eq:LT_betters_QEC}\\
    &\quad\times\Big(1\!-\!\cos\!\big(2\sum_{r=1}^R\chi^\text{ZI,im}(\anglevec{r})\big)\Big)}\geq 0.\notag
\end{flalign}
QEC+LT's advantage is large when coherent logical errors contribute substantially to the QEC-only logical infidelity.

Next, we compare with the QEC+LDD protocol (detailed in App.~\ref{app:dynamical_decoupling}), in which the QEC cycles are conjugated with an optimized, deterministic sequence of logical Pauli gates to refocus and suppress coherent logical errors. Although LDD optimizes over the same ensemble that LT averages, we find that the two protocols perform comparably for weak noise. This motivates our focus on LT, since it avoids a generally difficult optimization of the decoupling sequence. We provide intuition for these comparisons in App.~\ref{app:DD_LT_correspondence}.

% LDD may be preferred only in the special cases where the optimal sequence is easy to find, for the marginal advantage it offers over LT\textemdash for example, logical spin echo is near-optimal for positive, long-ranged inter-cycle correlations [App.~\ref{app:spin_echo}].

% We can build intuition for this near-equivalent performance using the weak-noise approximation of the exact single-cycle channel in Eq.~\eqref{eq:single_cyc_channel} by Eq.~\eqref{eq:approximate_induced_channel}. LT removes the logical Pauli off-diagonal term $\chi^{\mathrm{ZI},\mathrm{im}}\logicalZLZ$ from each single-cycle channel while LDD refocuses the coherent logical error  $\mathcal{R}_{\logicalZ}\!\big(\chi^{\mathrm{ZI},\mathrm{im}}\big)$ in each cycle so that their accumulation is suppressed, on average. These two actions are equivalent to the accuracy of Eq.~\eqref{eq:approximate_induced_channel}.

\subsubsection{Impact of correlated coherent errors after applying LT}
\label{sec:LT_correlation_effects}

Eq.~\eqref{eq:LIF_LT_approx_exp} provides a convenient starting point to assess noise correlation effects on $\LIF_\mathrm{LT}(R)$. Approximating $\LIF_\mathrm{LT}(R)$ using a second-order cumulant expansion yields:
\begin{equation}
    \LIF_\mathrm{LT}(R)\approx\frac{1-\exp\!\Big(\kappa_\mathrm{1,LT}(R)+\kappa_\mathrm{2,LT}(R)/2\Big)}{3}.\label{eq:LIF_LT_cumulant_expansion}
\end{equation}
The QEC+LT cumulants are related to the cumulants of the QEC-only case with the same encoding as:
\begin{subequations}\label{eq:LT_cumulants}
\begin{align}
   \kappa_\mathrm{1,LT}(R)&=\kappa_\mathrm{1}(R),\\
\kappa_\mathrm{2,LT}(R)&=\kappa_\mathrm{2}^\text{incoh}(R).
    \label{eq:second_cumulant_LT}
\end{align}
\end{subequations}
% From Eq.~\eqref{eq:cumulant_scalings}: $\kappa_\mathrm{1,LT}\!\sim\!R\sigma^{d_Z+1}$ and $\kappa_\mathrm{2,LT}\!\sim\!RR_c\sigma^{2d_Z+2}$.

With LT removing $\kappa^\mathrm{coh}_2(R)$, inter-cycle correlation effects on the QEC+LT protocol closely mirror those in an even-$d_Z$ code under QEC-only [Sec.~\ref{sec:features_common_across_odd_dZ_codes}]:
\begin{enumerate}
    \item Inter-cycle correlation effects are suppressed by LT; $\kappa_{2,\mathrm{LT}}(R)/\kappa_{1,\mathrm{LT}}(R)$ is suppressed by $\sigma^2$ relative to the QEC-only case.
    \item These correlations are beneficial since $\kappa_{2,\mathrm{LT}}(R)\ge0$. LT thus turns inter-cycle correlations into a resource, albeit a weak one.
    \item With the conventional encoding and positive noise correlations, $\LIF_\mathrm{LT}(R)$ decreases monotonically in inter-cycle correlation strength or range, to leading-order in the noise strength. 
% (Eq.~\eqref{eq:bare_model_second_cumulant_incoh} reduces to a polynomial in $\boldsymbol{\Sigma}$ with positive coefficients).
\end{enumerate}
Applying LT to an odd-$d_Z$ code reproduces the advantage of using an even-$d_Z$ code under long-ranged correlations.

\begin{figure}
    \centering
    \includegraphics[width=0.9\linewidth]{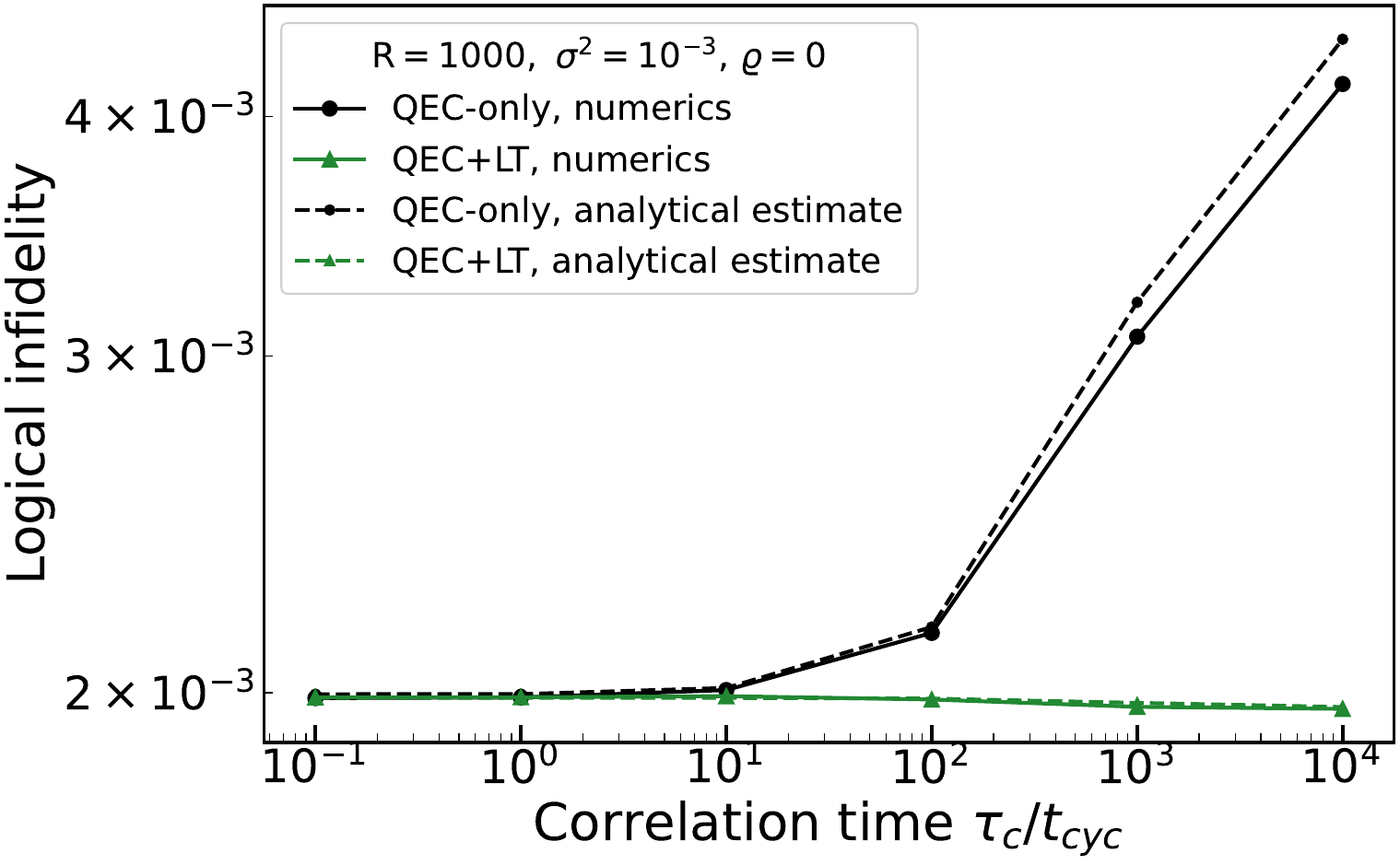}
\caption{%
\textbf{Fidelity improvement by applying logical Pauli twirling.} We plot the logical infidelity $\LIF(R)$ of the three-qubit phase-flip repetition code after $R=1000$ cycles of QEC-only (black, circles) and QEC+LT (green, triangles), for the noise model in Eq.~\eqref{eq:lorentzian_noise_correlations}. Noise strength $\sigma^2$ and inter-qubit correlation strength $\varrho$ are fixed, and inter-cycle correlation range $\tau_c/t_\mathrm{cyc}$ is varied.
Solid lines show numerical results, obtained by noise-averaging corresponding exact
expressions; error bars are within marker size. Dashed lines show analytical estimates from a second-order cumulant expansion, using
leading-order approximations from Eq.~\eqref{eq:bare_model_cumulants}: $\kappa_1(R)$ and $\kappa_2^\mathrm{coh}(R)$ for QEC-only, and
$\kappa_1(R)$ and $\kappa_2^\mathrm{incoh}(R)$ for QEC+LT.%
}
    \label{fig:LT_rep_3_code}
\end{figure}

Fig.~\ref{fig:LT_rep_3_code} illustrates these conclusions for the three-qubit phase-flip repetition code under the noise model of Eq.~\eqref{eq:lorentzian_noise_correlations}, with $\varrho=0$ and varying inter-cycle correlation range $\tau_c/t_\mathrm{cyc}$. For $\tau_c\ll t_\mathrm{cyc}$, QEC-only and QEC+LT perform similarly. For $\tau_c\gg t_\mathrm{cyc}$, the QEC-only logical infidelity grows as coherent logical errors accumulate across cycles, while LT suppresses this growth. In this regime, stronger and longer-ranged correlations reduce $\LIF_\mathrm{LT}(R)$, although negligibly.

\section{Combining PROSE Encoding and LT
\texorpdfstring{\protect\iconPECLT}{}}
\label{sec:comparison}

The previous section demonstrated that LT and PROSE encoding have complementary natures: LT suppresses coherent logical errors, while PROSE can be used to minimize the probability of incoherent logical errors. Combining them is thus a natural way to realize the benefits of both techniques. We show below that such a resulting QEC+LT+PROSE encoding protocol generally outperforms QEC-only, QEC+PROSE encoding, QEC+LT, and QEC+PT under stationary noise for a wide, practically relevant parameter regime (including the weak-noise, small-infidelity regime). 
% , including the weak-noise, small infidelity regime for stationary noise. 

% Comparisons against protocols combining PT with the other techniques are unnecessary: PT leaves the single-cycle logical channel with neither residual coherent errors for LT or LDD to suppress nor an eigenspace dependence for PROSE encoding to exploit [see App.~\ref{app:physical_Pauli_twirling}]. 
% We do not compare against protocols involving LDD since LT suppresses coherent logical errors comparably to LDD for weak noise while avoiding the search for an optimal decoupling sequence [see Sec.~\ref{sec:LT_versus_QEC_only}, App.~\ref{app:DD_LT_correspondence}]. We therefore take QEC+LT+PROSE encoding as the primary combined protocol here. Where the optimal decoupling sequence is known, LDD can replace LT as a supplement to PROSE encoding, providing a small additional gain.

\vspace{-2mm}
\subsection{Efficiently identifying the PROSE for QEC+LT}
\vspace{-2mm}

Before comparing with other techniques, we first outline when the PROSE for the QEC+LT protocol can be found efficiently, analogous to the discussion in Sec.~\ref{sec:efficiently_finding_the_PROSE}. For stationary noise, the PROSE can be found from a single-cycle comparison of stabilizer eigenspaces whenever the second cumulant is negligible compared to the first. For the QEC-only protocol, this corresponded to $R_c\sigma^{d_Z-1}\ll 1$. 

The use of LT expands the above regime where a single cycle can be used to find the PROSE, by removing the dominant coherent contribution to the second cumulant.  As a result, for QEC+LT the second cumulant is negligible compared to the first when the more forgiving condition  $R_c\sigma^{d_Z+1}\ll 1$ is satisfied (or more conservatively when $R\sigma^{d_Z+1}\ll 1$). The latter regime also captures the practically relevant range of $R$ over which the QEC+LT logical infidelity remains small (in this regime, $\kappa_{1,\mathrm{LT}}(R)\ll 1$ and thus, $\LIF_\mathrm{LT}(R)\ll1$). Thus, the upshot is that by using LT, it becomes easy (i.e., by studying just a single QEC cycle) to find the PROSE over a wider parameter regime. 
% PROSE for the QEC+LT protocol can thus be found tractably in practically relevant settings.

% In these regimes, LT and PROSE encoding suppress the logical infidelity complementarily: LT suppresses coherent logical errors, while PROSE encoding minimizes the noise-averaged probability of incoherent logical errors across stabilizer eigenspaces.

\vspace{-2mm}
\subsection{Advantages of combined LT+PROSE over other approaches}
\vspace{-2mm}
Now we compare QEC+LT+PROSE encoding to our other error suppression techniques for weak stationary noise in the wide parameter regime identified above where the PROSE for QEC+LT can be found efficiently i.e., $R_c\sigma^{d_Z+1}\ll 1$. We first illustrate our conclusions through a numerical example, then explain the observed hierarchy of protocols and establish its generality.

\begin{figure}[t]
    \centering
    \captionsetup[subfloat]{position=top,justification=raggedright,singlelinecheck=false}
    \subfloat[]{
        \includegraphics[width=\linewidth]{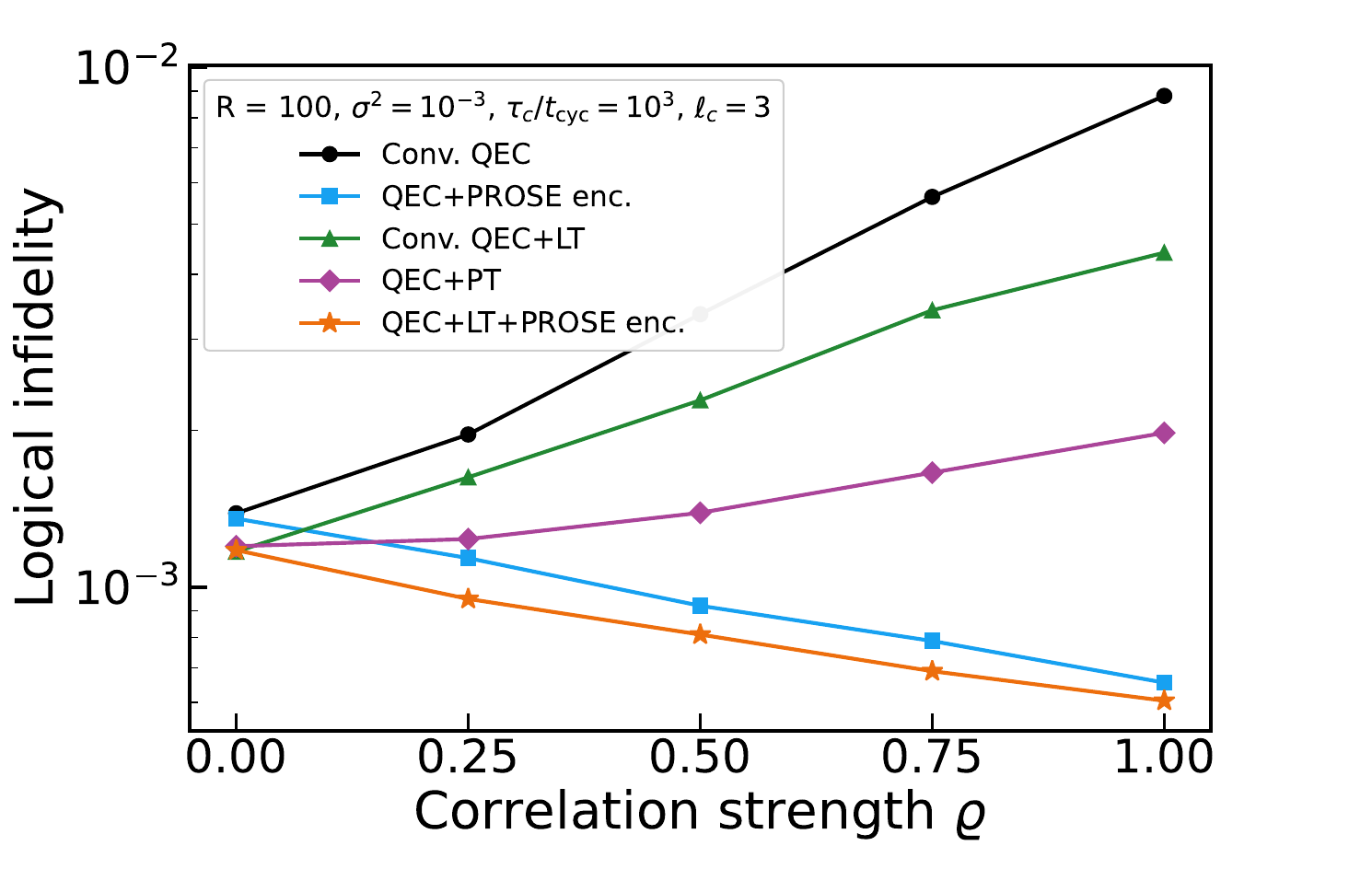}
        \label{fig:Surface_Code_FinalComparison}
    }
    \vspace{0.5em}
    \subfloat[]{
        \includegraphics[width=\linewidth]{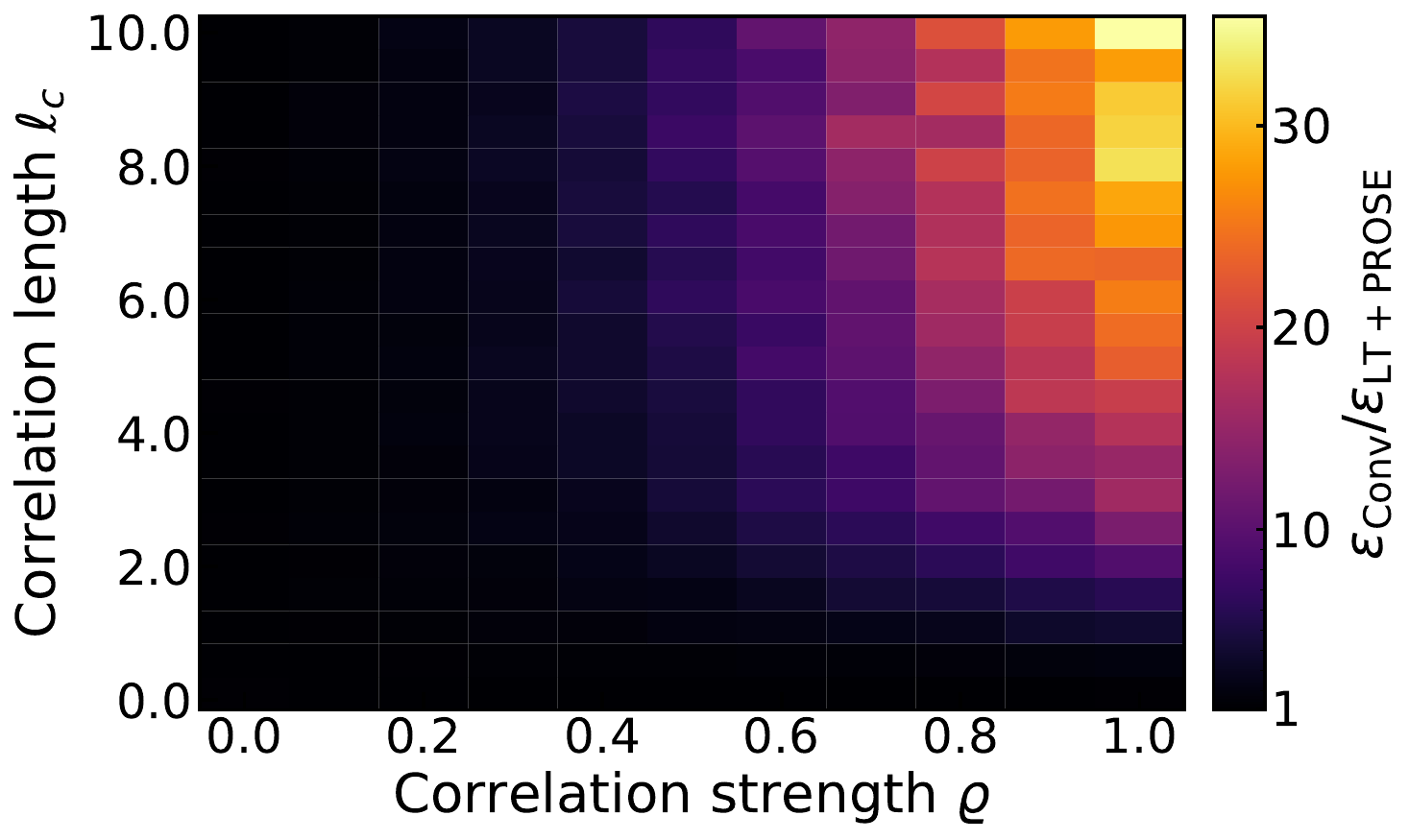}
        \label{fig:vary_varrho_lc}
    }
\caption{%
\textbf{Fidelity advantages of combining PROSE encoding and logical Pauli twirling.}
%Fidelity improvement on combining PROSE encoding and logical Pauli twirling, compared against other error suppression techniques; and fidelity ratio (conventional QEC/QEC+LT+PROSE) as noise parameters are varied.} 
Results here use the distance-$3$ rotated surface code under the noise model in Eq.~\eqref{eq:spatially_decaying_noise_correlations}, with noise strength $\sigma^2$, inter-cycle correlation range $\tau_c/t_\mathrm{cyc}$, inter-qubit correlation range $\ell_c$, and inter-qubit correlation strength $\varrho$.
(a)~Logical infidelity $\LIF(R)$ as a function of $\varrho$ for $R=100$, fixing the other parameters. For each $\varrho$, the PROSE for both QEC-only and QEC+LT is identified by a single-cycle comparison of encodings. We compare five protocols: conventional QEC-only, QEC+PROSE encoding, QEC+LT with conventional encoding, QEC+LT+PROSE encoding, and QEC+PT; QEC+LT+PROSE encoding performs best. Solid lines show numerical results, obtained by noise-averaging corresponding exact expressions; error bars are within the marker size.
(b)~Ratio of the logical infidelity of the conventional QEC-only protocol
to that of QEC+LT+PROSE encoding, as $\varrho$ and $\ell_c$ are varied. Results are obtained by noise-averaging exact expressions.
}
    \label{fig:Surface_Code_FinalComparison_Full}
\end{figure}

Figure~\ref{fig:Surface_Code_FinalComparison} presents this comparison for $100$ cycles of a distance-$3$ rotated surface code [see Fig.~\ref{fig:encoding_eigenspace_can_make_some_intra_cycle_correlations_a_resource}], with the noise specified by the covariance:
\vspace{-1mm}
\begin{equation}
    \hspace{-2mm}\boldsymbol{\Sigma}^{(r,r')}_{\ell\ell'}\!=\!\sigma^2
    \big((1-\varrho_{\ell\ell'})\delta_{\ell\ell'}+\varrho_{\ell\ell'}\big)
    \exp\Big(-\abs{r-r'}\frac{t_\mathrm{cyc}}{\tau_c}\Big).
    \label{eq:spatially_decaying_noise_correlations}
\end{equation}
In contrast to the correlation model of Eq.~\eqref{eq:lorentzian_noise_correlations}, we now consider the more realistic setting of an inter-qubit correlation strength $\varrho_{\ell\ell'}$ that decays with distance,
\vspace{-1mm}
\begin{equation}
    \varrho_{\ell\ell'}=\varrho\exp\left(-\frac{D(\ell,\ell')}{\ell_c}\right),
\end{equation}
with $D(\ell,\ell')$ the Euclidean distance between qubits on the surface code lattice and $\ell_c$ the correlation length. We set $\tau_c/t_\mathrm{cyc}=1000$ and $\ell_c=3$ in units of the nearest-neighbor spacing, and vary $\varrho$. 

Since $R_c\sigma^{d_Z-1}$ is sufficiently small for the above parameters, the PROSE for QEC-only matches that for QEC+LT (found from a single-cycle search) for each $\varrho$. We then compare five protocols over $100$ cycles: QEC-only with conventional encoding (all stabilizers $+1$), QEC+PROSE encoding, QEC+LT with conventional encoding, QEC+LT+PROSE encoding, and QEC+PT. As shown in the figure, QEC+LT+PROSE encoding performs at least as well as, and generally better than, every other technique considered.

% Moreover, in this example, for the two protocols that use PROSE encoding, the logical infidelity \emph{decreases} with increasing correlation strength. This reinforces the message of Sec.~\ref{sec:noise_correlations_as_a_resource}: with a suitable encoding, noise correlations become a resource. They can be exploited to drive the logical infidelity below its value for uncorrelated noise of equal strength.

Fig.~\ref{fig:vary_varrho_lc} compares conventional QEC-only to QEC+LT+PROSE encoding, varying $\varrho$ and $\ell_c$. The ratio of their infidelities exceeds one throughout and grows large for appreciable $\varrho$ or $\ell_c$, illustrating how dramatic the advantage provided by LT+PROSE can become.

We now explain these observations. While the advantage LT+PROSE encoding provides over QEC-only and QEC+LT using other encodings is more obvious, the advantage over QEC+PT is more subtle, which we discuss here. In the PT protocol, each QEC cycle is conjugated with a random sequence of Pauli gates on the physical qubits (in contrast, LT conjugates each cycle with random logical Paulis). Applying PT effectively twirls the noise on the physical qubits, tailoring coherent errors into stochastic Pauli errors [see App.~\ref{app:physical_Pauli_twirling} for a detailed discussion]. 

PT is equivalent to applying random logical Pauli gates while also randomizing the encoding eigenspace, each cycle. Like LT, PT thus suppresses the $\kappa^\mathrm{coh}_2(R)$ contribution present in the QEC-only case, leaving the QEC+PT infidelity first cumulant-dominated when $R_c\sigma^{d_Z+1}\ll 1$, just like the QEC+LT+PROSE encoding infidelity. 

The difference between the infidelities of the two protocols is then primarily captured by differences in their first cumulants. Since PT randomizes the encoding in each cycle, its first cumulant is the QEC-only first cumulant [Eq.~\eqref{eq:kappa1_BR}] averaged over encoding eigenspaces. This average simplifies under stationary noise, where the contribution of each cycle to the cumulant is independent of $r$, so
\begin{equation}
    \kappa_\mathrm{1,PT}(R)
    \!=\!
    -\frac{2R}{2^{n-1}}
    \!\!\!\sum_{\vec{g}\in\{0,1\}^{n-1}}
    \!\!\bigavg{\chi^\text{ZZ}(\anglevec{1},\vec{g})},
\end{equation}
with the dependence of $\chi^\mathrm{ZZ}$ on the encoding eigenspace labeled explicitly by the reference syndrome $\vec{g}$. PROSE encoding instead minimizes $\bigavg{\chi^\text{ZZ}(\anglevec{1},\vec{g})}$ over these eigenspaces (which generically is of course better than averaging over eigenspaces).  Hence QEC+LT+PROSE encoding performs at least as well as, and generally better than, QEC+PT, up to negligible second cumulant corrections, explaining the hierarchy in Fig.~\ref{fig:Surface_Code_FinalComparison}.

The above conclusions also apply to more general settings with stationary non-Gaussian noise, as none of the above comparisons rely on the error angles being Gaussian-distributed. We thus see that the combination of LT and PROSE encoding can offer significant advantages over other techniques for suppressing weak, correlated noise, for arbitrary choice of code.

\section{Discussion and outlook}
\label{sec:conclusion}

We have investigated correlated coherent $Z$ errors in stabilizer codes. For a fixed noise realization, we derived the exact logical channel induced by $R$ QEC cycles, revealing how the stabilizer eigenspace chosen as the codespace controls the interference of error amplitudes, and with it, the evolution of the logical qubit. Building on this, we developed a cumulant expansion of the noise-averaged logical infidelity, going beyond leading-order perturbation theory in the noise, and used it to uncover effects of Gaussian-correlated coherent errors that are universal across arbitrary stabilizer codes.

We then developed a practical error suppression strategy: encoding in the protected stabilizer eigenspace (PROSE) that minimizes the logical infidelity, which we showed can be identified efficiently in practically relevant cases. Strikingly, with PROSE encoding, noise correlations, even positive ones, can drive the logical infidelity below the uncorrelated baseline. This motivates a refined perspective: contrary to standard expectations, noise correlations can be exploited as a resource.

Finally, we showed that logical Pauli twirling (LT) complements PROSE encoding. For any weak, stationary noise (both Gaussian and non-Gaussian) and arbitrary codes, the combined protocol matches or outperforms every standard coherent error suppression technique considered here, establishing PROSE+LT as a broadly applicable route to suppressing correlated coherent errors.

Two extensions of this work naturally arise: generalizing the noise model and relaxing the assumption of ideal syndrome extraction. In the former direction, our cumulant-expansion formalism readily extends to more general noise models, such as coherent errors along two orthogonal axes. These generalizations will be the subject of upcoming work. In the latter direction, incorporating imperfections in the QEC gadget itself is an important next step. A complete treatment of the interplay between correlated coherent errors and imperfect syndrome extraction may reveal qualitatively new effects and more effective error suppression strategies.

% Two extensions of this work naturally arise: generalizing the noise model and relaxing the assumption of ideal syndrome extraction. The basic cumulant-expansion formalism we used can be extended to probe important directions of the former kind, including coherent errors along multiple axes, such as coherent $X$ and $Z$ errors, despite the non-commuting nature of these errors, which will be topic of upcoming work.

% Incorporating imperfections in the QEC gadget itself is also important, which generally demands modified correction procedures. A complete treatment of the interplay between correlated coherent errors and imperfect syndrome extraction may reveal qualitatively new effects and more effective error suppression strategies.

\section{Acknowledgments}
We thank Daniel Weiss, James Teoh, Shantanu Mundhada, Qile Su, and Gideon Lee for helpful discussions. This research was sponsored by the Army Research Office under Grant No. W911NF-23-1-0116. A.B. acknowledges support of the FRQNT through a doctoral scholarship.  A.C. acknowledges support from the Simons Foundation through a Simons Investigator Award (Grant No. 669487). This manuscript has been coauthored by UT-Battelle, LLC, under Contract No. DE-AC05-00OR22725 with the U.S. Department of Energy (DOE). This research used Oak Ridge Leadership Computing Facility's resources, which is a DOE Office of Science User Facility supported under Contract DE-AC05-00OR22725.

\appendix
\section{Computing the channel induced by a single QEC cycle}
\label{app:computing_the_single_cycle_channel}

In this section, we derive the channel induced by a single QEC cycle on data qubits supported on the codespace $\mathcal{C}_{\vec g}$ [Eq.~\eqref{eq:single_cyc_channel}] for a fixed noise realization. Our setting is an $\llbracket n,1\rrbracket$ stabilizer code with stabilizer group $\mathbb{S}$ generated by $G=\{\hat g_\ell\}_{\ell=1}^{n-1}$, and data qubits initialized in a state $\hat\rho^{(0)}$ supported on $\mathcal{C}_{\vec g}$. 

The data qubits are first subject to errors described by the superoperator in Eq.~\eqref{eq:expanded_noise_superoperator}, after which their state is:
\begin{align}
\hat{\rho}^{\mathrm{EA},1}(\anglevec{1})
&= \mathcal{N}(\anglevec{1})\!\big[\hat{\rho}^{(0)} \big] \label{eq:EA_state_1}\\
&= \sum_{\mathclap{\hat E,\hat E'\in\overline{\mathbb{P}}_{n,Z}}}
c_{\hat{E}}(\anglevec{1})\,c_{\hat{E'}}^*\!(\anglevec{1})\;
\hat E\,\hat{\rho}^{(0)}\,\hat E' \notag,
\end{align}
where $c_{\hat E}(\anglevec{r})$ are the error amplitudes defined in Eq.~\eqref{eq:cE_def}.

The stabilizer generators are then measured. The projector associated with the measured syndrome $\vec m$ is
\begin{equation}
    \hat\Pi_{\vec m}
    =
    \prod_{\ell=1}^{n-1}
    \frac{\hat I+(-1)^{m_\ell}\hat g_\ell}{2}.
\end{equation}
To characterize the average performance of the QEC protocol, we average over measured syndromes, obtaining:
\begin{align}
\hat{\rho}^{\mathrm{PM},1}(\anglevec{1})
&=
\sum_{\vec m}
\hat\Pi_{\vec m}\,
\hat{\rho}^{\mathrm{EA},1}(\anglevec{1})\,
\hat\Pi_{\vec m}
\label{eq:PMstate1}\\
&=
\sum_{\vec m}\,\,\,\,\,\,\,\,
\sum_{\mathclap{\hat E,\hat E'\in\overline{\mathbb{P}}_{n,Z}^{\vec m\oplus\vec g}}}
c_{\hat{E}}(\anglevec{1})c_{\hat{E'}}^*\!(\anglevec{1})
\hat E\hat\rho^{(0)}\hat E',
\notag
\end{align}
where $\overline{\mathbb{P}}_{n,Z}^{\vec m\oplus\vec g}$ denotes the set of $Z$ errors consistent with the measured outcome $\vec{m}$, defined as in Eq.~\eqref{eq:compatible_set} and $\oplus$ denotes addition modulo $2$. The measurement thus removes all terms in Eq.~\eqref{eq:EA_state_1} involving pairs of errors consistent with different syndromes.

After the syndrome measurement, a correction is applied conditioned on the measured syndrome. As discussed in the main text, we use a maximum-likelihood decoder designed for an uncorrelated, stochastic Pauli $Z$ error model, with phase-flip probabilities set by the marginal error probabilities of our noise model. Conditioned on the measured syndrome $\vec m$, the decoder applies a correction operator $\hat R_{\vec m\oplus\vec g}\in \overline{\mathbb{P}}_{n,Z}^{\vec m\oplus\vec g}$, which corrects the more probable class of errors within the syndrome-specific set $\overline{\mathbb{P}}_{n,Z}^{\vec m\oplus\vec g}$. After applying these corrections, we get:
\begin{align}
\hat{\rho}_{D}^{\mathrm{PC},1}(\anglevec{1})
&=
\sum_{\vec m}
\hat R_{\vec m\oplus\vec g}
\hat\Pi_{\vec m}
\hat{\rho}^{\mathrm{EA},1}(\anglevec{1})
\hat\Pi_{\vec m}
\hat R_{\vec m\oplus\vec g}^{\dagger}
\label{eq:AC_state_1}\\
&=
\sum_{\mathclap{
\substack{\vec s\\
\hat E,\hat E'\in\overline{\mathbb{P}}_{n,Z}^{\vec s}}}}
c_{\hat E}(\anglevec{1})c_{\hat E'}^*(\anglevec{1})
\hat R_{\vec s}\hat E\hat\rho^{(0)}\hat E'\hat R_{\vec s}^{\dagger}.
\notag
\end{align}

This post-correction state can be simplified. Fix a syndrome $\vec s$ and an error $\hat E\in\overline{\mathbb{P}}_{n,Z}^{\vec s}$. Since the correction operator $\hat R_{\vec s}$ is chosen from the same syndrome-specific set as $\hat E$, the product $\hat R_{\vec s}\hat E$ belongs to $\overline{\mathbb{N}_Z(\mathbb{S})}$. This set contains a single nontrivial logical operator, up to multiplication by $Z$ stabilizers. As discussed in the main text, for notational convenience, we choose the logical states so that this operator is the logical $Z$ operator $\logicalZ$. Thus any $\hat P\in\overline{\mathbb{N}_Z(\mathbb{S})}$ decomposes as:
\begin{equation}
\hat P=\hat S_{Z_{\hat P}}\,\logicalZ^{\,\nu_{\hat P}},
\label{eq:appendix_decomposition_rule}
\end{equation}
where $\hat S_{Z_{\hat P}}\in\mathbb{S}_Z$ and $\nu_{\hat P}=0$ or $1$.

Within each syndrome-specific set $\overline{\mathbb{P}}_{n,Z}^{\vec s}$, the correction maps every error, up to multiplication by a $Z$ stabilizer, either to the identity or to $\logicalZ$. In the former case the error is successfully corrected, whereas in the latter it produces a logical $Z$ error. With this identification:
\begin{align}
\hat{\rho}_{D}^{\mathrm{PC},1}(\anglevec{1})
=&\sum_{\vec s}
   \Big(
              \abs{A_{\vec s}(\anglevec{1})}^2\,\mathcal{I}
            +\abs{B_{\vec s}(\anglevec{1})}^2\,\logicalZsuperop
            \label{eq:DQ_evolution}\\[-2pt]
  &\hspace{9mm}
            +\Im\!\big[A_{\vec s}(\anglevec{1})B_{\vec s}^*(\anglevec{1})\big]\logicalZLZ
            \notag\\
  &\hspace{9mm}
            +\Re\!\big[A_{\vec s}(\anglevec{1})B_{\vec s}^*(\anglevec{1})\big]
            \{\logicalZ,\cdot\}_+
   \Big)\hat\rho^{(0)} .
\notag
\end{align}
The coefficients
\begin{subequations}\label{eq:AB_coeff_definitions}
\begin{align}
A_{\vec s}(\anglevec{1})
&=
\sum_{\hat E\in\overline{\mathbb{P}}_{n,Z}^{\vec s,\checkmark}}
(-1)^{\phi_{\hat R_{\vec s}\hat E}}\,
c_{\hat E}(\anglevec{1}),\\
B_{\vec s}(\anglevec{1})
&=
\sum_{\hat E\in\overline{\mathbb{P}}_{n,Z}^{\vec s,\times}}
(-1)^{\phi_{\hat R_{\vec s}\hat E}}\,
c_{\hat E}(\anglevec{1})
\end{align}
\end{subequations}
collect the amplitudes of the correctable and uncorrectable errors within each syndrome-specific set. Here
$\overline{\mathbb{P}}_{n,Z}^{\vec s,\checkmark}$ denotes the subset of $\overline{\mathbb{P}}_{n,Z}^{\vec s}$ containing errors that are successfully corrected, while $\overline{\mathbb{P}}_{n,Z}^{\vec s,\times}$ denotes the subset containing errors that produce a logical $Z$ error. For each $\hat P\in\overline{\mathbb{N}_Z(\mathbb{S})}$, the phase factor $(-1)^{\phi_{\hat P}}=\pm1$ is the eigenvalue of the stabilizer $\hat S_{Z_{\hat P}}$ on the chosen codespace:
\begin{equation}
    \hat S_{Z_{\hat P}}\ket{\psi}
    =
    (-1)^{\phi_{\hat P}}\ket{\psi},
    \qquad
    \forall\,\ket{\psi}\in\mathcal{C}_{\vec g},
\end{equation}
where $\hat S_{Z_{\hat P}}=\hat P\logicalZ^{\nu_{\hat P}}$ is the $Z$ stabilizer factor in Eq.~\eqref{eq:decomposition_rule}. These phases carry the entire dependence of the post-correction state on the encoding stabilizer eigenspace.

Since $\hat{\rho}_{D}^{\mathrm{PC},1}(\anglevec{1})$ is a valid density matrix for inputs supported on $\mathcal{C}_{\vec g}$, the anti-commutator term vanishes i.e.,
\begin{equation}
\sum_{\vec{s}}
\Re\!\big[A_{\vec{s}}(\anglevec{1})\, B_{\vec{s}}^*(\anglevec{1})\big] = 0,
\label{eq:trace_preserving_constraint}
\end{equation}
which we also verify explicitly in App.~\ref{app:imaginary_channel_coefficient}. From Eq.~\eqref{eq:DQ_evolution}, we can then extract the single-cycle channel induced on the codespace (Eq.~\eqref{eq:single_cyc_channel} in the main text):
\begin{equation}
\Lambda(\anglevec{1})
\!=\!
\chi^\text{II}(\anglevec{1})\,\mathcal{I}
\!+\!\chi^{\text{ZI},\mathrm{im}}(\anglevec{1})\,\logicalZLZ
\!+\!\chi^\text{ZZ}(\anglevec{1})\,\bar{\mathcal{Z}}_L ,
\label{eq:app_single_cyc_channel}
\end{equation}
with coefficients defined as:
\begin{subequations}
\begin{align}
\chi^\mathrm{II}(\anglevec{1})
&= \sum_{\vec s}\big|A_{\vec s}(\anglevec{1})\big|^2, \\
\chi^{\mathrm{ZI},\mathrm{im}}(\anglevec{1})
&= \Im\!\Big[\sum_{\vec s} A_{\vec s}(\anglevec{1})\, B_{\vec s}^*(\anglevec{1})\Big], \\
\chi^\mathrm{ZZ}(\anglevec{1})
&= \sum_{\vec s}\big|B_{\vec s}(\anglevec{1})\big|^2 .
\end{align}
\end{subequations}
These coefficients are expressed explicitly in terms of the error amplitudes in Eq.~\eqref{eq:explicit_process_matrix_elements} of the main text, after simplifying the phase factors using the relation:
\begin{equation}
    (-1)^{\phi_{\hat P}+\phi_{\hat P'}}
    =
    (-1)^{\phi_{\hat P\hat P'}},
    \label{eq:phase_combination}
\end{equation}
for $\hat P,\hat P'\in\overline{\mathbb{N}_Z(\mathbb{S})}$. This identity follows directly from the definition of the phases in Eq.~\eqref{eq:decomposition_rule}, since the $Z$ stabilizer factor in the decomposition of $\hat P\hat P'$ factorizes as
\begin{equation}
    \hat S_{Z_{\hat P\hat P'}}
    =
    \hat S_{Z_{\hat P}}\hat S_{Z_{\hat P'}} .
\end{equation}
We use this relation in several simplifications in the main text, most notably in deriving Eq.~\eqref{eq:bare_model_cumulants}.

\section{Computing the channel induced by $\textbf{R}$ QEC cycles}
\label{app:computing_the_R_cycle_channel}
Here, we shall show through induction that the exact $R$-cycle channel induced on the codespace $\mathcal{C}_{\vec{g}}$ is given by:
\begin{align}
\mathcal{E}(R,\vec{\theta}\,) 
&=\prod_{r=1}^R \Lambda(\anglevec{r})\label{eq:R_cycles_evolution_map}\\
&=\!\Xi^\text{II}(R,\vec{\theta}\,) \mathcal{I}
\!+\!\Xi^\text{ZI,im}(R,\vec{\theta}\,) \logicalZLZ
\!+\!\Xi^\text{ZZ}(R,\vec{\theta}\,) \bar{\mathcal{Z}}_L,\notag
\end{align}
where \(\Lambda(\anglevec{r})\) is the single-cycle channel induced by the \(r\)th QEC cycle, defined exactly as in Eq.~\eqref{eq:single_cyc_channel} but with the error angles of cycle \(1\) replaced by those of cycle \(r\). The corresponding channel coefficients are given by:
\begin{subequations}\label{eq:logical_coeffs_def}
\begin{align}
\Xi^\text{II}(R,\vec{\theta}\,) 
&=\frac{1+\Re\!\big(\Gamma(R,\vec{\theta}\,) \big)}{2},\\
\Xi^{\text{ZI},\mathrm{im}}(R,\vec{\theta}\,) 
&=\frac{\Im\!\big(\Gamma(R,\vec{\theta}\,) \big)}{2},\\
\Xi^\text{ZZ}(R,\vec{\theta}\,) 
&=\frac{1-\Re\!\big(\Gamma(R,\vec{\theta}\,) \big)}{2},
\end{align}
\end{subequations}
where
\begin{equation}
\Gamma(R,\vec{\theta}\,) =
\prod_{r=1}^R\!\Big(1-2\,\chi^\text{ZZ}(\anglevec{r})
+2i\,\chi^{\text{ZI},\mathrm{im}}(\anglevec{r})\Big),
\label{eq:appendix_Gamma_coefficient}
\end{equation}

The claim is immediate for $R=1$. Assume that it holds for a general $R$. The channel describing the evolution of the data qubits across $R+1$ QEC cycles is then the composition of the channel in Eq.~\eqref{eq:R_cycles_evolution_map} with the channel induced by the $(R+1)$th QEC cycle,
\begin{align}
\Lambda(\anglevec{R+1})
&=
\frac{1+x(\anglevec{R+1})}{2}\mathcal{I}
+\frac{y(\anglevec{R+1})}{2}\logicalZLZ
\notag\\
&\quad
+\frac{1-x(\anglevec{R+1})}{2}\logicalZsuperop.
\label{eq:appendix_per_cycle_channel}
\end{align}
We used the shorthand notation
\begin{align}
x(\anglevec{R+1})
&=
1-2\chi^\text{ZZ}(\anglevec{R+1}),\notag\\
y(\anglevec{R+1})
&=
2\chi^\text{ZI,im}(\anglevec{R+1}).
\end{align}
Then, using
\begin{equation}
\logicalZsuperop^2=\mathcal{I},
\quad
\logicalZsuperop\logicalZLZ
=
-\logicalZLZ,
\quad
\logicalZLZ^2=2(\logicalZsuperop-\mathcal{I}),
\end{equation}
we obtain
\begin{flalign}
\mathcal{E}(R\!+\!1,\vec{\theta})
&=
\Lambda(\anglevec{R+1})\mathcal{E}(R,\vec{\theta})
\notag\\
&=
\frac{
1\!+\!
    \Re\!\Big[
                            z(\anglevec{R+1})\Gamma(R,\vec{\theta}\,)
                \Big]
                }{2}\mathcal{I}
\notag\\
&\quad+
\frac{
\Im\!\Big[
                        z(\anglevec{R+1})\Gamma(R,\vec{\theta}\,)
            \Big]
            }{2}\logicalZLZ
\notag\\
&\quad+
\frac{
1\!\!-\!\!\Re\!\Big[
                            z(\anglevec{R+1})\Gamma(R,\vec{\theta}\,)
                \Big]
            }{2}\logicalZsuperop,\label{eq:R_plus_1_cycle_channel}
\end{flalign}
where $z(\anglevec{R+1})=x(\anglevec{R+1})+iy(\anglevec{R+1})$. Defining,
\begin{align}
\Gamma(R\!+\!1,\vec{\theta})
&=
z(\anglevec{R+1})\Gamma(R,\vec{\theta}\,)
\\
&=\prod_{r=1}^{R+1}
\Big(
            1\!\!-\!\!2\chi^\text{ZZ}(\anglevec{r})
            \!\!+\!\!2i\chi^\text{ZI,im}(\anglevec{r})
\Big)\notag,
\end{align}
the channel coefficients in Eq.~\eqref{eq:R_plus_1_cycle_channel} can be expressed in a form similar to Eq.~\eqref{eq:logical_coeffs_def}, completing the induction proof. We noise-average this channel and use it to compute the logical infidelity as in Eq.~\eqref{eq:LIF_in_terms_of_Gamma}.

\section{The coefficient $\chi^\mathrm{ZI}(\anglevec{r})$ is imaginary}
\label{app:imaginary_channel_coefficient}

Here, we show explicitly that the coefficient $\chi^\mathrm{ZI}(\anglevec{r})$, 
\begin{flalign}
    \hspace{-2.5mm}\chi^{\text{ZI}}(\anglevec{r})
        \!\!=\!\!
\sum_{\vec s}\!\!\!\sum_{\substack{\hat E\in\mathbb{\overline{P}}_{n,Z}^{\vec s,\checkmark}\\
\hat E'\in\mathbb{\overline{P}}_{n,Z}^{\vec s,\times}}}\!\!\!\!\!
(-1)^{\phi_{\hat{E}\hat{E}'}}c_{\hat E}(\anglevec{r})c^*_{\hat E'}(\anglevec{r}),
\label{eq:process_matrix_ZI_im_appendix}
\end{flalign}
is imaginary, or equivalently, $\sum_{\vec{s}}\Re\!\big[A_{\vec{s}}(\anglevec{1})\, B_{\vec{s}}^*(\anglevec{1})\big]\!\!=\!0$ [Eq.~\eqref{eq:trace_preserving_constraint}]. The coefficient can be re-expressed as:
\begin{flalign}
    \hspace{-2.5mm}\chi^{\text{ZI}}(\anglevec{r})
        \!=\!\!\!\!\!\!\!\!\!\!\!
\sum_{\substack{\hat{Z}_L\in \overline{\mathbb{{N}}_Z(\mathbb{S})}\setminus {\mathbb{S}}_Z}}\!\!\!\!\!\!\!\!\!\!(-1)^{\phi_{\hat{Z}_L}}
                                    \!\!\!\!\!\!\!\sum_{\vec{s},\hat E\in\mathbb{\overline{P}}_{n,Z}^{\vec s,\checkmark}}
\!\!\!\!
c_{\hat E}(\anglevec{r})\,c^*_{\hat{Z}_L\hat{E}}(\anglevec{r}).\label{eq:chi_ZI_im_re_expressed_first}
\end{flalign}
For a fixed $\hat Z_L$, the inner sum essentially runs over all correctable errors $\hat{E}$ and can be written as:
\begin{align}
S(\hat Z_L)=\sum_{\substack{
\hat E\in\mathbb{\overline P}_{n,Z}^{\checkmark}}}
\!\!\!\!c_{\hat E}(\anglevec{r})\,c^*_{\hat{Z}_L\hat{E}}(\anglevec{r}),
\label{eq:sum_S_Z_L}
\end{align}
where $\overline{\mathbb P}_{n,Z}^\checkmark=\bigcup_{\vec{s}}\overline{\mathbb P}_{n,Z}^{\vec{s},\checkmark}$ is the set of all correctable $Z$ errors. We shall show that this sum and $\chi^\mathrm{ZI}(\anglevec{r})$, is imaginary.

Using Eq.~\eqref{eq:cE_def}, each term in the sum can be written as
\begin{flalign}
c_{\hat E}(\anglevec r)c^*_{\hat Z_L\hat E}(\anglevec r)
\!\!=\!\!
(-i)^{w(\hat E)}i^{w(\hat Z_L\hat E)}
R_{\hat E,\hat Z_L\hat E}(\anglevec r),
\label{eq:term-phase}
\end{flalign}
where $R_{\hat E,\hat Z_L\hat E}(\anglevec r)\in\mathbb{R}$. Moreover,
\begin{equation}
w(\hat Z_L\hat E)
\!\!=\!\!
w(\hat Z_L)+w(\hat E)
-
2\left|\mathrm{supp}(\hat Z_L)\cap\mathrm{supp}(\hat E)\right|.
\end{equation}
Then,
\begin{align}
(-i)^{w(\hat E)}i^{w(\hat Z_L\hat E)}
&=
i^{w(\hat Z_L)-2|\mathrm{supp}(\hat Z_L)\cap\mathrm{supp}(\hat E)|}.
\label{eq:phase-parity}
\end{align}
Thus, every term in $S(\hat{Z}_L)$ is purely imaginary when $w(\hat{Z}_L)$ is odd and real when $w(\hat{Z}_L)$ is even. We shall show that for each even-weight $\hat{Z}_L$, the sum vanishes. 

In Eq.~\eqref{eq:sum_S_Z_L}, when summing over correctable $\hat{E}$, $\hat{Z}_L\hat{E}$ is uncorrectable and for even-weight $\hat{Z}_L$, the product of the error amplitudes of this correctable-uncorrectable error pair is real. Thus, the restricted sum in Eq.~\eqref{eq:sum_S_Z_L} is one half of the unrestricted sum:
\begin{equation}
S(\hat Z_L)
=
\frac{1}{2}
\sum_{\hat E\in\mathbb{\overline P}_{n,Z}}
c_{\hat E}(\anglevec r)c^*_{\hat Z_L\hat E}(\anglevec r).
\label{eq:restricted-half-unrestricted}
\end{equation}
Each term in this sum is given by:
\begin{align}
c_{\hat E}(\anglevec r)c^*_{\hat Z_L\hat E}(\anglevec r)
\!\!=\!\!\!\!
&\,\prod_{\ell\notin\mathrm{supp}(\hat E)}
\!\!\!\!\!\!\cos(\rotangle r\ell)
\!\!\!\!\prod_{\ell\in\mathrm{supp}(\hat E)}
\!\!\!\!\!\!\Big(-i\sin(\rotangle r\ell)\Big)
\notag\\
&\times\!\!\!\!\!\!\!\!\!\!
\prod_{\ell\notin\mathrm{supp}(\hat Z_L\hat E)}
\!\!\!\!\!\!\cos(\rotangle r\ell)
\!\!\!\!\!\!\!\!\prod_{\ell\in\mathrm{supp}(\hat Z_L\hat E)}
\!\!\!\!\!\!\Big(i\sin(\rotangle r\ell)\Big).
\end{align}
The contribution from the angle on qubit $\ell$ depends on if $\ell\in\mathrm{supp}(\hat E)$ and if $\ell\in\mathrm{supp}(\hat Z_L)$. In particular, if $\ell\in\mathrm{supp}(\hat Z_L)$, the index $\ell$ will be in exactly one of $\mathrm{supp}(\hat{E})$ and $\mathrm{supp}(\hat{Z}_L\hat{E})$. Then, the contribution from qubit $\ell$ is $\pm i\sin(\rotangle{r}{\ell})\cos(\rotangle{r}{\ell})$, with the sign determined by whether $\ell\in\mathrm{supp}(\hat E)$. Since $\mathrm{supp}(\hat Z_L)\neq\varnothing$, for some $m\in\mathrm{supp}(\hat Z_L)$, Eq.~\eqref{eq:restricted-half-unrestricted} can be written as:
\begin{align}
&S(\hat Z_L)
\!=
\frac{1}{2}
\left(i\sin(\rotangle{r}{m})\cos(\rotangle{r}{m})\right)
\label{eq:S_L_final}\\
&\times\Bigg(\!\!\!\!
\sum_{\substack{
\hat E\in\mathbb{\overline P}_{n,Z}\\
m\notin\mathrm{supp}(\hat E)}}
\!\!\!\!\!\!\tilde c_{\hat E}(\anglevec r)
\tilde c^*_{\hat Z_L\hat E}(\anglevec r)
\!-\!\!\!\!\!\!\!\!\!
\sum_{\substack{
\hat E\in\mathbb{\overline P}_{n,Z}\\
m\in\mathrm{supp}(\hat E)}}
\!\!\!\!\!\!\tilde c_{\hat E}(\anglevec r)
\tilde c^*_{\hat Z_L\hat E}(\anglevec r)
\!\!
\Bigg),\notag
\end{align}
where $\tilde{c}_{\hat{E}}(\anglevec{r})$ is the error amplitude with the factor from qubit $m$ omitted. For every $\hat{E}$ with $m\notin\mathrm{supp}(\hat{E})$, the error $\hat{Z}_m\hat{E}$ is in the second sum, and
\begin{equation}
\tilde c_{\hat E}(\anglevec r)
\tilde c^*_{\hat Z_L\hat E}(\anglevec r)
=
\tilde c_{\hat Z_m\hat E}(\anglevec r)
\tilde c^*_{\hat Z_L\hat Z_m\hat E}(\anglevec r),
\end{equation}
since the single-qubit factor of qubit $m$ does not appear in $\tilde{c}_{\hat{E}}(\anglevec{r})$. The two sums in Eq.~\eqref{eq:S_L_final} therefore cancel term by term, so $S(\hat{Z}_L)=0$ for every even-weight $\hat{Z}_L$ and $\chi^{\mathrm{ZI}}(\anglevec{r})$ is purely imaginary.

That $S(\hat{Z}_L)=0$ for every even-weight $\hat{Z}_L$ also implies that codes with only even-weight logical $Z$ representatives, and more generally with only even-weight logical operators in $\overline{\mathbb{N}_Z(\mathbb{S})}\setminus\mathbb{S}$, completely discretize coherent $Z$ errors on the data qubits. Averaged across syndrome outcomes, the residual logical errors are purely incoherent. This generalizes the observations of Ref.~\cite{huang_performance_2019} for even-distance repetition codes and surface codes to any code with only even-weight logical $Z$ representatives.

\section{Pauli-diagonal structure of the noise-averaged channels}
\label{app:noise_averaged_channel_is_Pauli_diagonal}

In this section we show that the noise-averaged $R$-cycle channel is logical-Pauli diagonal whenever the distribution of $\vec\theta$ is symmetric under $\vec\theta\mapsto-\vec\theta$. The same parity relations are used to simplify the cumulants to the form in Eq.~\eqref{eq:kappa_BR} for the Gaussian noise setting. 

Under sign inversion, the error amplitudes satisfy $c_{\hat E}(-\anglevec{r})=c_{\hat E}^*(\anglevec{r})$. It then follows from Eq.~\eqref{eq:explicit_process_matrix_elements} that $\chi^\text{ZZ}(\anglevec{r})$ is even and $\chi^{\text{ZI},\mathrm{im}}(\anglevec{r})$ is odd under inversion, and hence that
\begin{equation}
\Gamma(R,-\vec{\theta})=\Gamma^*(R,\vec{\theta}).
\end{equation}
Thus $\Im[\Gamma(R,\vec{\theta}\,)]$ is odd in $\vec\theta$ and averages to zero for any distribution symmetric about $\vec{\theta}=0$, as in the zero-mean Gaussian model of the main text. Since $\Xi^{\text{ZI},\mathrm{im}}(R,\vec{\theta})=\Im[\Gamma(R,\vec{\theta}\,)]/2$, the noise-averaged $R$-cycle channel,
\begin{align}
\overline{\mathcal{E}}(R)
&=
\bigavg{\Xi^\text{II}(R,\vec{\theta})}\mathcal{I}
+\bigavg{\Xi^{\text{ZI},\mathrm{im}}(R,\vec{\theta})}\logicalZLZ
\nonumber\\
&\quad
+\bigavg{\Xi^\text{ZZ}(R,\vec{\theta})}\bar{\mathcal{Z}}_L
\end{align}
is logical-Pauli diagonal. This does not imply that the coherent logical errors remaining after each QEC cycle are irrelevant: the coefficients $\chi^{\mathrm{ZI},\mathrm{im}}(\anglevec{r})$ still enter $\avg{\Gamma(R,\vec{\theta}\,)}$ and generally contribute to the logical infidelity. This only means that the logical Pauli off-diagonal coefficient vanishes in the complete noise-averaged $R$-cycle channel.

\section{When does the encoding choice matter?}
\label{app:when_does_the_eigenspace_matter}

Under coherent errors, both the induced channel and the logical infidelity may depend on the choice of 
encoding stabilizer eigenspace [see Sec.~\ref{sec:encoding_eigenspace_dependence}], which allows logical errors to be suppressed through PROSE encoding. A natural question is therefore when this dependence disappears. In this appendix, we shall show non-perturbatively that the noise-averaged $R$-cycle channel and the logical infidelity are independent of the encoding whenever the noise distribution is invariant under the per-qubit sign inversion
\begin{equation}
\bigl(\rotangle{1}{\ell},\ldots,\rotangle{R}{\ell}\bigr)
\mapsto
-\bigl(\rotangle{1}{\ell},\ldots,\rotangle{R}{\ell}\bigr)
\label{eq:appendix_symmetric_along_one_axis}
\end{equation}
for each qubit $\ell$ separately. This condition is weaker than the condition discussed in the main text (no inter-qubit correlations), though for zero-mean Gaussian noise the two are equivalent. When the error angles are independent and symmetrically distributed, the noise reduces to independent stochastic Pauli errors. The present result thus substantially generalizes the familiar fact that, for such errors, neither the induced channel nor the logical infidelity depends on the encoding eigenspace.

The coefficients of $\overline{\mathcal{E}}(R)$ [Eq.~\eqref{eq:logical_coeffs_def}] and $\LIF(R)$ [Eq.~\eqref{eq:LIF_in_terms_of_Gamma}] depend on the encoding eigenspace only through $\bigavg{\Gamma(R,\vec{\theta}\,)}$. Expanding the product in $\Gamma$ reduces this average to moments of the form:
\begin{equation}
M_{V,W}
=
\bigavg{
\prod_{r\in V}\chi^\text{ZZ}(\anglevec{r})
\prod_{r'\in W}\chi^{\text{ZI},\mathrm{im}}(\anglevec{r'})
},
\label{eq:appendix_generic_moment}
\end{equation}
where $V$ and $W$ are the disjoint sets of cycles contributing a $\chi^\text{ZZ}$ and a $\chi^{\text{ZI},\mathrm{im}}$ factor, respectively. Substituting the explicit coefficients [Eq.~\eqref{eq:explicit_process_matrix_elements}] expresses $M_{V,W}$ as a sum of contributions of the form:
\begin{equation}
(-1)^{\phi_{\prod_{q\in V\cup W}\hat E_q\hat E_q'}}
\Bigg\langle
\prod_{q\in V\cup W}
c_{\hat E_q}(\anglevec{q})\,c_{\hat E_q'}^*(\anglevec{q})
\Bigg\rangle ,
\label{eq:appendix_generic_contribution}
\end{equation}
where we have combined the phases using $(-1)^{\phi_{\hat P}+\phi_{\hat P'}}=(-1)^{\phi_{\hat P\hat P'}}$ for $\hat P,\hat P'\in\overline{\mathbb{N}_Z(\mathbb{S})}$ [Eq.~\eqref{eq:phase_combination}].  The only dependence of the term above on the encoding eigenspace is through the phase factor $(-1)^{\phi_{\prod_{q\in V\cup W}\hat E_q\hat E_q'}} $, which is the codespace associated eigenvalue of the $Z$ stabilizer appearing in the decomposition of the operator $\prod_{q\in V\cup W}\hat E_q\hat E_q'$ [Eq.~\eqref{eq:decomposition_rule}]. This operator lies in $\overline{\mathbb{N}_Z(\mathbb{S})}$ since the operators $\hat{E}_q$ and $\hat{E}_q'$ lie in the same syndrome-specific set: for $q\in V$, $\hat E_q,\hat E_q'\in\overline{\mathbb{P}}_{n,Z}^{\vec s_q,\times}$ for some syndrome $\vec s_q$, whereas for $q\in W$, $\hat E_q\in\overline{\mathbb{P}}_{n,Z}^{\vec s_q,\checkmark}$ and $\hat E_q'\in\overline{\mathbb{P}}_{n,Z}^{\vec s_q,\times}$.

We now show that every contribution to $M_{V,W}$ with a nontrivial phase vanishes under the noise average. By Eq.~\eqref{eq:cE_def}, each occurrence of $\hat Z_\ell$ in an error $\hat{E}_q$ contributes a factor $\sin\small(\rotangle{q}{\ell}\small)$ and each non-occurrence a factor $\cos\small(\rotangle{q}{\ell}\small)$. Hence, $\prod_{q}c_{\hat E_q}(\anglevec{q})\,c_{\hat E_q'}^*(\anglevec{q})$ is odd under inversion of $\vec\theta_\ell=(\rotangle{1}{\ell},\ldots,\rotangle{R}{\ell})$ whenever $\hat Z_\ell$ appears an odd number of times in $\{\hat E_q,\hat E_q'\}_{q\in V\cup W}$. If the distribution satisfied Eq.~\eqref{eq:appendix_symmetric_along_one_axis}, any such contribution averages to zero. A contribution survives only if every $\hat Z_\ell$ appears an even number of times in $\{\hat E_q,\hat E_q'\}_{q\in V\cup W}$, which forces
\begin{equation}
\prod_{q\in V\cup W}\hat E_q\hat E_q'=\hat I ,
\end{equation}
so that its phase factor is $(-1)^{\phi_{\hat I}}=1$. Every surviving term of every moment $M_{V,W}$ is thus independent of the encoding eigenspace, and so are $\bigavg{\Gamma(R,\vec{\theta}\,)}$, the noise-averaged $R$-cycle channel, and the logical infidelity, whenever Eq.~\eqref{eq:appendix_symmetric_along_one_axis} holds.

\section{Leading-order expansion of $\chi^\mathrm{ZI,im}(\anglevec{r})$}
\label{app:leading_order_expansions_of_single_cycle_channel_coefficients}

We shall derive the leading-order expansion of the single-cycle channel coefficient $\chi^\mathrm{ZI,im}(\anglevec{r})$ [Eq.~\ref{eq:chiZI_leading}] used throughout the main text, starting from Eq.~\eqref{eq:chi_ZI_im_re_expressed_first}:
\begin{flalign}
    \hspace{-2.5mm}\chi^{\text{ZI},\mathrm{im}}(\anglevec{r})
        \!\!=\!\!\Im\!\Bigl[\!\!\!\!\!\!\!\!
\sum_{\substack{\vec{s},\hat E\in\mathbb{\bar{P}}_{n,Z}^{\vec s,\checkmark}\\\hat{Z}_L\in \overline{\mathbb{{N}}_Z(\mathbb{S})}\setminus {\mathbb{S}}_Z}}
\!\!\!\!\!\!\!\!
(-1)^{\phi_{\hat{Z}_L}}\,c_{\hat E}(\anglevec{r})\,c^*_{\hat{Z}_L\hat{E}}(\anglevec{r})
\Bigr].\label{eq:chi_ZI_im_re_expressed}
\end{flalign}
In this expression, each product of error amplitudes scales as $\mathcal{O}(\theta^{w(\hat E)+w(\hat Z_L\hat E)})$, where $\theta$ is a typical error-angle magnitude. The inequality,
\begin{align}
    w(\hat{E})+w(\hat{Z}_L\hat{E})&=w(\hat{Z}_L)+2\bigl|\mathrm{supp}(\hat{E})\cap\mathrm{supp}(\hat{Z}_L\hat{E})\bigr|\notag\\
    &\geq d_Z \!+\!2\bigl|\mathrm{supp}(\hat{E})\cap\mathrm{supp}(\hat{Z}_L\!\hat{E})\bigr|,\label{eq:D2}
\end{align}
implies that these products are of order $\theta^{d_Z}$ or higher. The terms at order $\theta^{d_Z}$ are contributions from $w(\hat Z_L)=d_Z$ and $\mathrm{supp}(\hat E)\cap\mathrm{supp}(\hat Z_L\hat E)=\varnothing$; that is, $\mathrm{supp}(\hat Z_L)$ is partitioned between $\hat E$ and $\hat Z_L\hat E$. Whether these terms survive and contribute to $\chi^\mathrm{ZI,im}$ depends on the parity of $d_Z$.

First we discuss odd $d_Z$. In this case, these terms are purely imaginary and thus contribute to $\chi^\mathrm{ZI,im}$. Approximating $\chi^\mathrm{ZI,im}$ by these contributions gives:
\begin{flalign}
\chi^{\text{ZI},\mathrm{im}}(\anglevec{r})
&\approx \Im\Bigl[
\sum_{\substack{
\hat Z_L\in\overline{\mathbb{N}_Z(\mathbb{S})}\setminus{\mathbb{S}}_Z:
\\ \ w(\hat Z_L)=d_Z}}
(-1)^{\phi_{\hat Z_L}}
&&\label{eq:D3}\\
&\times\!\!\!\!\!\!\!\!\!\!\!\!\!\!\!\!
\sum_{\substack{
\hat E\in\mathbb{\overline P}_{n,Z}^{\checkmark}:\\
\mathrm{supp}(\hat E)\subsetneq\mathrm{supp}(\hat Z_L)}}
\!\!\!\!\!\!\!\!
\Big(\!\!\!\!\prod_{\ell\in\text{supp}(\hat{E})}\!\!\!\!-i\rotangle{r}{\ell}\Big)
\Big(\!\!\!\!\prod_{\ell\in\text{supp}(\hat{Z}_L\hat{E})}\!\!\!\!i\rotangle{r}{\ell}\Big)
\Bigr],\notag&&
\end{flalign}
with corrections of order $\mathcal{O}(\theta^{d_Z+2})$. Here, we have combined the sum over syndromes and correctable errors in syndrome-specific sets by defining $\mathbb{\bar{P}}^{\checkmark}_{n,Z}=\bigcup_{\vec{s}} \mathbb{\bar{P}}^{\vec{s},\checkmark}_{n,Z}$, the set of all correctable $Z$ errors. Further, since $\mathrm{supp}(\hat{E})\subset\mathrm{supp}(\hat{Z}_L)$,  Eq.~\eqref{eq:D3} simplifies to:
\begin{flalign}
\chi^{\text{ZI},\mathrm{im}}(\anglevec{r})
&\!=\!\! (-1)^{\frac{d_Z-1}{2}}\!\!\!\!\!\!\!\!\!\!
\sum_{\substack{
\hat Z_L\in\overline{\mathbb{N}_Z(\mathbb{S})}\setminus{\mathbb{S}}_Z:\\
w(\hat Z_L)=d_Z}}
\!\!\Bigg[(-1)^{\phi_{\hat Z_L}}
\Bigg(
\prod_{\ell\in\text{supp}(\hat Z_L)}\!\!\!\!\rotangle{r}{\ell}
\Bigg) && \notag\\
&\qquad\qquad\qquad
\times\Bigg(\!\!\!\!\!\!\!\!\!\!\!\!
\sum_{\substack{
\hat E\in\mathbb{\overline P}_{n,Z}^{\checkmark}:\\
\mathrm{supp}(\hat E)\subsetneq\mathrm{supp}(\hat Z_L)}}
\!\!\!\!\!\!\!\!\!\!\!\!\!\!\!\!
(-1)^{w(\hat E)}
\Bigg)\Bigg]. &&\label{eq:D4}
\end{flalign}

The expression above can be further simplified. For each weight-$d_Z$ operator $\hat Z_L$, an error $\hat E$ supported on $\mathrm{supp}(\hat Z_L)$ is paired with its complement $\hat Z_L\hat E$ in the inner sum. Since $d_Z$ is odd, exactly one of the pair has weight at most $(d_Z-1)/2$ and is the error corrected by the maximum likelihood decoder at low and comparable per-qubit marginals. Hence, the inner sum reduces to an alternating sum over all subsets of $\mathrm{supp}(\hat Z_L)$ of size at most $(d_Z-1)/2$:
\begin{equation}
    \sum_{\substack{
\hat E\in\mathbb{\overline P}_{n,Z}^{\checkmark}:\\
\mathrm{supp}(\hat E)\subsetneq\mathrm{supp}(\hat Z_L)}}
\!\!\!\!\!\!\!\!\!\!\!
(-1)^{w(\hat E)}=\sum_{w=0}^{(d_Z-1)/2}(-1)^w\binom{d_Z}{w}
\end{equation}
Evaluating this alternating binomial sum yields
\begin{equation}
    \hspace{-2mm}\chi^{\text{ZI},\mathrm{im}}(\anglevec{r})\approx f(d_Z)\!\!\!\!
\sum_{\substack{
\hat Z_L\in\overline{\mathbb{N}_Z(\mathbb{S})}\setminus{\mathbb{S}}_Z:
\\ \ w(\hat Z_L)=d_Z}}
\!\!\!\!\!\!\!\!
(-1)^{\phi_{\hat Z_L}}
\Big(\!\!\!\!\prod_{\ell\in\text{supp}(\hat{Z}_L)}\!\!\!\!\rotangle{r}{\ell}\Big),
\end{equation}
where
\begin{equation}
    f(d_Z)=\frac{d_Z+1}{2d_Z}\binom{d_Z}{(d_Z+1)/2}.
\end{equation}

In contrast to the odd-$d_Z$ case, for even $d_Z$, the order-$\theta^{d_Z}$ terms are real and do not contribute to $\chi^\mathrm{ZI,im}$. More broadly, as shown in App.~\ref{app:imaginary_channel_coefficient}, $\chi^{\mathrm{ZI},\mathrm{im}}$ receives contributions only from correctable/uncorrectable pairs whose product is an odd-weight logical $Z$ operator. The leading contribution thus comes from the product of amplitudes of errors $\hat{E}$ and $\hat{E}'$ such that $\hat{E}\hat{E}'$ is an odd-weight logical $Z$ operator $\hat{Z}_L$ of minimum weight and $\mathrm{supp}(\hat E)\cap\mathrm{supp}(\hat E')=\varnothing$. Approximating $\chi^{\mathrm{ZI},\mathrm{im}}(\anglevec{r})$ by these contributions, we get:
\begin{flalign}
\chi^{\text{ZI},\mathrm{im}}(\anglevec{r})
&= (-1)^{\frac{d_Z^\mathrm{odd}-1}{2}}\!\!\!\!\!\!\!\!\!\!\!\!\!\!
\sum_{\substack{
\hat Z_L\in\overline{\mathbb{N}_Z(\mathbb{S})}\setminus{\mathbb{S}}_Z:\\
w(\hat Z_L)=d_Z^\mathrm{odd}}}
\!\!\Bigg[(-1)^{\phi_{\hat Z_L}}\!\!\!
\Bigg(
\prod_{\ell\in\mathrm{supp}(\hat Z_L)}\!\!\!\!\rotangle{r}{\ell}
\Bigg) && \notag\\
&\qquad\qquad\qquad
\times\Bigg(\!\!\!\!\!\!\!\!\!\!\!\!
\sum_{\substack{
\hat E\in\mathbb{\overline P}_{n,Z}^{\checkmark}:\\
\mathrm{supp}(\hat E)\subsetneq\mathrm{supp}(\hat Z_L)}}
\!\!\!\!\!\!\!\!\!\!\!\!\!\!\!\!
(-1)^{w(\hat E)}
\Bigg)\Bigg], &&\label{eq:D9}
\end{flalign}
where
\begin{equation}
    \!\!\!\!d_Z^\mathrm{odd}\!=\!\min\{w(\hat Z_L)\!:\!\hat Z_L\in\overline{\mathbb{N}_Z(\mathbb{S})}\setminus\mathbb{S},\!w(\hat Z_L)\,\mathrm{is\,odd}\}.
\end{equation}
The expression above is the even-$d_Z$ analogue of Eq.~\eqref{eq:D4}. 

Here, the inner sum no longer simplifies. Unlike the odd-$d_Z$ case, some pairs have both members outside the weight range guaranteed to be correctable, and which member lies in $\mathbb{\overline P}_{n,Z}^{\checkmark}$ is code-dependent.

Equation~\eqref{eq:D9} nonetheless establishes the point used in the main text: for even-$d_Z$ codes, $\chi^{\mathrm{ZI},\mathrm{im}}(\anglevec{r})$ scales as $\theta^{d_Z^\mathrm{odd}}$, which is higher-order than $\theta^{d_Z}$, so the leading-order contributions to both cumulants of the logical infidelity of the QEC-only protocol for these even-$d_Z$ codes depends only on $\chi^{\mathrm{ZZ}}(\anglevec{r})$. 

This recovers the observation of App.~\ref{app:imaginary_channel_coefficient}: codes whose logical operators in $\overline{\mathbb{N}_Z(\mathbb{S})}\setminus\mathbb{S}_Z$ all have even weight leave only purely incoherent logical errors after each QEC cycle (upon averaging over syndromes).

\section{Accuracy of the cumulant expansion}
\label{app:accuracy_of_the_second_order_cumulant_expansion}

Here we assess the accuracy of the second-order cumulant expansion of the QEC-only logical infidelity used in the main text [Eq.~\eqref{eq:bare_LIF_in_terms_of_cumulants}]. The expansions for the other protocols can be analyzed similarly. The expansion involves two approximations: the exponential approximation of each single-cycle factor [Eq.~\eqref{eq:exp_approx_cycle}], and the truncation of the cumulant expansion at second-order. We quantify their errors in turn by comparing against a direct second-order cumulant expansion of the exact logical infidelity, and then by estimating the size of the neglected third cumulant.

We begin with the direct second-order cumulant expansion of the exact logical infidelity [Eq.~\eqref{eq:LIF_in_terms_of_Gamma}] yielding:
\begin{equation}
    \widetilde\LIF(R)
    \approx
    \frac{1-\Re\!\Big(\exp\!\big(\widetilde\kappa_{1}(R)
    +{\widetilde\kappa_{2}(R)}/{2}\big)\Big)}{3},
    \label{eq:direct_LIF_in_terms_of_cumulants}
\end{equation}
where the tilde distinguishes this expansion from the main-text one. For distributions symmetric about $\vec\theta=0$, the cumulants are related to those of the main text by
\begin{align}
\widetilde\kappa_1(R)=\kappa_1(R)&,\notag\\
\widetilde\kappa_2(R)=\kappa_2(R)
&-4\sum_{r}
\Bigavg{
\big(
\chi^{\text{ZZ}}(\anglevec{r})
\big)^2
}\label{eq:direct_vs_maintext_cumulants}\\
&+4\sum_{r}
\Bigavg{
\big(
\chi^{\text{ZI},\mathrm{im}}(\anglevec{r})
\big)^2
}.\notag
\end{align}
The first cumulants agree, while the second cumulant corrections in Eq.~\eqref{eq:direct_vs_maintext_cumulants} remove same-cycle contributions from $\kappa_2(R)$ [Eq.~\eqref{eq:kappa2_BR}], which the exponential approximation added. The exponential approximation essentially amounts to neglecting these second cumulant corrections, the dominant of which is
\begin{equation}
    \sum_{r}\Bigavg{\big(\chi^\mathrm{ZI,im}(\anglevec{r})\big)^2}
    \sim R\sigma^{2d_Z}.
    \label{eq:exp_approx_corrections}
\end{equation}
Here and below, $\sigma$ and $R_c$ are the scaling parameters of the main text: $\sigma$ sets the scale of the variances of the error angles, and $R_c$ is the number of cycles, out of $R$, over which the noise remains substantially correlated.

Next we turn to corrections to the second-order cumulant expansion from higher-order cumulants. Their size is estimated by the dominant third-cumulant contribution,
\begin{equation}
      \sum_{r,r',r''=1}^R\!\!\!\!
    \bigavg{\chi^\text{ZI,im}\!(\anglevec{r})
                       \chi^\text{ZI,im}\!(\anglevec{r'})
                       \chi^\text{ZZ}\!(\anglevec{r''})}_c
    \sim \!\!RR_c^2\sigma^{3d_Z+1}.
    \label{eq:third_cumulant_scaling}
\end{equation}
This contribution reduces the infidelity and within the accuracy of Eq.~\eqref{eq:approximate_induced_channel}, can be interpreted as the cancellation of accumulated coherent logical errors by an incoherent logical error.

We now compare these corrections to the size of the retained cumulants in the main-text expansion, which scale as
\begin{align}
    \kappa_1(R)&\sim R\sigma^{d_Z+1},\notag\\
    \kappa_2^\text{coh}(R)&\sim RR_c\sigma^{2d_Z}.
\end{align}
Their relative size sets the dominant contribution to $\LIF(R)$ and the scale against which corrections must be compared. This separates the comparison into two regimes.

The first regime is $\kappa_1(R)\gtrsim\kappa_2^\mathrm{coh}(R)$, equivalently $R_c\sigma^{d_Z-1}\lesssim1$, arising for sufficiently weak noise, short inter-cycle correlation range, or few cycles. Here, the first cumulant dominates the expansion. The second regime is $\kappa_2^\mathrm{coh}(R)\gtrsim\kappa_1(R)$, equivalently $R_c\sigma^{d_Z-1}\gtrsim1$. Although unusual from the perspective of a naive cumulant hierarchy, this does not by itself signal a breakdown of the expansion: $\kappa_2^\mathrm{coh}(R)$ captures the contribution of coherent logical errors, which do not enter the first cumulant. Note that in this regime, $\sigma\ll1$ implies $R_c\gg1$.

% The first regime is $\kappa_1(R)\gtrsim\kappa_2^\text{coh}(R)$, equivalently $R_c\sigma^{d_Z-1}\lesssim1$, arising for sufficiently weak noise, short inter-cycle correlation range, or few cycles. The correction to the second cumulant from Eq.~\eqref{eq:exp_approx_corrections} is suppressed relative to $\kappa_1(R)$ by $\sigma^{d_Z-1}$, while the third cumulant of Eq.~\eqref{eq:third_cumulant_scaling} is suppressed by $\big(R_c\sigma^{d_Z-1}\big)^2\sigma^2$. Both are therefore negligible in this regime provided $\sigma^2\ll1$.

% The second regime is $\kappa_2^\mathrm{coh}(R)\gtrsim\kappa_1(R)$, equivalently $R_c\sigma^{d_Z-1}\gtrsim1$. Although unusual from the perspective of a naive cumulant hierarchy, this does not by itself signal a breakdown of the expansion: $\kappa_2^\mathrm{coh}(R)$ captures the contribution of coherent logical errors, which do not enter the first cumulant. Here, the correction to the second cumulant is suppressed relative to $\kappa_2^\mathrm{coh}(R)$ by $1/R_c$, and the third cumulant by $R_c\sigma^{d_Z+1}$. Since $R_c\sigma^{d_Z-1}\gtrsim1$ here,  $\sigma\ll1$ implies $R_c\gg1$ and corrections from the exponential approximation are thus negligible. The third cumulant is in turn negligible if the noise is sufficiently weak that $R_c\sigma^{d_Z+1}\ll 1$. 

Then, comparing the corrections with the dominant contribution in the above two regimes shows that the third cumulant correction may be neglected if $R_c\sigma^{d_Z-1}\lesssim 1$ and $\sigma^2\ll 1$, or if $R_c\sigma^{d_Z-1}\gtrsim 1$ but $\sigma^2\ll 1$ such that $R_c\sigma^{d_Z+1}\ll 1$, establishing the accuracy of the direct second-order cumulant expansion. 

In both regimes, provided the noise is sufficiently weak ($\sigma\ll1$), the second-cumulant corrections in Eq.~\eqref{eq:exp_approx_corrections} may also be neglected, as expected, and the main-text second-order cumulant expansion closely approximates the direct one. This comparison with the direct cumulant expansion thus places the approximation in Eq.~\eqref{eq:exp_approx_cycle} on firm footing.

Since we are interested in the weak-noise regime, we adopt the main-text expansion because, to its order of accuracy, the single-cycle channel takes the transparent form of Eq.~\eqref{eq:approximate_induced_channel}, which separates the coherent and incoherent logical errors and provides valuable intuition. The main-text expansion thus captures the qualitative physics at the cost of a small quantitative error. If higher quantitative accuracy is required, the direct expansion in Eq.~\eqref{eq:direct_LIF_in_terms_of_cumulants} may be used instead.

\section{Logical dynamical decoupling}
\label{app:dynamical_decoupling}

Here we study supplementing QEC with dynamical decoupling (DD). DD alone can in some settings outperform QEC+DD~\cite{han_protecting_2025}, because QEC projects, and only partly corrects, coherent physical errors that DD could otherwise refocus away entirely. In practice, however, the noise may contain components that DD cannot suppress but QEC can partially correct, motivating their combination. We focus on one particular combination\textemdash namely logical dynamical decoupling (LDD). In LDD, an optimized deterministic sequence of logical Pauli gates is interleaved between QEC cycles to suppress residual coherent logical errors. LDD is the deterministic counterpart of the QEC+LT protocol of the main text, optimizing over logical Pauli sequences rather than averaging over them.

We shall first formally describe the QEC+LDD protocol and derive its logical infidelity $\LIF_\mathrm{LDD}(R)$. Then, we shall approximate it in the weak-noise regime, where it is nearly equivalent to $\LIF_\mathrm{LT}(R)$. We next identify settings where a logical spin-echo sequence is near-optimal and marginally outperforms QEC+LT, and finally briefly discuss the effect of noise correlations on $\LIF_\mathrm{LDD}(R)$.

\subsection{Protocol outline and computation of $\LIF_\mathrm{LDD}(R)$}

The residual coherent logical error after each QEC cycle lies along a fixed axis of the logical Bloch sphere, chosen here as the logical $Z$ axis. The goal of LDD is to modulate the sign of the rotation about this axis, so that coherent logical errors from different cycles interfere destructively. It therefore suffices to draw the logical Pauli applied in each cycle from $\{\hat{I},\hat{\bar X}_L\}$, with $\hat{\bar X}_L$ a representative logical $X$ operator (the QEC+LT protocol of the main text can also be restricted to the same set for this noise model).

A candidate decoupling sequence over \(R\) cycles can then be specified by a binary vector $\vec{\boldsymbol{u}}=(\boldsymbol{u}_1,\ldots,\boldsymbol{u}_R)\in\{0,1\}^R$, where if $\boldsymbol{u}_r=1$, a logical $X$ gate is inserted before the noise and after the correction step in cycle $r$ (these may again be absorbed into the correction steps). For a fixed noise realization, the single-cycle channel with the sequence $\vec{\boldsymbol{u}}$ applied is:
\begin{align}
\Lambda({\boldsymbol{u}}_r,\anglevec{r})\!=\!
\mathcal{\bar{X}}_\mathrm{L}^{\boldsymbol{u}_r}
\Big(
\chi^\text{II}(\anglevec{r})\mathcal{I}
&+\chi^{\text{ZI},\mathrm{im}}(\anglevec{r})\logicalZLZ\\
&+\chi^\text{ZZ}(\anglevec{r})\bar{\mathcal{Z}}_L
\Big)
\mathcal{\bar{X}}_\mathrm{L}^{\boldsymbol{u}_r},\notag
\end{align}
with $\mathcal{\bar X}_\mathrm{L}[\cdot]=\hat{\bar X}_\mathrm{L}[\cdot]\hat{\bar X}_\mathrm{L}$. Composing and noise-averaging these channels yields the channel describing the noise-averaged logical qubit evolution under $\vec{\boldsymbol{u}}$. We then apply the definition in Eq.~\eqref{eq:average_logical_infidelity} using this channel to compute the logical infidelity for this particular sequence:
\begin{equation}
    \LIF(R,\vec{\boldsymbol{u}})
    =
    \frac{
    1-\Bigavg{\Re\!\left(\Gamma(R,\vec{\boldsymbol{u}},\vec{\theta}\,)\right)}
    }{3},
    \label{eq:app_LIF_one_sequence}
\end{equation}
where
\begin{equation}
\Gamma(R,\vec{\boldsymbol{u}},\vec{\theta}\,) \!\!=\!\!
\!\prod_{r=1}^R\!\!\!
\Big(\!
1\!\!-\!2\chi^\text{ZZ}(\anglevec{r})
\!\!+\!2i(-1)^{{\boldsymbol{u}}_r}
\!\chi^{\text{ZI},\mathrm{im}}(\anglevec{r})
\!\Big).
\label{eq:Gamma_coefficient_LDD}
\end{equation}
The infidelity of the QEC+LDD protocol is obtained by minimizing $\LIF(R,\vec{\boldsymbol{u}})$ over the decoupling sequences,
\begin{equation}
    \LIF_\mathrm{LDD}(R)
    =
    \min_{\vec{\boldsymbol{u}}\in\{0,1\}^{R}}
    \LIF(R,\vec{\boldsymbol{u}}).
    \label{eq:app_LIF_LDD}
\end{equation}
In contrast, the QEC+LT infidelity, for the same code and encoding, is the average over these sequences:
\begin{equation}
    \LIF_\mathrm{LT}(R)
    =
    \avg{\LIF(R,\vec{\boldsymbol{u}})}_{\vec{\boldsymbol{u}}\in\{0,1\}^{R}}.
    \label{eq:app_LIF_LT_average}
\end{equation}

\subsection{$\LIF_\mathrm{LDD}(R)\approx\LIF_\mathrm{LT}(R)$ for weak noise}
\label{app:weak_noise_approximation_LDD_infidelity}

Now we discuss why optimizing over the logical Pauli ensemble does not provide a substantial advantage over averaging over it for weak noise. In this regime, we can approximate Eq.~\eqref{eq:app_LIF_one_sequence} by an exponential, neglecting higher-order terms as in Eq.~\eqref{eq:exp_approx_cycle}. Thus,
\begin{align}
    \LIF(R, \vec{\boldsymbol{u}})\approx\frac{1}{3}\Bigg(1-\Bigavg{
&\exp\!\left(
-2\sum_{r=1}^R
\chi^\text{ZZ}(\anglevec{r})
\right)\label{eq:app_LIF_approx_one_sequence}\\
&\times \cos\left(2\sum_{r=1}^R(-1)^{\boldsymbol{u}_r}
\chi^\text{ZI,im}(\anglevec{r})\right)
}\Bigg).\notag
\end{align}
Since $\LIF_\mathrm{LT}(R)$ is given by Eq.~\eqref{eq:LIF_LT_approx_exp} at the same accuracy,
\begin{flalign}
    \LIF(R,\vec{\boldsymbol{u}})\!-\!\LIF_\mathrm{LT}(R)\!\!&\approx\!\!\frac{1}{3}\Bigavg{\exp\!\Big(
-2\sum_{r=1}^R
\chi^\text{ZZ}(\anglevec{r})
\Big)\label{eq:app_DD_LT_bound}\\
    &\times\!\!\Big(1\!-\!\cos\!\big(2\sum_{r=1}^R(-1)^{\boldsymbol{u}_r}\chi^\text{ZI,im}(\anglevec{r})\big)\Big)}\geq 0.\notag
\end{flalign}
Thus, at this accuracy, the average over the logical Pauli ensemble must lie close to the optimum, so
\begin{align}
    \LIF_\mathrm{LDD}(R)\!\approx\!\LIF_\mathrm{LT}(R)
    \!=\!\frac{1-\Bigavg{
                                            \exp\!\Big(
                                                                    -2\sum_{r=1}^R
                                                                        \chi^\text{ZZ}(\anglevec{r})
                                                            \Big)}}{3}.\label{eq:app_LDD_approximates_LT}
\end{align}
Note that agreement holds for arbitrary noise distributions, only relying on the assumption of weak noise.

\begin{figure}
    \centering
    \includegraphics[width=\linewidth]{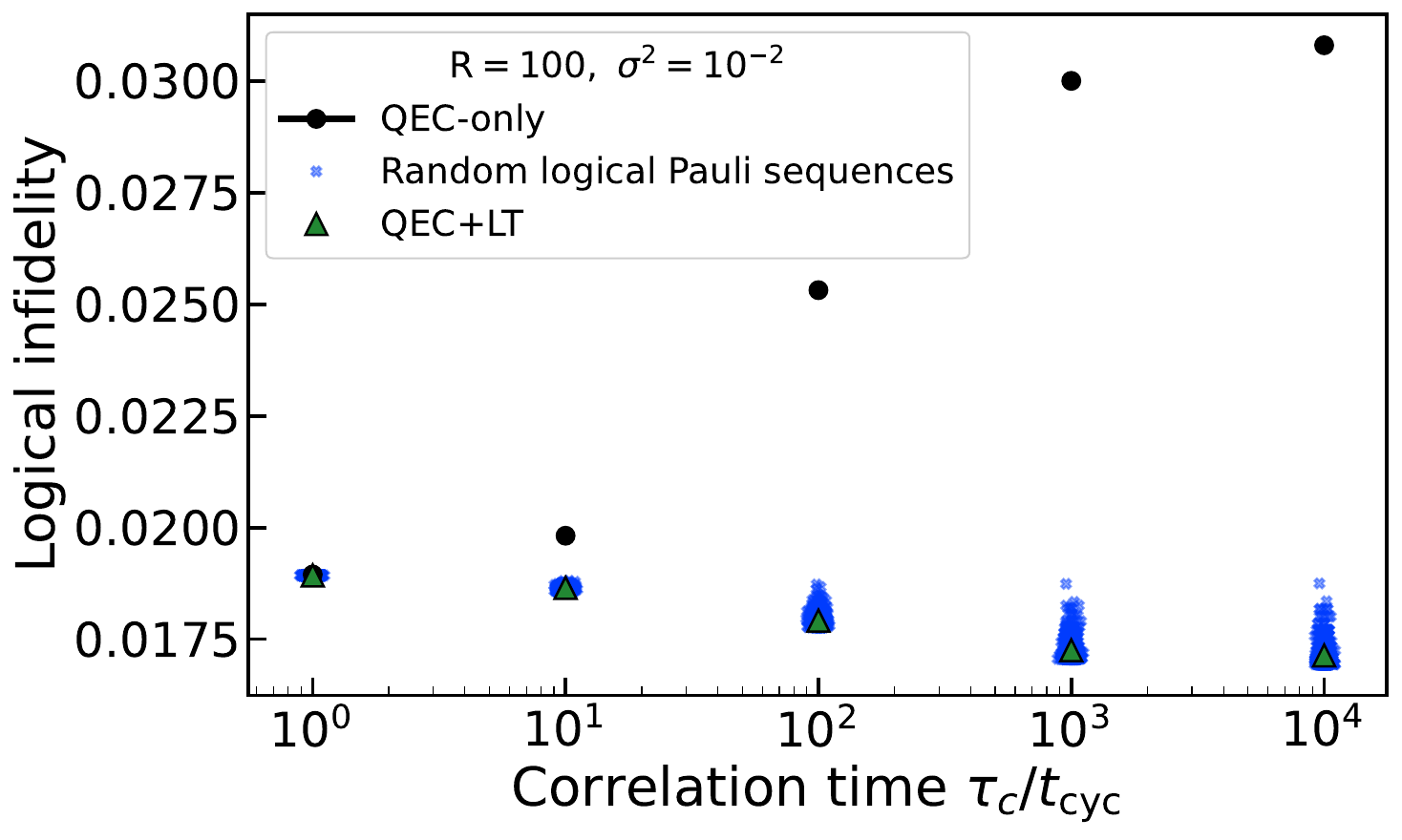}
\caption{%
\textbf{Logical infidelity of the three-qubit phase-flip repetition code under QEC-only, random logical Pauli sequences, and QEC+LT, as the inter-cycle correlation range is varied.}
We plot the logical infidelity $\LIF(R)$ of the three-qubit phase-flip repetition code for the noise model in Eq.~\eqref{eq:appendix_lorentzian_noise_correlations}, comparing QEC-only (black, circles), $10^3$ individual realizations of QEC with a random logical Pauli sequence conjugating each cycle (blue), and QEC+LT (green, triangles). Noise strength $\sigma^2$ and inter-qubit correlation strength $\varrho$ are fixed, while the inter-cycle correlation range $\tau_c/t_\mathrm{cyc}$ is varied. For short inter-cycle correlation ranges, the coherent logical errors cancel on average and all protocols perform similarly. As the correlation range increases, coherent logical errors accumulate constructively under QEC-only, whereas typical random logical Pauli sequences suppress this accumulation and concentrate near the QEC+LT value. Data is obtained by noise-averaging the exact
expression in Eq.~\eqref{eq:app_LIF_one_sequence}; error bars are within marker size.%
}
    \label{fig:LT_versus_LDD}
\end{figure}

Figure~\ref{fig:LT_versus_LDD} provides numerical evidence for this conclusion using a $3$-qubit repetition code under Gaussian noise with covariance
\begin{equation}
    \boldsymbol{\Sigma}^{(r,r')}_{\ell\ell'}=\sigma^2
    \exp\Big(-\abs{r-r'}\frac{t_\mathrm{cyc}}{\tau_c}\Big),
    \label{eq:appendix_lorentzian_noise_correlations}
\end{equation}
fixing noise strength $\sigma^2$ and varying the inter-cycle correlation range set by $\tau_c/t_\mathrm{cyc}$. For small $\tau_c/t_\mathrm{cyc}$, coherent logical errors contribute negligibly and all protocols agree. As $\tau_c/t_\mathrm{cyc}$ increases, these errors interfere constructively and raise the QEC-only infidelity (see Sec.~\ref{sec:features_common_across_odd_dZ_codes}), while typical logical Pauli sequences suppress this accumulation and concentrate near the QEC+LT value. This confirms both that QEC+LT is comparable to the optimum and that the infidelity has small variance over the sequence ensemble.

This near-equivalence motivates our focus on LT in the main text: since the two perform comparably, there is little benefit to identifying the optimal sequence, which may depend on $R$ and be hard to determine. The small ensemble variance further implies that only a few random sequences are needed to approach QEC+LT performance, as observed in Ref.~\cite{hashim_randomized_2021} for non-QEC settings.

\subsection{Intuition for why QEC+LDD and QEC+LT agree in the weak-noise regime}
\label{app:DD_LT_correspondence}

The approximation of the exact single-cycle channel in Eq.~\eqref{eq:approximate_induced_channel} for a fixed noise realization provides intuition for this agreement. It is given by:
\begin{align}
\Lambda(\anglevec{r})
&\!=\!
\chi^\text{II}(\anglevec{r})\mathcal{I}
\!+\!\chi^{\text{ZI},\mathrm{im}}(\anglevec{r})\logicalZLZ
\!+\!\chi^\text{ZZ}(\anglevec{r})\bar{\mathcal{Z}}_L
\label{eq:appendix_approximate_single_cycle_channel}\\
&\!\approx\!
\mathcal{R}_{\logicalZ}\!\big(\chi^{\text{ZI},\mathrm{im}}(\anglevec{r})\big)
\left[
\chi^\text{II}(\anglevec{r})\,\mathcal{I}
+
\chi^\text{ZZ}(\anglevec{r})\,\bar{\mathcal Z}_L
\right].
\notag
\end{align}
LDD refocuses coherent logical errors $\mathcal{R}_{\logicalZ}(\chi^{\mathrm{ZI},\mathrm{im}}(\anglevec{r}))$ so that their accumulation across cycles is suppressed, whereas LT removes the logical Pauli off-diagonal term $\chi^{\text{ZI},\mathrm{im}}(\anglevec{r})\logicalZLZ$ in each cycle. To the accuracy of the above approximation, these actions are near-equivalent.

This intuition clarifies two points. First, it shows that the near-equivalence is not tied to a specific noise distribution. Second, it identifies the mechanism: stabilizer measurements separate the scales of the incoherent and coherent logical errors. For odd-$d_Z$ codes, $\chi^\mathrm{ZI,im}\!\!\!\sim\!\!\!\theta^{d_Z}$ while $\chi^\mathrm{ZZ}\!\!\!\sim\!\!\!\theta^{d_Z+1}$. The contribution of the coherent logical error $\mathcal{R}_{\logicalZ}$ to the logical Pauli-diagonal coefficient $(\chi^\mathrm{ZI,im})^2$ is higher-order than $\chi^\mathrm{ZZ}$ when $d_Z\geq2$ (for even-$d_Z$ the separation is stronger still). Thus suppressing $\mathcal{R}_{\logicalZ}$ becomes near-equivalent to removing $\mathcal{H}$. This also explains why the conclusion fails outside QEC settings ($d_Z\!=\!1$), and decoupling and twirling act differently.

\subsection{Spin-echo on the logical qubit is near-optimal for positive, long-ranged inter-cycle correlations}
\label{app:spin_echo}
Identifying the optimal decoupling sequence for an arbitrary noise correlation structure is generally difficult, but the problem simplifies for positive, long-ranged inter-cycle correlations, where the logical spin-echo sequence
\begin{equation}
    \vec{\boldsymbol{u}}_{\rm SE}
    =
    (0,1,0,1,\ldots)
\end{equation}
is near-optimal. This regime matters most for odd $d_Z$, where coherent logical errors add constructively, substantially raising QEC-only infidelity [Figs.~\ref{fig:odd_even_distinction_rep_code} and \ref{fig:LT_versus_LDD}]. 

We first outline the optimum that logical spin-echo approaches. For a fixed noise realization, the single-cycle channel can be re-written as:
\begin{align}
\Lambda(\anglevec{r})
&\!=\!
\mathcal{R}_{\logicalZ}\!\big(\varphi(\anglevec{r})\big)
\left[
p_{\mathrm I}(\anglevec{r})\mathcal{I}
+
p_{\mathrm Z}(\anglevec{r})\bar{\mathcal Z}_L
\right],
\end{align}
with the exact parameters
\begin{align}
    p_\mathrm{Z}(\anglevec{r})
    &=
    \frac{
    1\!-\!
    \sqrt{
    \big(\chi^\mathrm{II}(\anglevec{r})\!-\!\chi^\mathrm{ZZ}(\anglevec{r})\big)^2
    \!+
    \big(2\chi^\mathrm{ZI,im}(\anglevec{r})\big)^2
    }
    }{2},
    \notag\\
    \varphi(\anglevec{r})
    &=
    \frac{1}{2}
    \arctan\!\left(
    \frac{
    2\chi^\mathrm{ZI,im}(\anglevec{r})
    }{
    \chi^\mathrm{II}(\anglevec{r})-\chi^\mathrm{ZZ}(\anglevec{r})
    }
    \right),
    \label{eq:app_exact_single_cycle_parameters}
\end{align}
where $p_\mathrm{I}(\anglevec{r})=1-p_\mathrm{Z}(\anglevec{r})$ and the arctan branch is $(-\pi/2,\pi/2]$. The logical infidelity with a decoupling sequence $\vec{\boldsymbol{u}}$ is then given by:
\begin{align}
    \LIF(R,\vec{\boldsymbol{u}})
    &=
    \frac{1}{3}
    \Bigg(1-
    \left\langle
    \left(
    \prod_{r=1}^R
    \left(1-2p_Z(\anglevec{r})\right)
    \right)
    \right.\\
    &\left.
    \qquad\qquad\qquad\times \cos\!\left(
    \sum_{r=1}^R
    2(-1)^{\boldsymbol{u}_r}\varphi(\anglevec{r})
    \right)\right\rangle
    \Bigg).\notag
\end{align}
$\LIF(R,\vec{\boldsymbol{u}})$ be bounded from below as:
\begin{equation}
    \LIF(R,\vec{\boldsymbol{u}})
    \geq
    \frac{
    1-
    \Bigavg{
    \prod_{r=1}^R
    \left(1-2p_Z(\anglevec{r})\right)
    }
    }{3},\label{eq:lower_bound}
\end{equation}
reflecting that these sequences refocus only the residual coherent logical errors, at best canceling them entirely.

We find that, as expected, the logical spin-echo sequence approaches this optimum when the error angles vary slowly between adjacent cycles. Concretely, if
\begin{equation}
    \frac{1}{n}
    \left\langle
    \norm{\anglevec{r+1}-\anglevec{r}}_2^2
    \right\rangle
    =
    \delta_\mathrm{corr}
    \ll 1,
    \label{eq:positive_corr_condition_for_DD}
\end{equation}
with $\norm{\cdots}_2$ the Euclidean norm, the logical spin-echo infidelity approaches the bound with corrections of order $\mathcal{O}(\sqrt{\delta_\mathrm{corr}})$. This is the QEC analogue of the standard regime in which spin-echo is effective: noise that is strongly and positively correlated over timescales longer than the pulse spacing, here set by the cycle duration $t_\mathrm{cyc}$ since logical $X$ pulses are applied after every other cycle.

\begin{figure}
    \centering
    \includegraphics[width=\linewidth]{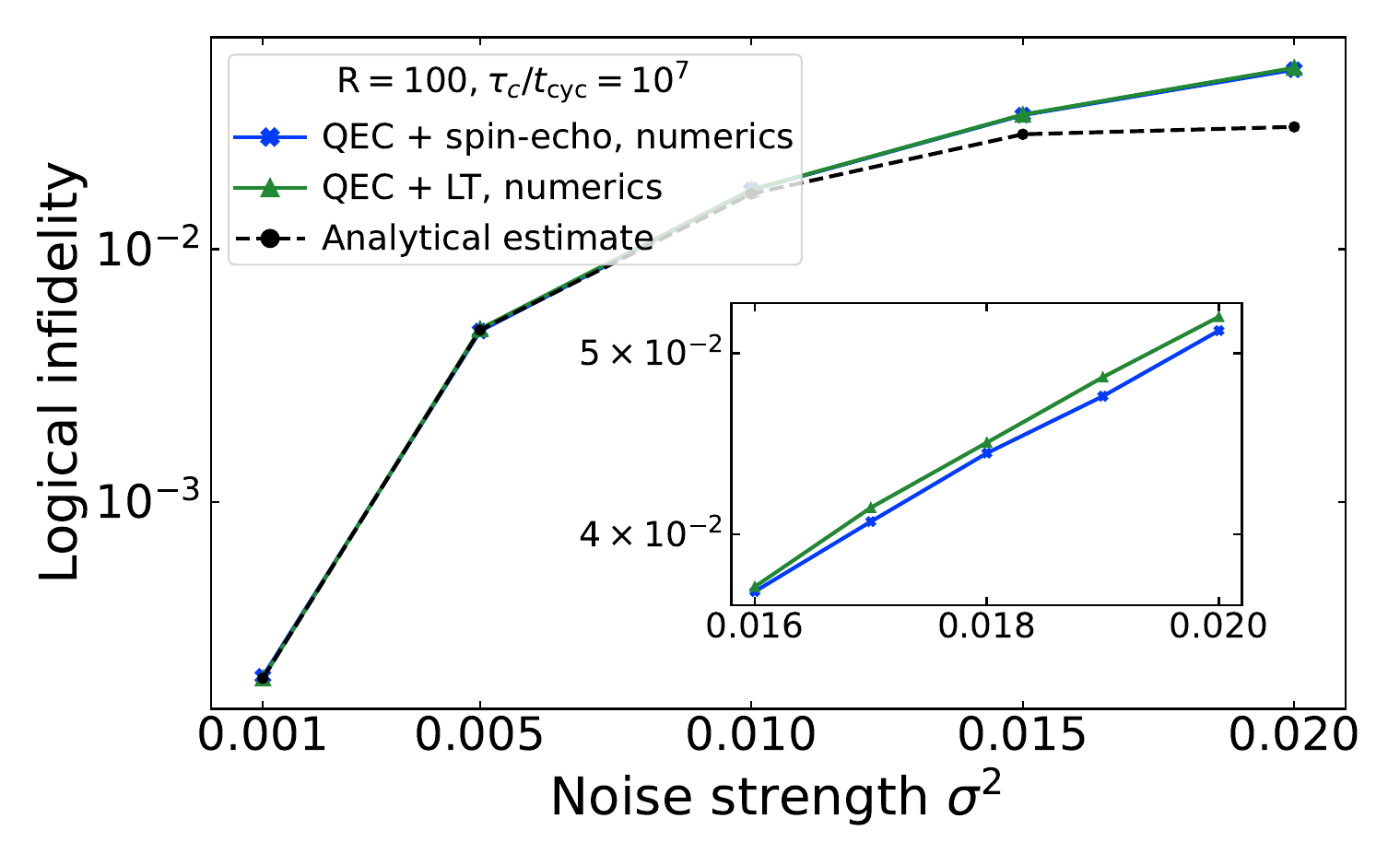}
    \caption{
Near-equivalent performance of QEC+logical spin-echo and QEC+LT for positive, long-ranged inter-cycle correlations. We show the logical infidelity of the three-qubit phase-flip repetition code over $100$ QEC cycles under the noise model of Eq.~\eqref{eq:appendix_lorentzian_noise_correlations}, with $\tau_c/t_{\rm cyc}= 10^7$. Varying the noise strength $\sigma^2$ shows the two infidelities are nearly identical, even beyond the regime where the second-order cumulant approximation, which is the same for both protocols, remains accurate. The inset resolves the large-$\sigma^2$ regime, where QEC+spin-echo yields a modest improvement over QEC+LT. Solid lines show numerical results, obtained by noise-averaging corresponding exact
expressions; error bars are within marker size. Dashed lines show analytical estimates from a second-order cumulant expansion, using
leading-order approximations from Eq.~\eqref{eq:bare_model_cumulants}: $\kappa_1(R)$ and $\kappa_2^\mathrm{coh}(R)$ for QEC-only, and
$\kappa_1(R)$ and $\kappa_2^\mathrm{incoh}(R)$ for QEC+LT.}
    \label{fig:DD_LT_correspondence}
\end{figure}

Figure~\ref{fig:DD_LT_correspondence} confirms this near-optimality of the logical spin-echo sequence for $100$ cycles of the three-qubit repetition code under Eq.~\eqref{eq:appendix_lorentzian_noise_correlations} with $\tau_c/t_\mathrm{cyc}=10^7$. QEC+spin-echo and QEC+LT agree closely, and varying $\sigma$ shows the correspondence persists beyond the regime where the shared second-order cumulant approximation of Eq.~\eqref{eq:app_LDD_approximates_LT} remains accurate. The inset resolves the large-$\sigma$ regime, where logical spin-echo marginally outperforms QEC+LT.

\subsection{Effect of noise correlations after applying LDD}

Now we turn to the effects of noise correlations on $\LIF_\mathrm{LDD}(R)$. For weak noise, $\LIF_\mathrm{LDD}(R)\approx\LIF_\mathrm{LT}(R)$, so noise correlations affect QEC+LDD similar to QEC+LT [see Sec.~\ref{sec:LT_correlation_effects}]. In brief: LDD suppresses the coherent logical errors, so inter-cycle correlations affect $\LIF_\mathrm{LDD}(R)$ only weakly, and their residual effect is beneficial ($\kappa_2^\mathrm{incoh}(R)\geq 0$.) For the conventional encoding, $\LIF_\mathrm{LDD}(R)$ monotonically decreases with the strength and range of inter-cycle correlations.

\begin{figure}
    \centering
    \includegraphics[width=\linewidth]{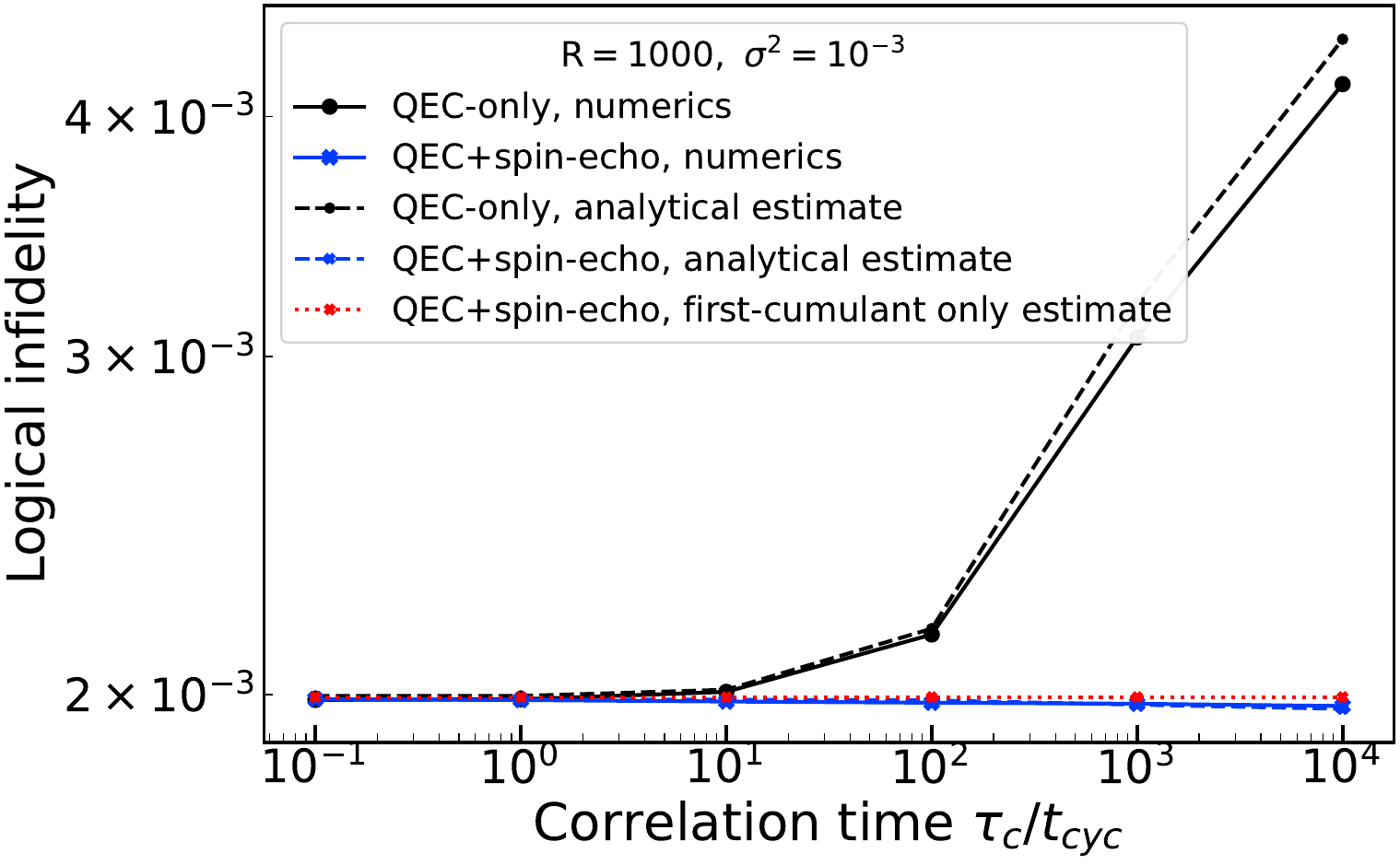}
\caption{%
\textbf{Logical infidelity improvement under QEC+logical spin-echo relative to QEC-only for the three-qubit phase-flip repetition code and positive, long-ranged correlations, as the inter-cycle correlation range is varied.}
We plot the infidelity $\LIF(R)$ of the three-qubit phase-flip repetition code after $R=1000$ QEC cycles for the noise model in Eq.~\eqref{eq:appendix_lorentzian_noise_correlations}, comparing QEC-only (black, circles) and QEC+logical spin-echo (blue, crosses) as the inter-cycle correlation range $\tau_c/t_\mathrm{cyc}$ is varied. Solid lines show numerical results, obtained by noise-averaging corresponding exact
expressions; error bars are within marker size. Dashed lines show analytical estimates from a second-order cumulant expansion, using
leading-order approximations from Eq.~\eqref{eq:bare_model_cumulants}: $\kappa_1(R)$ and $\kappa_2^\mathrm{coh}(R)$ for QEC-only, and
$\kappa_1(R)$, with and without, $\kappa_2^\mathrm{incoh}(R)$ for QEC+LDD.%
}
    \label{fig:DD_rep_3_code}
\end{figure}

Figure~\ref{fig:DD_rep_3_code} illustrates these conclusions using the three-qubit phase-flip repetition code and logical spin-echo, which is near-optimal for $\tau_c\gg t_\mathrm{cyc}$ where coherent logical errors dominate. For $\tau_c\ll t_\mathrm{cyc}$, coherent errors do not accumulate and the two protocols agree; for $\tau_c\gg t_\mathrm{cyc}$, they accumulate and raise the QEC-only infidelity, which logical spin-echo suppresses. In the latter regime, stronger and longer-ranged correlations further reduce the QEC+logical spin-echo infidelity. Comparing the leading-order cumulant estimate with and without the second-cumulant contribution confirms that this reduction stems from the enhanced cancellation of incoherent logical errors across cycles, as in QEC+LT, rather than from logical spin-echo better suppressing coherent errors as one would expect in non-QEC settings.

\section{Physical Pauli twirling}
\label{app:physical_Pauli_twirling}

Logical Pauli twirling removes the logical Pauli off-diagonal coefficient of each single-cycle channel but leaves the diagonal coefficients unchanged. Within these coefficients, the amplitudes of different uncorrectable errors in each syndrome-specific set interfere [Eq.~\eqref{eq:explicit_process_matrix_elements}], which can significantly increase the logical infidelity under strong intra-cycle correlations (e.g., Fig.~\ref{fig:conventional_encoding_random_CSS_code}). Twirling at the level of the physical qubits removes this effect.

\subsection{Protocol outline and computation of $\LIF_\mathrm{PT}(R)$}

In the QEC+PT protocol, a random $n$-qubit Pauli gate $\hat P^{(r)}\in\overline{\mathbb P}_n$ is inserted before the noise layer in each cycle $r$ and applied again after the correction step, with the correction step necessarily modified so that the QEC gadget itself acts as before. These gates can be absorbed into state preparation and the correction steps, and thus carry no additional overhead.

The effect of this randomization can be clarified by decomposing each Pauli gate as~\cite{pelchat_degenerate_2012}
\begin{align}
    \hat P=\hat D^{\vec{g}'}\hat{P}_L\hat S,
    \label{eq:decomposition_rule_n_qubit_Paulis}
\end{align}
where $\hat S\in\mathbb S$, $\hat P_L\in\{\hat I,\hat{\bar X}_L,\logicalZ,i\hat{\bar X}_L\logicalZ\}$ is a logical Pauli \footnote{If the logical states are defined such that the logical Paulis are not physical Paulis, then the $\hat{P}_L$ factor in the decomposition is from a set of four physical Paulis, including $\hat{I}$, that preserve the codespace and have the same commutation relations as the logical Paulis.}, $\vec g'\!=\!(g_1',\ldots,g_{n-1}')$ with $g_i'\!\in\!\{0,1\}$, and $\hat D^{\vec{g}'}=\prod_{i=1}^{n-1} \hat D_i^{g_i'}$, with $n$-qubit Paulis $\hat D_i$ satisfying
\begin{equation}
\begin{aligned}
    \{\hat D_i,\hat g_i\}&=0,\\
    [\hat D_i,\hat g_j]&=0,\qquad j\neq i,\\
    [\hat D_i,\hat{P}_L]&=0,\qquad
    \hat{P}_L\in\{\hat I,\hat{\bar X}_L,\logicalZ,
    i\hat{\bar X}_L\logicalZ\}.
\end{aligned}
\end{equation}
The three factors act distinctly on data qubits supported on the codespace $\mathcal C_{\vec g}$: $\hat D^{\vec g'}$ maps them to $\mathcal C_{\vec g\oplus\vec g'}$ while preserving the logical information, $\hat P_L$ changes the logical information while preserving the codespace, and $\hat S$ contributes only an irrelevant global phase. The correction step must then be updated to use $\vec{g}\oplus\vec{g}'$ as the reference syndrome. After the correction step, the second Pauli layer returns the data qubits to $\mathcal C_{\vec g}$. Thus, this evolution is codespace-preserving. The channel induced on the codespace for a fixed noise realization and Pauli layer is
\begin{align}
\Lambda(\anglevec{r},\hat{P}^{(r)})
=
\mathcal{P}_L
\Big(
&\chi^{\mathrm{II}}(\anglevec{r},\vec g')\mathcal I
+\chi^{\mathrm{ZI},\mathrm{im}}(\anglevec{r},\vec g')\logicalZLZ\notag\\
&+\chi^{\mathrm{ZZ}}(\anglevec{r},\vec g')\bar{\mathcal Z}_L
\Big)
\mathcal{P}_L ,
\label{eq:PT_single_cycle_for_one_layer}
\end{align}
with $\mathcal P_L[\cdot]=\hat P_L[\cdot]\hat P_L$ and coefficients as in Eq.~\eqref{eq:explicit_process_matrix_elements}, but with the encoding-dependent phase factors evaluated on $\mathcal C_{\vec g\oplus\vec g'}$.

Each cycle thus conjugates the single-cycle channel by a random logical Pauli while shifting the data qubits to a random stabilizer eigenspace, both determined by the applied Pauli gate. Moreover, the decomposition in Eq.~\eqref{eq:decomposition_rule_n_qubit_Paulis} is a bijection between $\overline{\mathbb P}_n$ and the triples $(\vec g',\hat P_L,\hat S)$. Concretely, the $4^n$ Pauli gates partition into $2^{n-1}$ equally sized classes labeled by $\vec g'$, within each of which $\hat P_L\hat S$ ranges once over all $4\times 2^{n-1}$ possibilities. A uniformly random $\hat P^{(r)}\in\overline{\mathbb P}_n$ therefore corresponds to independent, uniformly random $\vec g'$, $\hat P_L$, and $\hat S$. This implies that PT is equivalent to uniformly randomizing the encoding eigenspace in each cycle while conjugating each cycle by an independent, uniformly random logical Pauli. This is the result most relevant to the main text.

To quantify the effect of this randomization, we measure performance after averaging over first the applied random Pauli layers and then, over the noise. Since the Pauli layers are independent across cycles, each single-cycle channel may be averaged over its layer independently. This average further factorizes over $\hat D^{\vec g'}$, $\hat P_L$, and $\hat S$, of which only the first two are nontrivial. 

The $\hat P_L$ average removes the logical Pauli off-diagonal coefficient, as in LT. The $\vec g'$ average then modifies the logical Pauli-diagonal coefficients, which depend on the eigenspace only through the phases $(-1)^{\phi_{\hat E\hat E'}(\vec g')}$ present in terms involving distinct errors $\hat E\neq\hat E'$ from the same syndrome-specific set [Eq.~\eqref{eq:first_cumulant_opened_up}]. For $\hat{E}\neq\hat{E}'$, the decomposition of $\hat E\hat E'$ according to Eq.~\eqref{eq:decomposition_rule} involves a nontrivial $Z$ stabilizer, with $(-1)^{\phi_{\hat E\hat E'}(\vec g')}$ as its eigenvalue in the codespace $\mathcal{C}_{\vec{g}\oplus\vec{g}'}$. This eigenvalue averages to zero over the eigenspaces, and thus all cross terms between different errors are removed. This leaves the physical Pauli-twirled channel for a single noise realization as:
\begin{equation}
\Lambda_{\mathrm{PT}}(\anglevec{r})
=
\chi_{\mathrm{PT}}^{\mathrm{II}}(\anglevec{r})\mathcal I
+
\chi_{\mathrm{PT}}^{\mathrm{ZZ}}(\anglevec{r})\bar{\mathcal Z}_L,
\label{eq:single_cyc_channel_PT_after_Pauli_avg}
\end{equation}
with
\begin{subequations}\label{eq:channel_coefficients_PT}
\begin{align}
\chi_{\mathrm{PT}}^{\mathrm{II}}(\anglevec{r})
&=\sum_{\hat E\in\overline{\mathbb P}_{n,Z}^{\checkmark}}
\abs{c_{\hat E}(\anglevec{r})}^2,
\\
\chi_{\mathrm{PT}}^{\mathrm{ZZ}}(\anglevec{r})
&=\sum_{\hat E\in\overline{\mathbb P}_{n,Z}^{\times}}
\abs{c_{\hat E}(\anglevec{r})}^2 ,
\end{align}
\end{subequations}
where $\overline{\mathbb P}_{n,Z}^{\checkmark}=\bigcup_{\vec s}\overline{\mathbb P}_{n,Z}^{\vec s,\checkmark}$ and $\overline{\mathbb P}_{n,Z}^{\times}=\bigcup_{\vec s}\overline{\mathbb P}_{n,Z}^{\vec s,\times}$ are the sets of all correctable and uncorrectable $Z$ errors. This channel is logical Pauli diagonal and independent of the encoding eigenspace. 

Equivalently, it is the channel induced by the QEC gadget with the noise superoperator of Eq.~\eqref{eq:noise_single_cycle} replaced by its twirled counterpart,
\begin{align}
    \widetilde{\mathcal N}\!\big(\anglevec{r}\big)[\cdot]
    &=\frac{1}{4^n}\sum_{\mathcal P^{(r)}\in\overline{\mathbb P}_n}
    \mathcal P^{(r)}
    \Big(\prod_{\ell=1}^{n}\mathcal R_{Z_\ell}\!\big(\theta_{\ell}^{(r)}\big)\Big)
    \mathcal P^{(r)}[\cdot] \notag\\
    &=\sum_{\hat E\in\overline{\mathbb P}_{n,Z}}
    \abs{c_{\hat E}(\anglevec{r})}^2\hat E[\cdot]\hat E^\dagger.
\end{align}
As expected, PT thus converts coherent $Z$ errors on the physical qubits into stochastic Pauli $Z$ errors with probabilities $|c_{\hat E}(\anglevec{r})|^2$. The coefficients in Eq.~\eqref{eq:channel_coefficients_PT} can thus be interpreted as the total probabilities of correctable and uncorrectable errors in cycle $r$. Composing the single-cycle channels and averaging over the noise allows us to compute the physical Pauli- and noise-averaged infidelity,
\begin{equation}
    \LIF_\mathrm{PT}(R)=\frac{1-\avg{\Gamma_\mathrm{PT}(R,\vec{\theta}\,) }}{3},
\end{equation}
where 
\begin{equation}
    \Gamma_\mathrm{PT}(R,\vec{\theta}\,) =
\prod_{r=1}^R\Big(
1-2\chi^\text{ZZ}_\mathrm{PT}(\anglevec{r})
\Big).
\label{eq:Gamma_PT}
\end{equation}

\subsection{Comparison with conventional QEC-only, QEC+LT, and QEC+LDD}

With $\LIF_\mathrm{PT}(R)$ in hand, we now compare with QEC-only, QEC+LT, and QEC+LDD to determine when applying PT provides an advantage. Since $\LIF_\mathrm{PT}(R)$ is independent of the encoding eigenspace while the QEC-only infidelity is not, we compare against the conventional QEC-only protocol (all stabilizers $+1$).

The result of this comparison is that in the weak-noise regime with positive noise correlations, QEC+PT performs at least as well as, and generally better than, conventional QEC-only, up to corrections of order $\mathcal{O}(R\sigma^{d_Z+3})$ and $\mathcal{O}(RR_c\sigma^{2d_Z+2})$ to the corresponding cumulants. QEC+PT also outperforms QEC+LT and QEC+LDD with the conventional encoding in the same setting. We emphasize that this advantage is relative to the conventional encoding only: applying PROSE encoding results in an improvement over PT [see Sec.~\ref{sec:comparison}].

We now discuss the mechanism underlying the advantage of QEC+PT over conventional QEC-only. The comparisons with QEC+LT and QEC+LDD under the same encoding follow along the same lines. A second-order cumulant expansion of $\LIF_\mathrm{PT}(R)$ yields
\begin{equation}
    \LIF_\mathrm{PT}(R)\approx\frac{1-\exp\!\big(\kappa_\mathrm{1,PT}(R)+\kappa_\mathrm{2,PT}(R)/2\big)}{3},
    \label{eq:second_order_cumulant_PT}
\end{equation}
with cumulants
\begin{subequations}\label{eq:PT_cumulants}
\begin{align}
    \kappa_\mathrm{1,PT}(R)
    &=-2\sum_{r=1}^R\avg{\chi_\mathrm{PT}^\text{ZZ}(\anglevec{r})},
    \label{eq:first_cumulant_PT}\\
    \kappa_\mathrm{2,PT}(R)
    &=4\bigavg{\Big(\sum_{r=1}^R\chi_\mathrm{PT}^\text{ZZ}(\anglevec{r})\Big)^2}_c.
    \label{eq:second_cumulant_PT}
\end{align}
\end{subequations}
For sufficiently weak noise, these cumulants can be approximated to leading order in the noise strength as
\begin{align}
    \kappa_{\mathrm{1,PT}}(R)&
    \!\approx\!
                    -2\sum_{r=1}^R\!\!\!\!\!\!\!
                        \sum_{\substack{\hat{E}\in\overline{\mathbb{P}}_{n,Z}^\times:\\w(\hat{E})=(d_Z+1)/2}}
                        \!\!\!\!\Bigavg{\!\!\!\!
                        \prod_{\ell\in\mathrm{supp}(\hat{E})}\!\!\!\!\!\!\!\big(\rotangle{r}{\ell}\big)^2},\label{eq:PT_leading_order_cumulants}\\
    \kappa_{\mathrm{2,PT}}(R)&
    \!\approx 4\!\!\!
        \sum_{r,r'=1}^R\!\!\!\!\!\!\!\!\!\!\!\!
\sum_{\substack{\hat{E},\hat{E}'\in\overline{\mathbb{P}}_{n,Z}^\times\\w(\hat{E}),w(\hat{E}')=(d_Z+1)/2}}\!\!\!\!\!\!\!\!\!\!\!\!
        \Bigavg{\!\!\!\!\prod_{\ell\in\mathrm{supp}(\hat{E})}\!\!\!\!\!\!\!\big(\rotangle{r}{\ell}\big)^2\!\!\!\!\!\!\!\prod_{\ell'\in\mathrm{supp}(\hat{E}')}\!\!\!\!\!\!\!\big(\rotangle{r'}{\ell'}\big)^2}_c,\notag
\end{align}
which scale as $R\sigma^{d_Z+1}$ and $RR_c\sigma^{2d_Z+2}$, respectively.

Using this approximation of $\LIF_\mathrm{PT}(R)$ and the corresponding approximation of $\LIF(R)$, we obtain
\begin{align}
    \LIF(R)-\LIF_\mathrm{PT}(R)
    &\approx
    \frac{\exp\big(\kappa_\mathrm{1}(R)+\kappa_\mathrm{2}(R)/2\big)}{3}\label{eq:LIF_difference_PT} \\
    &\quad\times
    \Big(\exp\big(\Delta\kappa_1(R)+\Delta\kappa_2(R)/2\big)-1\Big),\notag
\end{align}
with $\Delta\kappa_\alpha(R)=\kappa_{\alpha,\mathrm{PT}}(R)-\kappa_{\alpha}(R)$. The sign of $\Delta\kappa_1(R)+\Delta\kappa_2(R)/2$ controls the comparison. The two cumulants scale differently with $\sigma$, $R$, and $R_c$, and either may dominate, since for the QEC-only case the second-order approximation remains accurate even when $\kappa_2(R)$ dominates [App.~\ref{app:accuracy_of_the_second_order_cumulant_expansion}]. A conclusion valid across all regimes where this expansion remains accurate therefore requires comparing each difference separately.

We first examine the second cumulant difference. Since $\kappa_{\mathrm{2,PT}}(R)$ is of the same order as the terms neglected in the leading-order approximation to $\kappa_2(R)$,
\begin{equation}
    \Delta\kappa_2(R)\approx-\kappa_\mathrm{2}^\text{coh}(R)\geq0,
\end{equation}
with $\kappa_\mathrm{2}^\text{coh}(R)$ as defined in Eq.~\eqref{eq:bare_model_second_cumulant_coh}. This difference captures the accumulation of coherent logical errors in the QEC-only protocol, which PT removes. The first cumulant difference,
\begin{align}
\Delta\kappa_{1}(R)
&= \!2\sum_{r=1}^R\!\!\!\!\!\!\!\!\!\!\!\!\!\!\!\!
\sum_{\substack{\vec{s},\\\hat E\neq\hat E'\in\overline{\mathbb{P}}_{n,Z}^{\vec s,\times}:\\
w(\hat E),w(\hat E')=(d_Z+1)/2}}
\!\!\!\!\!\!\!\!\!\!\!\!\!\!\!\!
\Bigavg{
\Big(\!\!\!\!\prod_{\ell\in\mathrm{supp}(\hat E)}\!\!\!\!\rotangle{r}{\ell}\Big)
\Big(\!\!\!\!\prod_{\ell'\in\mathrm{supp}(\hat E')}\!\!\!\!\rotangle{r}{\ell'}\Big)},
\label{eq:cumulant_diffs_PT}
\end{align}
collects the terms in $\kappa_1(R)$ involving pairs of distinct dominant uncorrectable errors, which PT removes. For positive intra-cycle correlations, $\Delta\kappa_{1}(R)$ reduces to a polynomial in the entries of $\boldsymbol{\Sigma}$ with non-negative coefficients, so $\Delta\kappa_1(R)\geq0$. With both differences non-negative, Eq.~\eqref{eq:LIF_difference_PT} gives $\LIF(R)\gtrsim\LIF_\mathrm{PT}(R)$, establishing the advantage of applying PT in this common setting.

\subsection{Effect of noise correlations after applying PT}

\begin{figure}
    \centering
    \includegraphics[width=\linewidth]{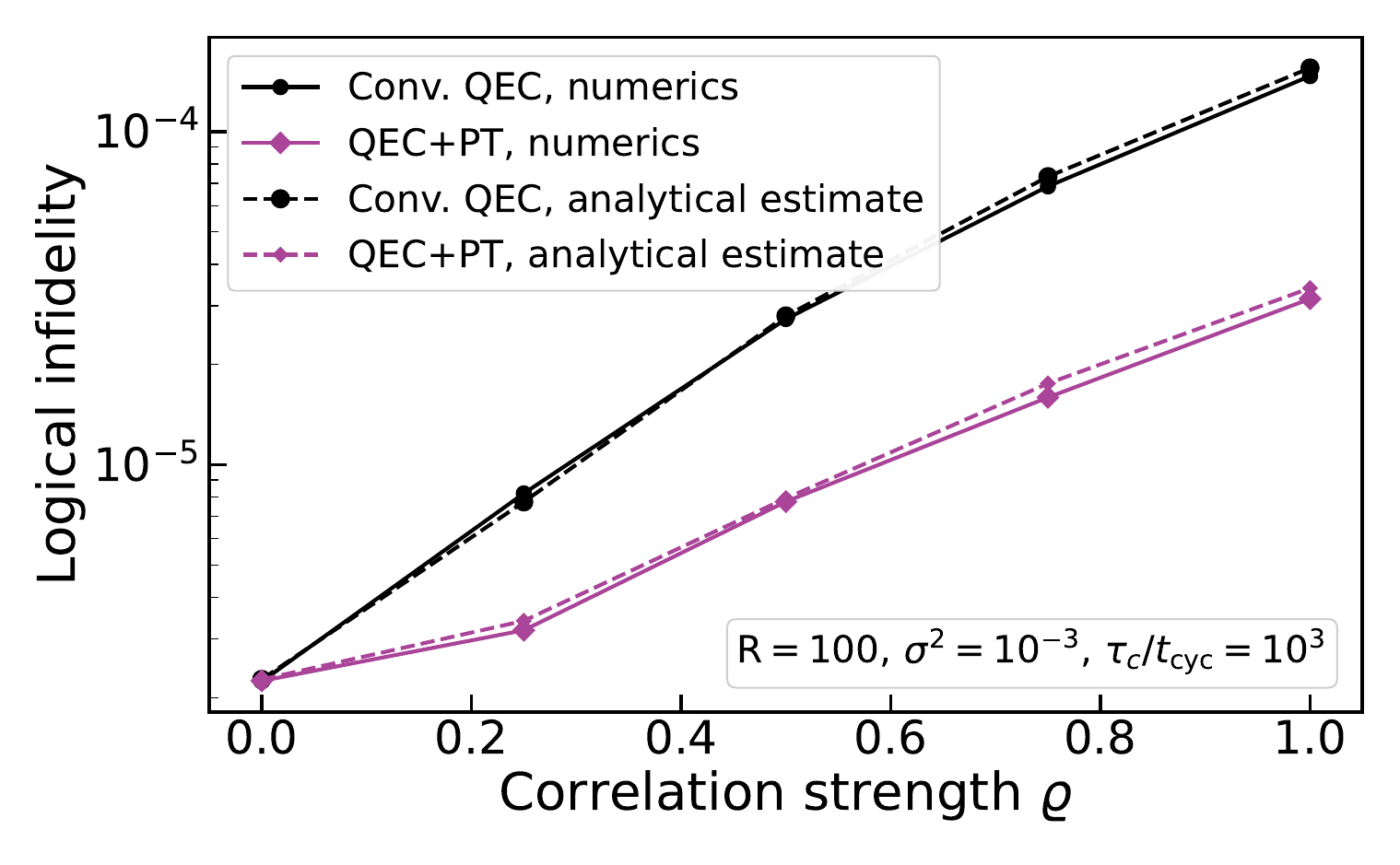}
\caption{%
\textbf{Logical infidelity of an $\llbracket 9,1\rrbracket$ CSS code under QEC-only and QEC+PT, as the inter-qubit correlation strength is varied, showing the advantage of applying PT under positive noise correlations.}
We plot the infidelity $\LIF(R)$ after $R=100$ cycles, comparing QEC-only (black, circles) and QEC+PT (pink, squares) for one of the CSS codes from Fig.~\ref{fig:conventional_encoding_random_CSS_code} and the same noise model. Solid lines show numerical results, obtained by noise-averaging exact expressions; error bars are within marker size. Dashed lines show analytical estimates from a second-order cumulant expansion, using the leading-order approximations from Eq.~\eqref{eq:bare_model_cumulants} for QEC-only and Eq.~\eqref{eq:PT_leading_order_cumulants} for QEC+PT.%
}
    \label{fig:PT_CSS_Code}
\end{figure}

The second-order cumulant expansion of $\LIF_\mathrm{PT}(R)$ also reveals how applying PT reshapes noise correlation effects. At leading-order in the noise strength, the effect of intra-cycle correlations is captured by $\kappa_{1,\mathrm{PT}}(R)$ [Eq.~\eqref{eq:PT_leading_order_cumulants}]. Comparing against the uncorrelated counterpart $\boldsymbol{\overline \Sigma}^{(r,r')}_{\ell\ell'}=\delta_{rr'}\delta_{\ell\ell'}\boldsymbol{\Sigma}_{\ell\ell}^{(r,r)}$ (corresponding to independent stochastic Pauli errors), the Gaussian product inequality \cite{frenkel_pfaffians_nodate,russell_new_2022} for squares of zero-mean jointly Gaussian variables implies
\begin{equation}
    \Bigavg{\!\!\!\!
                        \prod_{\ell\in\mathrm{supp}(\hat{E})}\!\!\!\!\!\!\!\big(\rotangle{r}{\ell}\big)^2}\geq\!\prod_{\ell\in\mathrm{supp}(\hat{E})}\!\!\!\!\Bigavg{\big(\rotangle{r}{\ell}\big)^2}.
                        \label{eq:Gaussian_product_inequality}
\end{equation}
Intra-cycle correlations therefore increase the magnitude of $\kappa_\mathrm{1,PT}(R)$, and hence $\LIF_\mathrm{PT}(R)$, at leading-order; for positive correlations, this increase is monotonic in the correlation strength. Applying PT thus partly suppresses intra-cycle correlation effects by removing the interference between error amplitudes, but also rules out exploiting these correlations as a resource to reduce the infidelity below the uncorrelated limit — which PROSE encoding allows [see Fig.~\ref{fig:encoding_eigenspace_can_make_some_intra_cycle_correlations_a_resource}].

Inter-cycle correlations enter at leading-order through $\kappa_{2,\mathrm{PT}}(R)$. Their effect is suppressed, but is, in contrast to the intra-cycle case, beneficial. As in QEC+LDD and QEC+LT, due to the suppression of coherent logical errors, inter-cycle correlations negligibly affect $\LIF_\mathrm{PT}(R)$ relative to $\LIF(R)$, and its effect is reversed in sign: these correlations reduce the logical infidelity relative to the Markovianized noise model [Eq.~\eqref{eq:markovianization}] at the leading-order. Under positive noise correlations, this decrease is monotonic in the correlation strength and range. Inter-cycle correlations are thus, as after applying LDD or LT, a weakly beneficial resource.

Fig.~\ref{fig:PT_CSS_Code} illustrates the above conclusions by comparing QEC+PT against conventional QEC-only for one of the CSS codes of Fig.~\ref{fig:conventional_encoding_random_CSS_code} under the noise model of Eq.~\eqref{eq:lorentzian_noise_correlations}. At $\varrho=0$, the cross terms between distinct errors in $\kappa_1(R)$ — precisely the terms PT removes — vanish on noise averaging, so the difference between $\LIF(R)$ and $\LIF_\mathrm{PT}(R)$ arises only from PT suppressing coherent logical errors. For the parameters shown, these errors contributed negligibly to QEC-only and PT gives only a small improvement for $\varrho=0$. As $\varrho$ increases, positive intra-cycle correlations significantly increase $\LIF(R)$. They affect $\LIF_\mathrm{PT}(R)$ less since PT removes the interference between the amplitudes of different errors, but still increase it monotonically.

\section{$\llbracket9,1\rrbracket$ CSS Code used in Fig.~\ref{fig:conventional_encoding_random_CSS_code}}
\label{app:numerical_details}

The three random $\llbracket 9,1\rrbracket$ CSS codes used in Fig.~\ref{fig:conventional_encoding_random_CSS_code} all have $d_Z=5$ and $d_X=1$. We specify their check matrices in the standard binary representation over $\mathbb F_2$, where a row $h=(h_1,\ldots,h_9)$ of $H_X$ denotes the stabilizer generator $\prod_{j:h_j=1}\hat X_j$, and similarly for $H_Z$. 

Code-$1$, re-used in Figs.~\ref{fig:encoding_eigenspace_search} and \ref{fig:PT_CSS_Code}, is given by:
\begin{equation}
\resizebox{0.88\columnwidth}{!}{$
H_X^{(1)}\!\!=\!\!\left(\begin{smallmatrix}
0&1&0&0&0&1&1&0&1\\
1&1&0&0&0&1&1&1&1\\
1&0&1&1&1&1&1&1&0\\
0&0&1&1&1&0&0&1&1\\
0&1&0&0&0&1&1&0&0
\end{smallmatrix}\right)\!\!,
H_Z^{(1)}\!\!=\!\!\left(\begin{smallmatrix}
0&0&1&0&1&0&0&0&0\\
0&0&1&0&1&1&1&0&0\\
0&0&0&1&1&1&1&0&0
\end{smallmatrix}\right).
$}
\end{equation}
Codes $2$ and $3$ are specified by:
\begin{equation}
\resizebox{0.88\columnwidth}{!}{$
H_X^{(2)}\!\!=\!\!\left(\begin{smallmatrix}
1&1&1&0&1&0&0&1&0\\
0&0&1&0&1&0&0&1&0\\
0&1&1&0&1&0&1&0&0\\
1&1&0&0&1&0&1&1&0\\
0&1&1&0&1&1&1&1&1
\end{smallmatrix}\right)\!\!,
H_Z^{(2)}\!\!=\!\!\left(\begin{smallmatrix}
0&0&1&1&0&1&1&1&0\\
0&0&0&1&0&1&0&0&1\\
0&0&0&0&0&1&0&0&1
\end{smallmatrix}\right).
$}
\end{equation}
\begin{equation}
\resizebox{0.88\columnwidth}{!}{$
H_X^{(3)}\!\!=\!\!\left(\begin{smallmatrix}
1&0&0&0&0&1&0&0&1\\
1&0&1&1&0&1&0&1&1\\
0&0&1&0&0&1&0&0&0\\
1&0&0&0&0&0&1&0&1\\
1&0&1&1&0&1&1&0&1
\end{smallmatrix}\right)\!\!,
H_Z^{(3)}\!\!=\!\!\left(\begin{smallmatrix}
0&1&0&0&0&0&0&0&0\\
0&0&0&0&1&0&0&0&0\\
1&1&0&0&1&0&0&0&1
\end{smallmatrix}\right).
$}
\label{eq:appendix_Hz_css_checks}
\end{equation}

\bibliography{references}
\end{document}